\pdfoutput=1
 \documentclass[
	aps, pra, superscriptaddress, twocolumn,
	10pt
	floatfix, 
    nofootinbib,
	tightenlines
]{revtex4-1}
\usepackage[final]{graphicx}
\usepackage{times,bbm,amsmath,amssymb}
\usepackage{epsfig,color}
\usepackage{xcolor}
\usepackage{hyperref}
\hypersetup{
    colorlinks = true
}
\usepackage{cleveref}
\usepackage{microtype}

\usepackage{float,siunitx}
\usepackage[caption = false]{subfig}

\usepackage[greek,english]{babel}
\usepackage{thumbpdf,enumerate}
\usepackage{booktabs}
\usepackage{sidecap}
\usepackage[scaled=.8]{couriers}
\usepackage{multirow}
\usepackage{placeins}
\usepackage{relsize}
\usepackage{pst-grad,bm}
\usepackage{epigraph}
\usepackage{gensymb}
\usepackage{longtable}
\usepackage{ulem} 
\usepackage{acronym}
\usepackage{physics}

\usepackage{easyReview}

\DeclareUnicodeCharacter{0301}{\'{e}}

\begin{document}

\address{Dipartimento di Fisica, Sapienza Universit\`{a} di Roma, Piazzale Aldo Moro 5, I-00185 Roma, Italy}

\vspace{10pt}

\title{Generation and characterization of polarization-entangled states \\ using quantum dot single-photon sources}

\author{Mauro Valeri}
\address{Dipartimento di Fisica, Sapienza Universit\`{a} di Roma, Piazzale Aldo Moro 5, I-00185 Roma, Italy}

\author{Paolo Barigelli}
\address{Dipartimento di Fisica, Sapienza Universit\`{a} di Roma, Piazzale Aldo Moro 5, I-00185 Roma, Italy}

\author{Beatrice Polacchi}
\address{Dipartimento di Fisica, Sapienza Universit\`{a} di Roma, Piazzale Aldo Moro 5, I-00185 Roma, Italy}

\author{Giovanni Rodari}
\address{Dipartimento di Fisica, Sapienza Universit\`{a} di Roma, Piazzale Aldo Moro 5, I-00185 Roma, Italy}

\author{Gianluca De Santis}
\address{Dipartimento di Fisica, Sapienza Universit\`{a} di Roma, Piazzale Aldo Moro 5, I-00185 Roma, Italy}

\author{Taira Giordani}
\address{Dipartimento di Fisica, Sapienza Universit\`{a} di Roma, Piazzale Aldo Moro 5, I-00185 Roma, Italy}

\author{Gonzalo Carvacho}
\email{gonzalo.carvacho@uniroma1.it}
\address{Dipartimento di Fisica, Sapienza Universit\`{a} di Roma, Piazzale Aldo Moro 5, I-00185 Roma, Italy}

\author{Nicol\`o Spagnolo}
\address{Dipartimento di Fisica, Sapienza Universit\`{a} di Roma, Piazzale Aldo Moro 5, I-00185 Roma, Italy}

\author{Fabio Sciarrino}
\address{Dipartimento di Fisica, Sapienza Universit\`{a} di Roma, Piazzale Aldo Moro 5, I-00185 Roma, Italy}

\begin{abstract}
Single-photon sources based on semiconductor quantum dots find several applications in quantum information processing due to their high single-photon indistinguishability, on-demand generation, and low multiphoton emission. In this context, the generation of entangled photons represents a challenging task with a possible solution relying on the interference in probabilistic gates of identical photons emitted at different pulses from the same source. In this work, we implement this approach via a simple and compact design that generates entangled photon pairs in the polarization degree of freedom.
We operate the proposed platform with single photons produced through two different pumping schemes, the resonant excited one and the longitudinal-acoustic phonon-assisted configuration. 
We then characterize the produced entangled two-photon states by developing a complete model taking into account relevant experimental parameters, such as the second-order correlation function and the Hong-Ou-Mandel visibility. 
Our source shows long-term stability and high quality of the generated entangled states, thus constituting a reliable building block for optical quantum technologies.
\end{abstract}

\maketitle

\section{Introduction}
\label{sec:1}

The generation of entangled states of light is fundamental to several quantum information applications ranging from quantum communication \cite{vajner2022quantum,PhysRevA.76.012307,doi:10.1126/sciadv.abe8905}, quantum sensing and metrology \cite{giovannetti2011advances,PhysRevA.94.012101}, quantum networks \cite{Carvacho:22,lu2021quantum,Lodahl_2018,PhysRevApplied.19.064083} and quantum computing \cite{doi:10.1126/science.270.5234.255,o2007optical,PhysRevA.57.120}.
Spontaneous parametric down-conversion (SPDC)-based sources can be employed to obtain high-fidelity entangled photons. However, this technology presents a trade-off between the brightness and quality of the output states. Indeed, due to the probabilistic nature of the process, increasing source brightness is inherently accompanied by higher multiphoton terms, and thus the generation rate must be kept below a certain threshold; such limitation can be overcome by exploiting quantum dot (QD) sources \cite{wang2019towards,senellart}. In the last twenty years, QD single-photon sources (SPSs) have been widely exploited for secure quantum communication \cite{waks2002quantum,Bozzio2022} due to their simultaneous characteristics of high purity, brightness and single-photon indistinguishability. Moreover, QD-based SPSs represent a favorable approach for measured-based quantum computing \cite{zwerger2016measurement}, optical quantum networks \cite{lu2021quantum}, generation of photonic cluster states \cite{lindner2009proposal} and high-performance Boson Sampling \cite{wang2017high}. 
In particular, it has been recently demonstrated that SPSs provide a significant improvement in quantum key distribution (QKD) schemes due to their low multi-photon contribution, thus minimizing information leakage to malicious eavesdroppers \cite{brassard2000limitations}. For this reason, QD sources have been mainly adopted in standard BB84 protocols \cite{waks2002quantum,rau2014free,takemoto2015quantum}, representing a natural solution to multiphoton-based attack as well as QKD protocols based on entangled photon sources (EPS) \cite{basso2021quantum,schimpf2021quantum,basso2022daylight}. 

Furthermore, a crucial element for the development of multi-partite protocols \cite{PhysRevA.59.1829,doi:10.1126/sciadv.abe0395} is represented by Greenberger-Horne-Zeilinger (GHZ) states \cite{Greenberger1989}. Their generation has been achieved in different experimental setups up to a recent record of 12 photons \cite{zhong201812} using SPDC. The combination of QD sources and GHZ states is promising for the scaling-up of complex quantum networks and, thus, particular attention must be devoted to this research line. As of today, GHZ states have been realized by using a semiconductor QD and integrated photonics to demonstrate scalable heralded controlled-NOT operations \cite{PhysRevLett.126.140501, pont2022highfidelity}.

Entanglement generation by means of QD-based SPSs has been investigated by employing different degrees of freedom and configurations, such as structured photons \cite{suprano2022orbital}, polarization-encoded GHZ states based on temporal-delay fiber loops \cite{istrati2020sequential,li2020multiphoton,cogan}, spin-photon entanglement \cite{de2012quantum,schwartz2016deterministic}, and time-distributed photon-number entangled states \cite{wein2022photon}. Designing QDs enabled by null fine structure splitting (FSS) to directly generate pair of entangled photons via biexciton-exciton radiative cascade \cite{basso2021quantum,wang2019demand,liu2019solid} represents one of the most common applications and efficient methods. However, the quality of the entangled states is degraded when FSS is present. Therefore, the development of alternative schemes to generate entanglement for this case is pivotal for further extending the study of QD-based protocols. Nowadays, the generation of entangled states exploiting non-entangled emission by QDs has been demonstrated in a few works \cite{Fattal_bellpol_dot} and represents a practical possibility to scale QD-based GHZ states \cite{li2020multiphoton}. 
The single-photon emission by exciton-based QD can be achieved in two different regimes, the first one with the pumping laser being resonant with the exciton energy level and the non-resonant excitation \cite{senellart2017high}.
On one hand, the main advantage of non-resonant excitation resides in the fact that single photons emitted can be easily separated from the excitation laser through spectral filtering. On the other hand, resonant fluorescence (RF) can be adopted to improve single-photon indistinguishability, eliminating dephasing and time jitter \cite{he2013demand}, at the price of more challenging suppression of the residual pump laser light. Also, due to the cross-polarization filtering procedure of the residual pump, the Firs Lens Brightness is at least halved \cite{wang2019towards}.

\begin{figure*}[ht!]
    \centering
    \includegraphics[width=\textwidth]{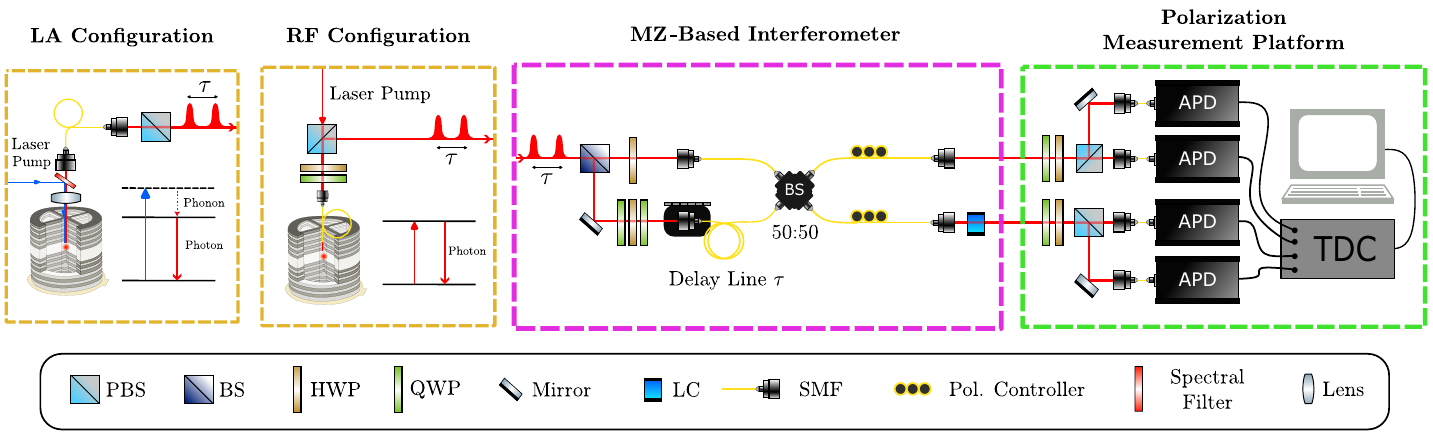}
    \caption{Schematic representation of the experimental setup. Streams of single photons are generated by the QD in LA or RF configuration (yellow box); a polarizing beam-splitter (PBS) is used to transmit only horizontal polarized photons. The MZ-Based interferometer (Purple box) is composed of two beam splitters (BS) and a delay line to probabilistically allow the interference between a wavepacket emitted at time $t$ with the consecutive emitted one at $(t+\tau)$. Furthermore, a set of half-wave plates (HWP) and quarter-wave plates (QWP) are placed along the internal arms in order to prepare orthogonal polarization states interfering at the second BS. At the output of the MZI, the combination of manual polarization paddles (Pol. controller) and a liquid crystal (LC) are used to prepare one of the four possible Bell states. Pair of entangled photons are detected by post-selecting the events in which two photons exit from different outputs of the MZI. To record coincidence events we employ standard polarization measurements (Green box) composed by HWP, QWP, single-mode fibers (SMF), single photon detectors (APDs) and a time-to-digital converter (TDC) used to analyse the detectors' signals.}
    \label{fig:source}
\end{figure*}

In this work, we design and fully characterize an optical scheme to generate polarization-entangled pairs of photons using a semiconductor quantum dot device emitting highly indistinguishable single-photon streams. Two quantum dots were considered for the study: one coherently controlled via RF excitation, and the other operated under non-resonant excitation assisted by longitudinal-acoustic phonons (LA) \cite{senellart2017high}. In our design, the entangled pairs are obtained by employing interference between two photons generated in two consecutive pulses by the QD source, and, therefore, the overall quality of the generated states depends on the indistinguishability of the exploited photons. This feature has been extensively discussed in terms of Hong-Ou-Mandel (HOM) effect and multiphoton contributions \cite{ollivier2021hong}. The resulting performance of the polarization-entangled source is quantified by directly measuring the brightness and through a CHSH test \cite{clauser1969proposed}
for both QD excitation schemes. 

Finally, to determine the quality of the entanglement that can be generated, we present a model involving several experimental parameters of our system. We tested the time-stability and the robustness of the generated entangled states while these parameters varied. In this way, we identified the quantities which mainly affect the quality of the generated entangled state, i.e. the degree of indistinguishability between consecutive emissions and the multi-photon component of the QD source itself. 

\section{Experimental platform}
\label{sec:exp_setup}

\begin{figure*}[ht!]
    \centering
    \includegraphics[width= \textwidth]{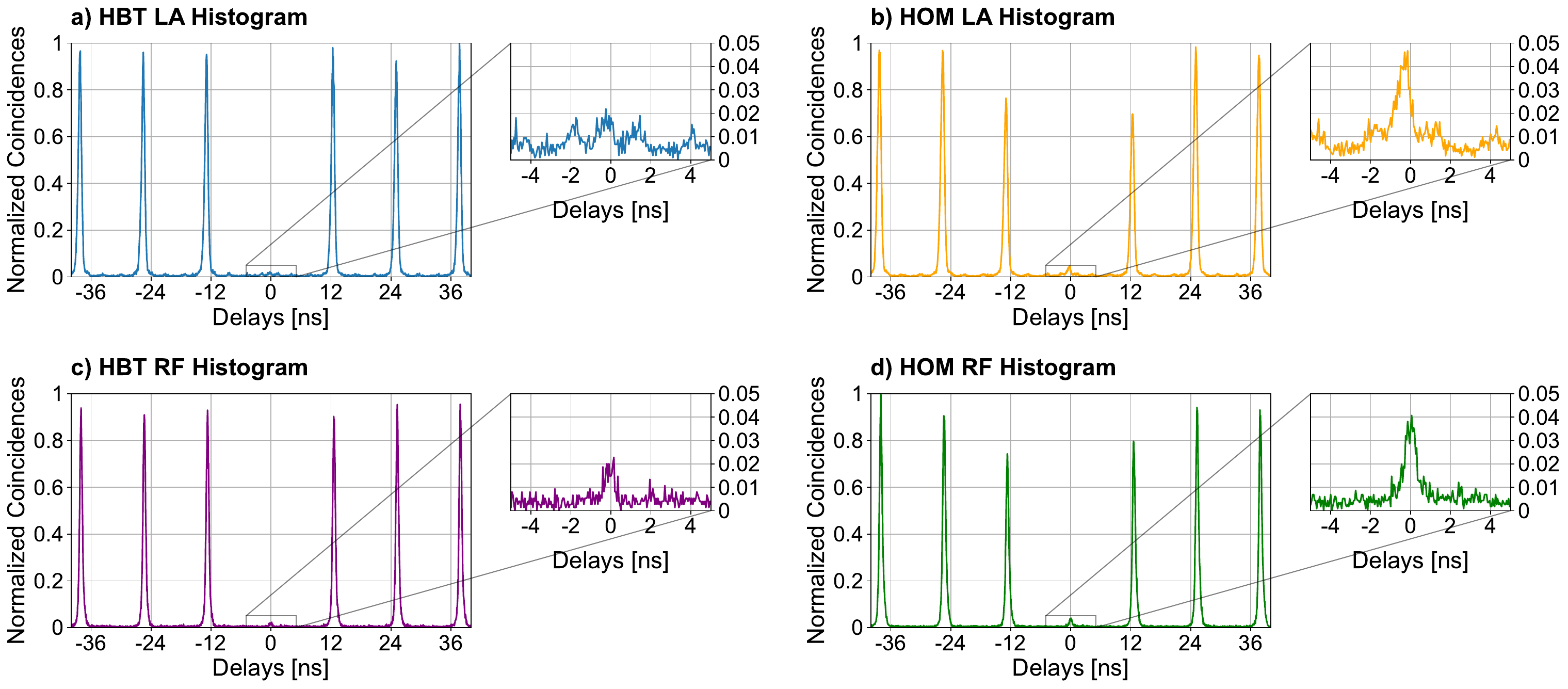}
    \caption{Normalized correlation histograms between the two outputs of the QD$^\text{LA}$ and QD$^\text{RF}$ sources in the case of HBT and HOM setups for measuring  the second order correlation function $g^{(2)}(0)$ (panel a and c) and the HOM visibility $v_{m}$ (panel b and d). The normalization is computed with respect to the highest peak, having unitary height. }
    \label{fig:corr}
\end{figure*}

\begin{table}[hb!]
    \centering        
\begin{tabular}{c c c c}
\multicolumn{4}{c}{Transmittivity}  \\
    \hline
    & & Partial & Total \\
    \hline

\begin{tabular}{c}  QDSPS (LA) \end{tabular}	& \begin{tabular}{c} Polarized First lens Brightness\\
Quandela QFiber system \\
 \end{tabular}  &
\begin{tabular}{c} 20\% \\ 
52\% \\
\end{tabular} &
10.4\% \\

\hline \hline

\begin{tabular}{c}  QDSPS (RF) \end{tabular}	& \begin{tabular}{c} Polarized First lens Brightness \\
Quandela QFiber system \\
\end{tabular} & 
\begin{tabular}{c} 11\% \\
65\% \\
\end{tabular} &
7.2\% \\

\hline \hline
Fiber link & \begin{tabular}{c}
    QDSPS (LA) to MZI \\
    QDSPS (RF) to MZI \\ 
\end{tabular} & & \begin{tabular}{c}
    57 $\%$ \\
    50 $\%$ \\
\end{tabular}  \\
 \hline \hline
\begin{tabular}{c}  MZI \end{tabular}	& \begin{tabular}{c} Air-to-fiber coupling \\
fiber-BS transmission \\ \end{tabular} 
& 
\begin{tabular}{c} 80\% \\ 63\% \\ \end{tabular} 
& 
50\% \\

\hline \hline 

\end{tabular}\
\caption{Efficiency analysis of the experimental apparatus. The Polarized First Lens Brightness in both configurations takes into account the polarization selection.}
\label{tab:losses} 
\end{table}

The experimental platform is depicted in Fig.\ref{fig:source}. Both quantum dot single-photon sources (QDSPS) are commercial Quandela \emph{e-Delight} systems (yellow dotted line panel). The semiconductor devices consist of a InGaAs matrix placed in a nanoscale electrically controlled micropillar cavity \cite{somaschi2016near} kept at cryogenic temperature (around 4K) by an \emph{Attocube-Attodry800} He-closed cycle cryostat. The QDSPS are optically excited by a 79 MHz-pulsed laser in two different excitation schemes: the first device works in a longitudinal phonon-assisted configuration (LA) \cite{thomas2021bright}, while emission for the second device is achieved resonantly in the so-called RF one \cite{loredo2019generation}.\\ 
The LA optical excitation (QD$^{\text{LA}}$) is obtained by blue-detuning the laser pump at 927.2 nm, and produces single photons at 927.8 nm. 
The emitted photons are coupled into a single-mode fiber (SMF) through a free-space confocal microscope mounted atop of the cryostat shroud and spectrally separated from the residual pumping laser with spectral filters. 
We report in Tab.~\ref{tab:losses} an efficiency analysis for our platform. In detail, the overall transmission efficiency of the QDSPS is $\eta^{\text{LA}}_{\text{QDSPS}}=10.4\%$, which includes the Polarized First Lens Brightness (20\%) and the transmission through the Quandela Qfiber system (52\%); the fiber link (FL) connecting the QDSPS to the source, including mating sleeves, has a transmission efficiency of $\eta^{LA}_{FL}=57\%$.

The RF optical excitation (QD$^{\text{RF}}$) employs a laser pump at a wavelength centered at 928.05 nm, enabling resonant single photon generation by exciton emission. For this second QDSPS, we achieve the single photon collection by means of a single-mode fiber (SMF) located almost in direct contact with the sample inside the cryostat. Photons are separated from the residual pumping laser in a cross-polarization configuration. The brightness of the source depends on the coupling efficiency into the SMF, the losses due to the cross-polarization scheme amounting at least to 50\% of the emitted photons \cite{ollivier2020reproducibility}, the efficiency of the RF Quandela Qfiber system (65\%), and the detection efficiency (35\%). Therefore, without accounting for the unknown coupling efficiency of the SMF inside the cryostat, we estimate a Polarized First Lens Brightness of $11\%$; the overall transmission efficiency for the RF excitation thus amounts to $\eta^{\text{RF}}_{\text{QDSPS}}=7.2\%$. 
The fiber link between the QDSPS and the source, including mating sleeves, has a transmission efficiency equal to $\eta^{RF}_{FL}=50\%$.

The train of single photons is sent to an unbalanced Mach-Zehnder-based Interferometer (MZI) for the generation of entangled photon pairs in the polarization degree of freedom (purple dotted line panel in Fig.\ref{fig:source}). A first in-bulk balanced beam splitter (BS) divides the photons into two paths, with a relative temporal delay corresponding to the difference between two consecutive emissions ($\sim 12 ns$) achieved through a specifically tuned fiber delay line. A photon entering the shorter path is prepared in vertical polarization (V), while the photon in the longer path is prepared in horizontal polarization (H). A second in-fiber 50/50 BS recombines the two signals. In this way, when two consecutive photons - namely head and tail - take different internal paths of the MZI, if the tail photon takes the shorter path it then recombines in the BS with the head one. Precise temporal overlap between the photon pairs, on the output BS, is achieved by finely adjusting the delay line along the longer path. Finally, when post-selecting on the cases when photons take different output paths, i.e. on two-fold coincidences, the obtained theoretical output state is $\ket{\psi^{(\phi)}}=\frac{1}{\sqrt{2}}(\ket{HV}+e^{i\phi}\ket{VH})$.
A liquid crystal sets the phase $\phi$, while polarization controllers are used to prepare one of the four states of the Bell basis. The measured coincidence rate exiting the MZI is $R^{measured}_{\text{CC}}  = 1$ kHz ($0.5$ kHz) for QD$^\text{LA}$ (QD$^\text{RF}$). This value is compatible with the transmission efficiency of the apparatus ($\eta_{\text{MZI}}=50\%$).
Indeed, considering a near-unity generation rate inside the QD cavity ($R_{\text{QD}} \sim 79$ MHz), the probability to detect a zero-delay coincidence at the output of the MZI is first reduced by a factor 4 from the first BS -- i.e. the success probability of obtaining the head photon in the delayed path and the tail photon in the shorter path of the MZI -- and then halved by the post-selection process, thus providing a final expected coincidence rate: $R_{\text{CC}}=\frac{1}{8} \eta^2_{\text{QDSPS}}\eta^{2}_{FL} \eta^2_{\text{MZI}} \eta^2_{d} R_{\text{QD}}$, compatible with the loss budget. The quality of the generated entangled state mainly depends on the indistinguishability of consecutively emitted photons as well as the presence of multi-photon components. These characteristics can be quantified and studied, respectively, through the visibility $v$ of the HOM effect, and the second-order correlation function $g^{(2)}(0)$. 
\section{Results}
\label{sec:exp_results}

\begin{figure}[h!]
    \centering
    \includegraphics[width=0.5\textwidth]{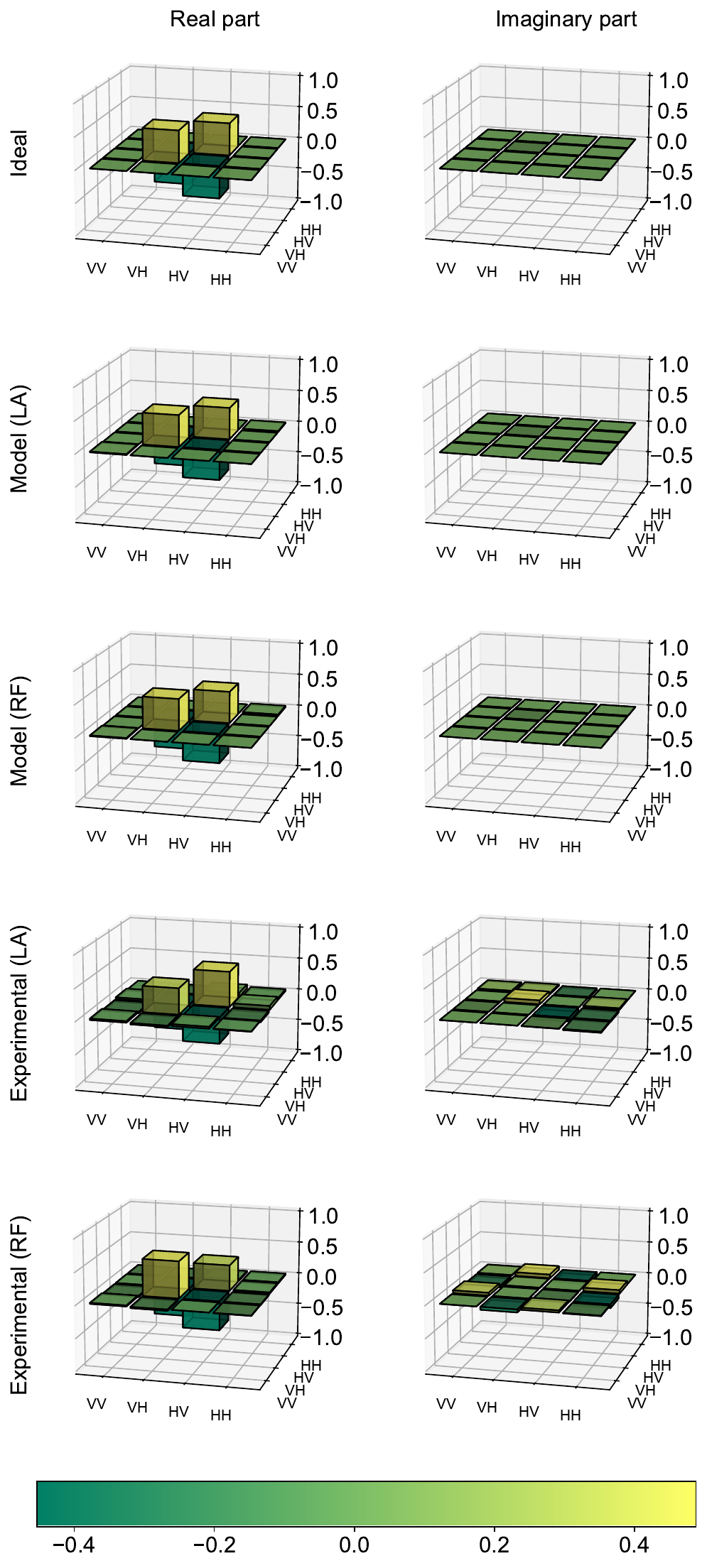}
    \caption{Quantum state tomography of the source. Real and imaginary parts of the measured density matrix in LA and RF configurations are compared with the ideal singlet state and the one extracted by the model. Note that the latter is computed considering the experimental parameters of the setup (see Supplementary Table I).}
    
    \label{fig:tomo}
\end{figure}

\subsection{Theoretical Model}
When realizing an experimental platform for the generation of entangled states we necessarily have to take into account multiple factors that preclude the realization of an ideal maximally entangled pure state. In what follows, we describe a model which identifies the main parameters of our experimental apparatus and relate them to the quality of the output entangled state. A detailed investigation of such quantities is reported in the Supplementary Information. 

The first parameter to be considered is the degree of indistinguishability between the two interfering photons generating the entangled state, which depends on the overlap of their wavepackets. This overlap is commonly quantified by measuring the visibility of HOM effect. Indeed, when two partially distinguishable photons impinge on the inputs of a balanced beam-splitter there is a non-zero probability $P_{cc}$ of obtaining a coincidence event at the two outputs of the BS, which is at most equal to $P_{cc} = 1/2$ for fully distinguishable photons. The HOM visibility $v$ is then defined as:

\begin{equation}
    \label{hom}
    v = 1 - 2P_{cc},
\end{equation}

This quantity is equal to one for perfectly indistinguishable photons, that is, no coincidence counts are detected at the output of the BS; this is the typical signature of the HOM effect.
However, the experimentally measured HOM visibility ($v_m$) does not match with the true photon indistinguishability ($v$), namely the single photon trace purity \cite{ollivier2021hong}. This is due to the possible presence of additional photons (defined as noise photons) traveling together with the SPS signal.
Despite the quantum dots employed in the experiment show a high probability $p_1$ of single-photon generation for each pump pulse ($p_1 \sim 1$ for resonant excitation, $p_1 \le 1$ for phonon-assisted excitation \cite{thomas2021bright}), a residual probability $p_2$ of noise-photon component exists, such that $p_0+p_1+p_2=1$, where $p_0$ is the probability of no-excitation. Moreover, in the RF regime, the emitted single-photon state is a coherent superposition with the vacuum state \cite{istrati2020sequential}, i.e. $\sqrt{1-q} e^{i \phi_q}\ket{0} + \sqrt{q}\ket{1}$ with $ q \in [0,1] ,\phi_q \in \mathbb{R}$ (Eq.~7 of Supplementary Information). The parameter $q$ can in principle be varied by tuning the laser pump intensity \cite{loredo2019generation} and ideally is set to near unity. In our model, the only difference between LA and RF is the presence of parameter $q$ for RF. However, the model for LA can be recovered by setting $q=1$, so the considerations made are valid for both QDSPSs.

To quantify the impact of noisy photons, a measurement of the second-order correlation $g^{(2)}(0)$ can be performed through the Hanbury-Brown-Twiss (HBT) setup. The probability of having more than one photon $p_{2}$ is then computed through the relation $g^{(2)}(0)=\frac{2p_2}{(p_1+2p_2)^2}$, as detailed in Supplementary Note 1.

Furthermore, the HOM visibility is reduced if the employed BS is not perfectly symmetric, that is, with reflectivity $R =1/2$. In particular, according to \cite{ollivier2021hong}, the relation between multiphoton emission and photon indistinguishability is given by:
\begin{equation}\label{Model_g2_vs_Hom_ollivier}
    v_{m} = 4RT\Bigl( 1+v-\frac{1+v}{1-v_{sn}}g^{(2)}(0)\Bigr) - 1,
\end{equation}

where $T$ ($R$) is the transmissivity (reflectivity) of the beam-splitter and $v_{sn}$ is the overlap between the single photon signal and the noise photons. Therefore, considering the additional photon distinguishable from the QD signal, i.e. $v_{sn}=0$ and a 50:50 BS, it is possible to compute a more precise estimation of the photon indistinguishability as:

\begin{equation}
    \label{hom_corr}
    v = \frac{v_m + g^{(2)}(0)}{1-g^{(2)}(0)}.
\end{equation}
 
The partial indistinguishability and the use of non-ideal BSs in the interferometer cause the state produced by the two-photon interference ($\hat{\rho}_{11}$) to be different from the ideal pure state ($\hat{\rho}_{\psi}=\ket{\psi^{(\phi)}}\bra{\psi^{(\phi)}}$), that is $\hat{\rho}_{11} \neq \hat{\rho}_{\psi}$. In particular, the distinguishability among the photons has different origins and its effect on the final state of the source is to introduce non-ideal terms, that acts as dephasing and white noise. Indeed, the presence of noise photons not only changes the estimation of HOM visibility, but also provides a contribution to the final state, that we indicate as $\hat{\rho}_{12}$, $\hat{\rho}_{02}$ and $\hat{\rho}_{l}$ (Eq.~(2) of Supplementary Information). Such multi-photon component gives rise to noise terms such as $\ket{HH}\bra{HH}$ and $\ket{VV}\bra{VV}$, as well as to terms $\ket{HV}\bra{HV}$ and $\ket{VH}\bra{VH}$ in the final state. The probability of each term depends on the generation mechanism, described in detail in the Supplementary Information. 
Furthermore, to take into account optical components imperfections or any other constant noise, like darkcounts, a contribution of white noise $\hat{\rho}_{wn}=\mathbb{I} / 4$ is added to the experimental state since these terms provide a uniform contribution to the coincidence counts in all the bases. Therefore, the expected experimental state has the form:
\begin{equation}
\label{eq:state_source_n}
\hat{\rho}=c_{wn}\hat{\rho}_{\text{exp}} + (1-c_{wn})\frac{\mathbb{I}}{4}
\end{equation}
where
\begin{equation}
\label{eq:state_source}
    \hat{\rho}_{\text{exp}} = c_{11} \hat{\rho}_{11} + c_{12} \hat{\rho}_{12} + c_{02} \hat{\rho}_{02} + c_{l}\hat{\rho}_{l}
\end{equation}
The coefficients $c_{02},c_{11},c_{12},c_{l}$ depend on several parameters such as the overall transmissivity $\eta$ and the percentage of a multi-photon signal expressed through the probability $p_0,p_1,p_2$ to have zero, one and two photons emission respectively (see Supplementary Information). 

Using Eq.~\eqref{eq:state_source_n}, it is possible to compute the maximum degree of entanglement achievable with our experimental setup. Furthermore, the quality of the entanglement can be asserted in a device-independent manner by using a Bell-CHSH test \cite{clauser1969proposed}. According to its formulation, the presence of entanglement between two stations (A and B) is a necessary condition to violate the classical bound of the Bell-CHSH inequality: 
\begin{equation}
S = E(A_0,B_0)+E(A_0,B_1)+E(A_1,B_0)-E(A_1,B_1) \leq 2
\end{equation}
where $E(A_i,B_j)$ (with $i,j=0,1$) is the expectation value of the correlator between two possible dichotomic measurements made in the subsystems A and B. It is well known that maximally entangled states, like the Bell states, can reach $S = 2\sqrt{2}$ which is the maximal violation of the Bell-CHSH inequality allowed within quantum mechanics.
For simplicity, we consider the so-called singlet Bell state ($\phi=\pi$, so that $\ket{\psi^{(\phi)}}=\ket{\psi^{(-)}}$). To achieve $S = 2\sqrt{2}$ with a singlet state, one can choose as measurements $\hat{A}_0=\hat{\sigma}_z,\,\hat{A}_1=\hat{\sigma}_x$ and $\hat{B}_0=(\hat{\sigma}_z+\hat{\sigma}_x)/\sqrt{2},\,\hat{B}_1=(\hat{\sigma}_z-\hat{\sigma}_x)/\sqrt{2}$, where $\hat{\sigma}_z$ and $\hat{\sigma}_x$ are the standard Pauli operators.

The state in Eq.~\eqref{eq:state_source_n} corresponds to a lower violation of the Bell inequality with respect to its maximum value of $2 \sqrt{2}$ achievable via a singlet state. Here the expectation values of the correlators over the modelled state have been computed as $E(A_i,B_j)=\mathrm{Tr}[\hat{\rho}_\text{exp} \hat{A}_i \otimes \hat{B}_j]$ with $i,j=0,1$. As a result this relation shows the strongest dependence on the parameters $g^{(2)}(0)$ and $c_{wn}$, i.e. the fraction of white noise. Conversely, the highest achievable value of $S$ does not depend significantly from non-ideal splitting ratios of the BSs.

Similar results hold when considering the fidelity with respect to the ideal singlet state $\hat{\rho}_{\psi^{(-)}}$, defined as:
\begin{equation}
\mathcal{F} = \mathrm{Tr}[\sqrt{\sqrt{\hat{\rho}_{\psi^{(-)}}} \hat{\rho}_{exp}\sqrt{\hat{\rho}_{\psi^{(-)}}}}]^{2}
\end{equation}

In Supplementary information is described a complete model of the main parameters affecting the maximum achievable Bell-CHSH violation $S$ and $\mathcal{F}$ over the experimentally generated state $\hat{\rho}$. 

\begin{figure}[h]
    \centering
    \includegraphics[width=0.5\textwidth]{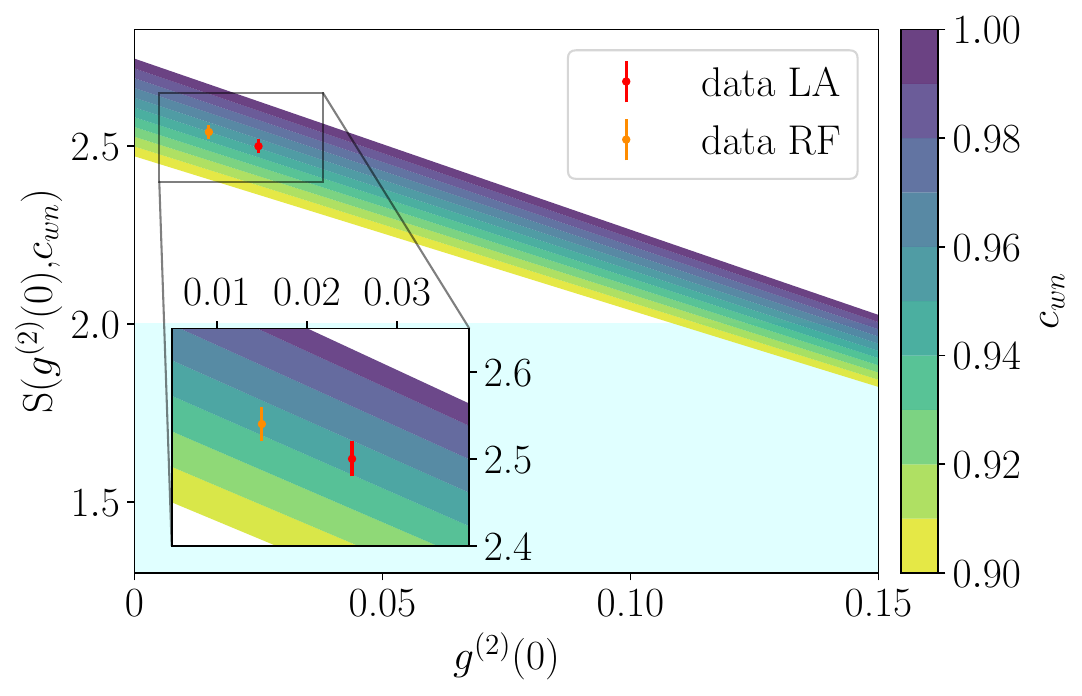}
    \caption{Maximum achievable value of the Bell-CHSH quantity as a function of the $g^{(2)}(0)$ and the fraction of white noise due to state polarization imperfection and darkcounts ($c_{wn}$). The experimental points are $(0.025 \pm 0.002,2.50 \pm 0.02)$ for LA (red) and $(0.015 \pm 0.002,2.54 \pm 0.02)$ for RF (orange) without subtracting accidental coincidences which are compatible with $c_{\text{wn}}\sim 0.95$. For simplicity, RF and LA data have been plotted together due to the very low difference in the overall transmission model.}
    \label{fig:chsh}
\end{figure}

\begin{figure}[h]
    \centering
    \includegraphics[width=0.5\textwidth]{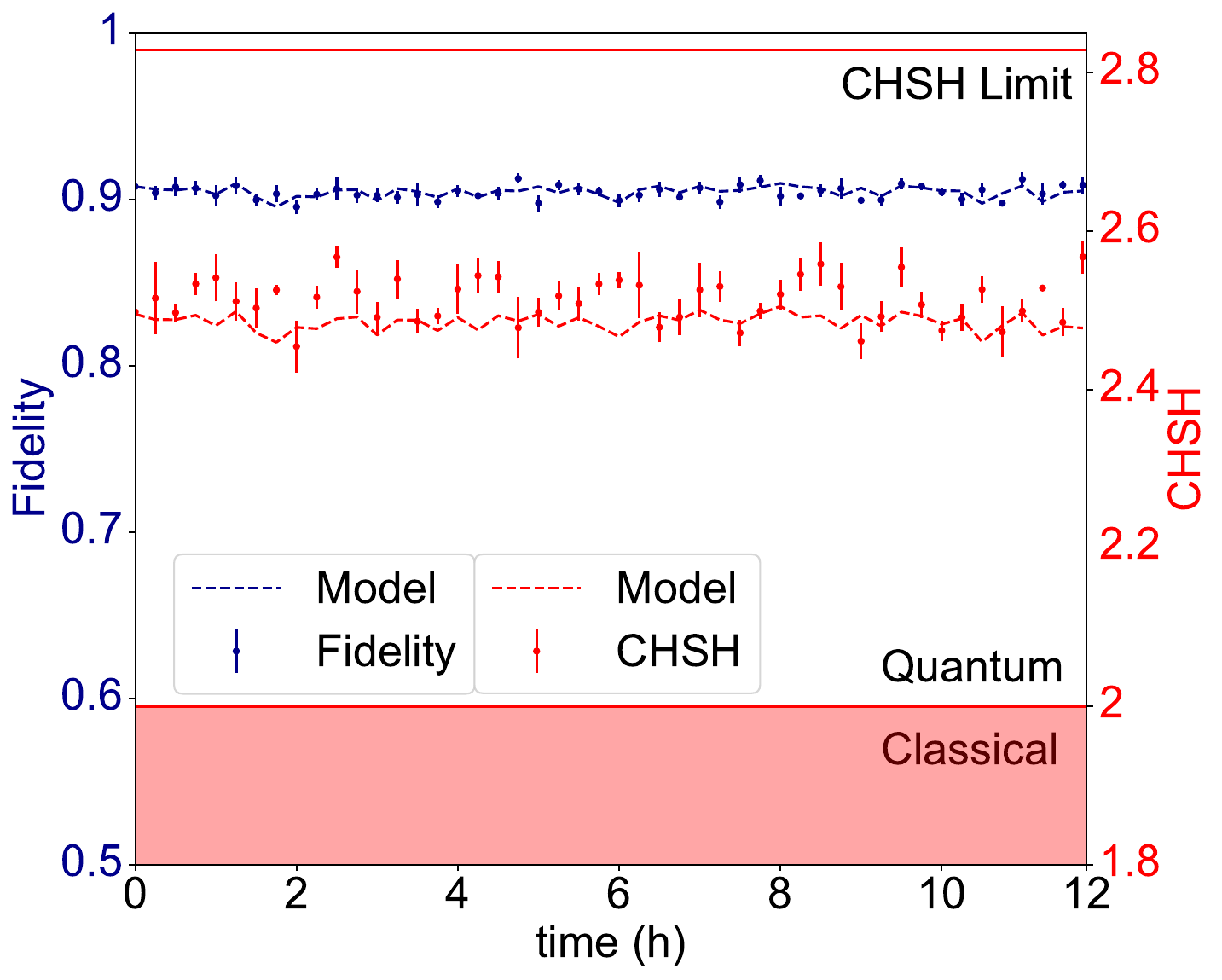}
    \caption{Performances of the LA-based source as a function of the time. Long-term stability of the source is monitored in terms of fidelity (left vertical axis) and CHSH violation (right vertical axis) during 12 hours of operation. The dotted lines correspond to the theoretical prediction with $c_{\text{wn}}=0.95$. Note that Classical/Quantum regions are referred to the Bell's parameter while the solid red lines stands for the minimum/maximum value of the CHSH parameter.}
    \label{fig:time}
\end{figure}
\begin{figure*}[ht!]
    \centering
    \includegraphics[width=0.99\textwidth]{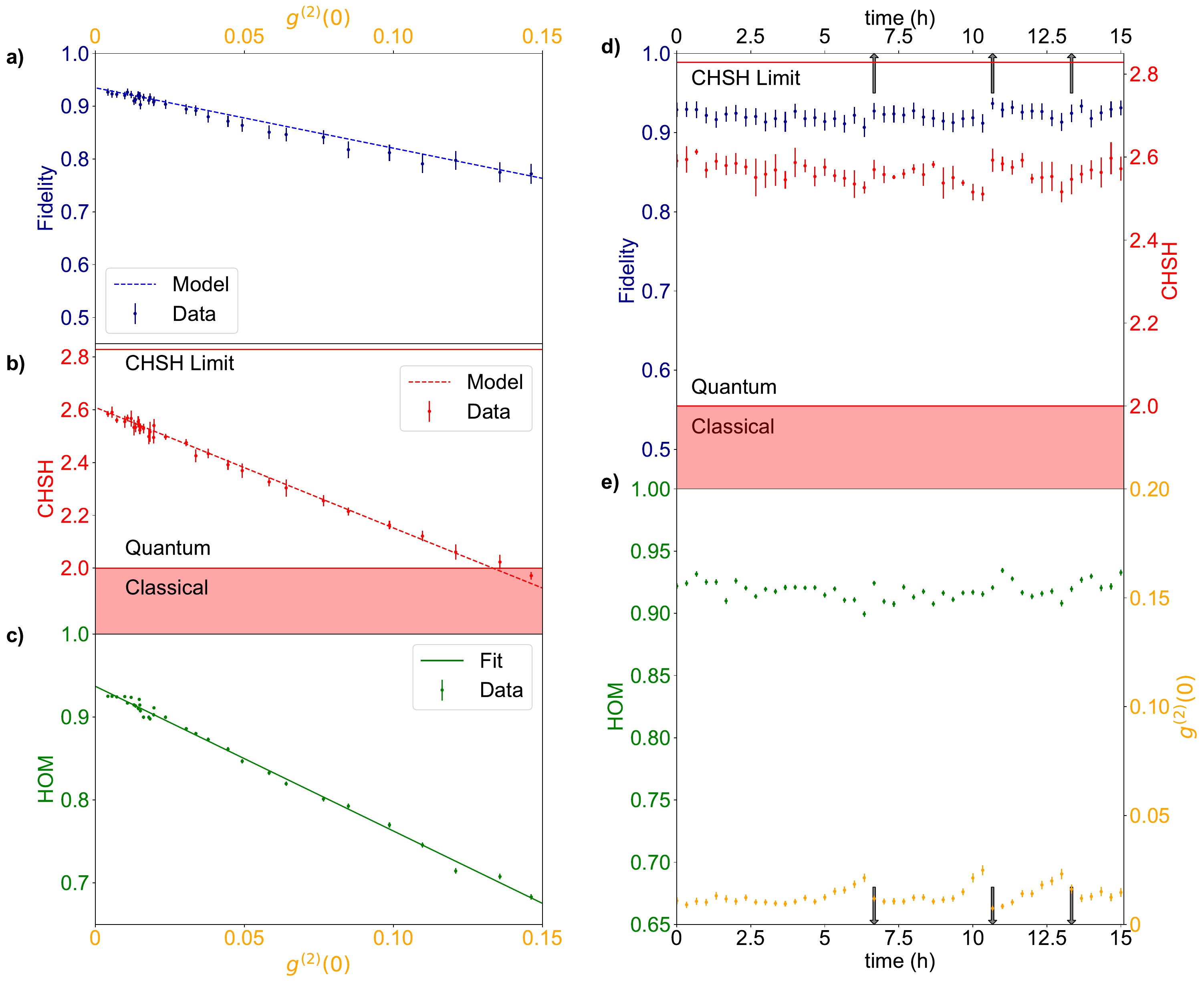}
    \caption{Stability analysis of the RF-based source. \textbf{(a-c)} In this configuration, behavior of the fidelity (panel a), CHSH violation (panel b), and HOM visibility (panel c) in function of the $g^{(2)}(0)$ is shown. Note that in panel c the error-bars are lower than the markers. Its increasing over time is caused by the instability of the pump laser suppression mechanism. Furthermore, as shown in panels a-b, the trends of the best achievable values of CHSH and fidelity predicted by our model (Eqs.~(27), (29) of Supplementary information) are in agreement with this behavior when $g^{(2)}(0)$ changes. 
    \textbf{(d-e)} This effect can be mitigated by performing a real-time optimization of the suppression. As soon as $g^{(2)}(0)$ deviates much from its mean value, we manually act on the QD source to recover proper suppression of the laser pump. The optimization stages are indicated with vertical arrows.}
    \label{fig:timeRF_opt}
\end{figure*}

\subsection{Performances}\label{sec:perf}
The measured two-photon coincidence rate at the output of the source is $R^{\text{LA}}_{m}=1$ kHz ($R^{\text{RF}}_{m}=0.5$ kHz). The APDs used have a detection efficiency of around $35\%$, from which it is possible to infer the generated rate by the source of $R^{\text{LA}}_{g} \sim 8$ kHz ($R^{\text{RF}}_{g} \sim 4$ kHz). Such rate can be improved by a factor 4 replacing the passive demultiplexer, i.e. the first BS of the MZI, with deterministic choice of the path for the consecutive photons, sending deterministically the head (tail) photon in the longer (shorter) path. This active demultiplexing can be realized, for example, by adopting an electro-optic modulator and a polarizing beam splitter (PBS) \cite{munzberg2022fast}. As explained previously, the quality of the produced state is evaluated in terms of $g^{(2)}(0)$ and the indistinguishability of consecutive photons.
The former is computed with an HBT setup -- obtained by blocking one arm of the interferometer -- while the latter by using the HWP to rotate the photon polarization in the shorter path from vertical to horizontal (the same as the photon in the longer path). 
Fig.~\ref{fig:corr} shows the time correlation histograms for both measures: QD$^\text{LA}$ provides $v_m^{\text{LA}}=(90.3\pm0.3)\%$ and $g^{(2)}_{\text{LA}}(0)=(1.2\pm0.1)\%$, while for QD$^\text{RF}$ the measured values are $v_m^{\text{RF}}=(91.8\pm0.2)\%$ and $g^{(2)}_{\text{RF}}(0)=(1.6\pm0.2)\%$. According to Eq.~\eqref{hom_corr}, the HOM visibility corrected by $g^{(2)}(0)$ are found to be $v^{\text{LA}}=(92.7 \pm 0.3)\%$ and $v^{\text{RF}}=(94.9 \pm 0.3)\%$.
Polarization measurement stages are placed at the two outputs of the MZI interferometer in order to characterize in detail the final state of the source. After fine-tuning of the liquid crystal for setting the singlet Bell state, a quantum state tomography and the CHSH test have been made. The resulting density matrix is shown in Fig.~\ref{fig:tomo}, which correspond to a fidelity $F^{\text{LA}}_{raw}=(90\pm1)\%$ and $F^{\text{RF}}_{raw}=(92\pm1)\%$ with respect to the ideal singlet state $\hat{\rho}_{\psi^{(-)}}$.
Considering the accidental counts measured in the middle of two pulses of the time-correlation histogram (e.g. at delay $6$ ns), it is possible to extract only the noise contribution due to detector dark counts which affect such fidelities. By subtracting this contribution, we obtain a fidelity of $F^{\text{LA}}=(92\pm1)\%$ and $F^{\text{RF}}=(95\pm1)\%$. As previously discussed, non-ideal contributions which limit the fidelity derive from multi-photon components, white noise and actual distinguishability of single photons emitted by the QD sources, according to Eq.~\eqref{eq:state_source_n}. 
Finally, we measured the CHSH parameter obtaining $S^{\text{LA}}_{raw}=(2.50\pm0.02)$ from raw data, and $S^{\text{LA}}=(2.58\pm0.02)$ subtracting dark counts in the case of QD$^\text{LA}$. 
Similarly, the use of QD$^\text{RF}$ provides $S^{RF}_{raw}=(2.54\pm0.02)$ and $S^{\text{RF}}=(2.63\pm0.02)$.
In Fig.~\ref{fig:chsh} we compare the measured values of $S^{LA}_{raw}$ and $S^{RF}_{raw}$ with the expected values of the CHSH parameter according to the proposed model (see Supplementary Information). The plot shows how the most influential parameters considered in the model reduce the maximum achievable $S$ by our experimental scheme. It also predicts the maximum value of $S$ obtainable given the value of $g^{(2)}(0)$ and $c_{wn}$.  As expected, the observed experimental values do not exceed the maximum values predicted by the model. 

Fig.~\ref{fig:time} shows the stability of the produced quantum state in time, demonstrating no drop in performance over 12 hours without user intervention in the case of longitudinal phonon-assisted excitation. Such a feature is relevant when the source is used for long duration measurement protocols. Conversely, the RF approach showed less stability, having a drop in performance after few hours (Figs.~\ref{fig:timeRF_opt}d-e). As shown by the Fig.~\ref{fig:timeRF_opt}a-c this reduced stability is related to an increasing $g^{(2)}(0)$ while the variation of other parameters have reduced or no effects. According to ref.\cite{ollivier2021hong}, a linear fit made with the parameters $v_{m}$ and $g^{(2)}(0)$ returns the value of the photon superposition equal to $v=0.937\pm0.001$. The predictions of the theoretical model are compatible with the experimental data considering an uniform noise quantified by $c_{wn}\sim 0.95$. The increasing $g^{(2)}(0)$ is related to the non-perfect elimination of the pump laser by the polarization-based suppression system. In fact, Figs.~\ref{fig:timeRF_opt}d-e show how the performance was recovered by resetting that suppression whenever it showed a decrease. 

\section{Conclusions}
\label{sec:5}

We demonstrated a simple design to generate pairs of entangled photons in polarization by employing a quantum dot device emitting single photons streams. The source is based on a MZI-like configuration, having long-term stability and providing high quality of the entangled pair. Its characterization in both LA and RF configurations demonstrated a high fidelity of the produced state with respect to the ideal single Bell state, i.e. $F^{\text{LA}}_{raw}=(90\pm1)\%$ and $F^{\text{RF}}_{raw}=(92\pm1)\%$, and certified the presence of non locality by violating CHSH inequality achieving $S^{\text{LA}}_{raw}=(2.50\pm0.02)$ and $S^{\text{RF}}_{raw}=(2.54\pm0.02)$, respectively.
Entangled-based quantum communication protocols using quantum dot have so far been demonstrated only with EPS sources \cite{basso2021quantum}. The performances of our source enable such implementations also adopting QDSPS as well. In fact, despite the presence of post-selection on coincidence events, the security of QKD protocols is not further affected because the same post-selection procedure is repeated on the algorithms based on entangled states \cite{Basso_Basset_2023}. Furthermore, we provided a model that can distinguish the different types of noise contributing to such kinds of sources. 
Therefore, this work represents a significant step forward in the generation of entanglement states via QDSPS exciton emission. 
Indeed, similar source-designs have been considered for such purposes \cite{istrati2020sequential,li2020multiphoton}. The results demonstrate that the main factors limiting our source are $g^{(2)}(0)$, HOM visibility and white noise. In presence of high indistinguishability and low multiphoton emission, our design shows great stability in generating high quality entangled photon pairs in  polarization, despite possible non-idealities of the parameters which constitute the source structure (e.g. splitting ratio of BSs) or long-time measurements. Thus, the measured CHSH values together with the tomography process, confirm the high purity of the employed quantum dot device in emitting highly indistinguishable single photons.\\

\section*{ACKNOWLEDGEMENTS}
We thank Petr Stepanov, Niccolo Somaschi and {\em Quandela} for all the support provided. This work is supported by the ERC Advanced grant QU-BOSS (Grant Agreement No. 884676) and by the PNRR MUR project PE0000023-NQSTI.


\begin{thebibliography}{51}%
\makeatletter
\providecommand \@ifxundefined [1]{%
 \@ifx{#1\undefined}
}%
\providecommand \@ifnum [1]{%
 \ifnum #1\expandafter \@firstoftwo
 \else \expandafter \@secondoftwo
 \fi
}%
\providecommand \@ifx [1]{%
 \ifx #1\expandafter \@firstoftwo
 \else \expandafter \@secondoftwo
 \fi
}%
\providecommand \natexlab [1]{#1}%
\providecommand \enquote  [1]{``#1''}%
\providecommand \bibnamefont  [1]{#1}%
\providecommand \bibfnamefont [1]{#1}%
\providecommand \citenamefont [1]{#1}%
\providecommand \href@noop [0]{\@secondoftwo}%
\providecommand \href [0]{\begingroup \@sanitize@url \@href}%
\providecommand \@href[1]{\@@startlink{#1}\@@href}%
\providecommand \@@href[1]{\endgroup#1\@@endlink}%
\providecommand \@sanitize@url [0]{\catcode `\\12\catcode `\$12\catcode
  `\&12\catcode `\#12\catcode `\^12\catcode `\_12\catcode `\%12\relax}%
\providecommand \@@startlink[1]{}%
\providecommand \@@endlink[0]{}%
\providecommand \url  [0]{\begingroup\@sanitize@url \@url }%
\providecommand \@url [1]{\endgroup\@href {#1}{\urlprefix }}%
\providecommand \urlprefix  [0]{URL }%
\providecommand \Eprint [0]{\href }%
\providecommand \doibase [0]{http://dx.doi.org/}%
\providecommand \selectlanguage [0]{\@gobble}%
\providecommand \bibinfo  [0]{\@secondoftwo}%
\providecommand \bibfield  [0]{\@secondoftwo}%
\providecommand \translation [1]{[#1]}%
\providecommand \BibitemOpen [0]{}%
\providecommand \bibitemStop [0]{}%
\providecommand \bibitemNoStop [0]{.\EOS\space}%
\providecommand \EOS [0]{\spacefactor3000\relax}%
\providecommand \BibitemShut  [1]{\csname bibitem#1\endcsname}%
\let\auto@bib@innerbib\@empty
\bibitem [{\citenamefont {Vajner}\ \emph {et~al.}(2022)\citenamefont {Vajner},
  \citenamefont {Rickert}, \citenamefont {Gao}, \citenamefont {Kaymazlar},\
  and\ \citenamefont {Heindel}}]{vajner2022quantum}%
  \BibitemOpen
  \bibfield  {author} {\bibinfo {author} {\bibfnamefont {D.~A.}\ \bibnamefont
  {Vajner}}, \bibinfo {author} {\bibfnamefont {L.}~\bibnamefont {Rickert}},
  \bibinfo {author} {\bibfnamefont {T.}~\bibnamefont {Gao}}, \bibinfo {author}
  {\bibfnamefont {K.}~\bibnamefont {Kaymazlar}}, \ and\ \bibinfo {author}
  {\bibfnamefont {T.}~\bibnamefont {Heindel}},\ }\href {\doibase
  10.1002/qute.202100116} {\bibfield  {journal} {\bibinfo  {journal} {Advanced
  Quantum Technologies}\ }\textbf {\bibinfo {volume} {5}},\ \bibinfo {pages}
  {2100116} (\bibinfo {year} {2022})}\BibitemShut {NoStop}%
\bibitem [{\citenamefont {Ma}\ \emph {et~al.}(2007)\citenamefont {Ma},
  \citenamefont {Fung},\ and\ \citenamefont {Lo}}]{PhysRevA.76.012307}%
  \BibitemOpen
  \bibfield  {author} {\bibinfo {author} {\bibfnamefont {X.}~\bibnamefont
  {Ma}}, \bibinfo {author} {\bibfnamefont {C.-H.~F.}\ \bibnamefont {Fung}}, \
  and\ \bibinfo {author} {\bibfnamefont {H.-K.}\ \bibnamefont {Lo}},\ }\href
  {\doibase 10.1103/PhysRevA.76.012307} {\bibfield  {journal} {\bibinfo
  {journal} {Phys. Rev. A}\ }\textbf {\bibinfo {volume} {76}},\ \bibinfo
  {pages} {012307} (\bibinfo {year} {2007})}\BibitemShut {NoStop}%
\bibitem [{\citenamefont {Schimpf}\ \emph
  {et~al.}(2021{\natexlab{a}})\citenamefont {Schimpf}, \citenamefont {Reindl},
  \citenamefont {Huber}, \citenamefont {Lehner}, \citenamefont {Silva},
  \citenamefont {Manna}, \citenamefont {Vyvlecka}, \citenamefont {Walther},\
  and\ \citenamefont {Rastelli}}]{doi:10.1126/sciadv.abe8905}%
  \BibitemOpen
  \bibfield  {author} {\bibinfo {author} {\bibfnamefont {C.}~\bibnamefont
  {Schimpf}}, \bibinfo {author} {\bibfnamefont {M.}~\bibnamefont {Reindl}},
  \bibinfo {author} {\bibfnamefont {D.}~\bibnamefont {Huber}}, \bibinfo
  {author} {\bibfnamefont {B.}~\bibnamefont {Lehner}}, \bibinfo {author}
  {\bibfnamefont {S.~F. C.~D.}\ \bibnamefont {Silva}}, \bibinfo {author}
  {\bibfnamefont {S.}~\bibnamefont {Manna}}, \bibinfo {author} {\bibfnamefont
  {M.}~\bibnamefont {Vyvlecka}}, \bibinfo {author} {\bibfnamefont
  {P.}~\bibnamefont {Walther}}, \ and\ \bibinfo {author} {\bibfnamefont
  {A.}~\bibnamefont {Rastelli}},\ }\href {\doibase 10.1126/sciadv.abe8905}
  {\bibfield  {journal} {\bibinfo  {journal} {Science Advances}\ }\textbf
  {\bibinfo {volume} {7}},\ \bibinfo {pages} {eabe8905} (\bibinfo {year}
  {2021}{\natexlab{a}})},\ \Eprint
  {http://arxiv.org/abs/https://www.science.org/doi/pdf/10.1126/sciadv.abe8905}
  {https://www.science.org/doi/pdf/10.1126/sciadv.abe8905} \BibitemShut
  {NoStop}%
\bibitem [{\citenamefont {Giovannetti}\ \emph {et~al.}(2011)\citenamefont
  {Giovannetti}, \citenamefont {Lloyd},\ and\ \citenamefont
  {Maccone}}]{giovannetti2011advances}%
  \BibitemOpen
  \bibfield  {author} {\bibinfo {author} {\bibfnamefont {V.}~\bibnamefont
  {Giovannetti}}, \bibinfo {author} {\bibfnamefont {S.}~\bibnamefont {Lloyd}},
  \ and\ \bibinfo {author} {\bibfnamefont {L.}~\bibnamefont {Maccone}},\ }\href
  {\doibase 10.1038/nphoton.2011.35} {\bibfield  {journal} {\bibinfo  {journal}
  {Nature photonics}\ }\textbf {\bibinfo {volume} {5}},\ \bibinfo {pages} {222}
  (\bibinfo {year} {2011})}\BibitemShut {NoStop}%
\bibitem [{\citenamefont {Huang}\ \emph {et~al.}(2016)\citenamefont {Huang},
  \citenamefont {Macchiavello},\ and\ \citenamefont
  {Maccone}}]{PhysRevA.94.012101}%
  \BibitemOpen
  \bibfield  {author} {\bibinfo {author} {\bibfnamefont {Z.}~\bibnamefont
  {Huang}}, \bibinfo {author} {\bibfnamefont {C.}~\bibnamefont {Macchiavello}},
  \ and\ \bibinfo {author} {\bibfnamefont {L.}~\bibnamefont {Maccone}},\ }\href
  {\doibase 10.1103/PhysRevA.94.012101} {\bibfield  {journal} {\bibinfo
  {journal} {Phys. Rev. A}\ }\textbf {\bibinfo {volume} {94}},\ \bibinfo
  {pages} {012101} (\bibinfo {year} {2016})}\BibitemShut {NoStop}%
\bibitem [{\citenamefont {Carvacho}\ \emph {et~al.}(2022)\citenamefont
  {Carvacho}, \citenamefont {Roccia}, \citenamefont {Valeri}, \citenamefont
  {Basset}, \citenamefont {Poderini}, \citenamefont {Pardo}, \citenamefont
  {Polino}, \citenamefont {Carosini}, \citenamefont {Rota}, \citenamefont
  {Neuwirth}, \citenamefont {da~Silva}, \citenamefont {Rastelli}, \citenamefont
  {Spagnolo}, \citenamefont {Chaves}, \citenamefont {Trotta},\ and\
  \citenamefont {Sciarrino}}]{Carvacho:22}%
  \BibitemOpen
  \bibfield  {author} {\bibinfo {author} {\bibfnamefont {G.}~\bibnamefont
  {Carvacho}}, \bibinfo {author} {\bibfnamefont {E.}~\bibnamefont {Roccia}},
  \bibinfo {author} {\bibfnamefont {M.}~\bibnamefont {Valeri}}, \bibinfo
  {author} {\bibfnamefont {F.~B.}\ \bibnamefont {Basset}}, \bibinfo {author}
  {\bibfnamefont {D.}~\bibnamefont {Poderini}}, \bibinfo {author}
  {\bibfnamefont {C.}~\bibnamefont {Pardo}}, \bibinfo {author} {\bibfnamefont
  {E.}~\bibnamefont {Polino}}, \bibinfo {author} {\bibfnamefont
  {L.}~\bibnamefont {Carosini}}, \bibinfo {author} {\bibfnamefont {M.~B.}\
  \bibnamefont {Rota}}, \bibinfo {author} {\bibfnamefont {J.}~\bibnamefont
  {Neuwirth}}, \bibinfo {author} {\bibfnamefont {S.~F.~C.}\ \bibnamefont
  {da~Silva}}, \bibinfo {author} {\bibfnamefont {A.}~\bibnamefont {Rastelli}},
  \bibinfo {author} {\bibfnamefont {N.}~\bibnamefont {Spagnolo}}, \bibinfo
  {author} {\bibfnamefont {R.}~\bibnamefont {Chaves}}, \bibinfo {author}
  {\bibfnamefont {R.}~\bibnamefont {Trotta}}, \ and\ \bibinfo {author}
  {\bibfnamefont {F.}~\bibnamefont {Sciarrino}},\ }\href {\doibase
  10.1364/OPTICA.451523} {\bibfield  {journal} {\bibinfo  {journal} {Optica}\
  }\textbf {\bibinfo {volume} {9}},\ \bibinfo {pages} {572} (\bibinfo {year}
  {2022})}\BibitemShut {NoStop}%
\bibitem [{\citenamefont {Lu}\ and\ \citenamefont {Pan}(2021)}]{lu2021quantum}%
  \BibitemOpen
  \bibfield  {author} {\bibinfo {author} {\bibfnamefont {C.-Y.}\ \bibnamefont
  {Lu}}\ and\ \bibinfo {author} {\bibfnamefont {J.-W.}\ \bibnamefont {Pan}},\
  }\href {\doibase 10.1038/s41565-021-01033-9} {\bibfield  {journal} {\bibinfo
  {journal} {Nature Nanotechnology}\ }\textbf {\bibinfo {volume} {16}},\
  \bibinfo {pages} {1294} (\bibinfo {year} {2021})}\BibitemShut {NoStop}%
\bibitem [{\citenamefont {Lodahl}(2017)}]{Lodahl_2018}%
  \BibitemOpen
  \bibfield  {author} {\bibinfo {author} {\bibfnamefont {P.}~\bibnamefont
  {Lodahl}},\ }\href {\doibase 10.1088/2058-9565/aa91bb} {\bibfield  {journal}
  {\bibinfo  {journal} {Quantum Science and Technology}\ }\textbf {\bibinfo
  {volume} {3}},\ \bibinfo {pages} {013001} (\bibinfo {year}
  {2017})}\BibitemShut {NoStop}%
\bibitem [{\citenamefont {Li}\ \emph {et~al.}(2023)\citenamefont {Li},
  \citenamefont {Li}, \citenamefont {Cao}, \citenamefont {Yin},\ and\
  \citenamefont {Peng}}]{PhysRevApplied.19.064083}%
  \BibitemOpen
  \bibfield  {author} {\bibinfo {author} {\bibfnamefont {B.}~\bibnamefont
  {Li}}, \bibinfo {author} {\bibfnamefont {Y.-H.}\ \bibnamefont {Li}}, \bibinfo
  {author} {\bibfnamefont {Y.}~\bibnamefont {Cao}}, \bibinfo {author}
  {\bibfnamefont {J.}~\bibnamefont {Yin}}, \ and\ \bibinfo {author}
  {\bibfnamefont {C.-Z.}\ \bibnamefont {Peng}},\ }\href {\doibase
  10.1103/PhysRevApplied.19.064083} {\bibfield  {journal} {\bibinfo  {journal}
  {Phys. Rev. Appl.}\ }\textbf {\bibinfo {volume} {19}},\ \bibinfo {pages}
  {064083} (\bibinfo {year} {2023})}\BibitemShut {NoStop}%
\bibitem [{\citenamefont
  {DiVincenzo}(1995)}]{doi:10.1126/science.270.5234.255}%
  \BibitemOpen
  \bibfield  {author} {\bibinfo {author} {\bibfnamefont {D.~P.}\ \bibnamefont
  {DiVincenzo}},\ }\href {\doibase 10.1126/science.270.5234.255} {\bibfield
  {journal} {\bibinfo  {journal} {Science}\ }\textbf {\bibinfo {volume}
  {270}},\ \bibinfo {pages} {255} (\bibinfo {year} {1995})},\ \Eprint
  {http://arxiv.org/abs/https://www.science.org/doi/pdf/10.1126/science.270.5234.255}
  {https://www.science.org/doi/pdf/10.1126/science.270.5234.255} \BibitemShut
  {NoStop}%
\bibitem [{\citenamefont {O'Brien}(2007)}]{o2007optical}%
  \BibitemOpen
  \bibfield  {author} {\bibinfo {author} {\bibfnamefont {J.~L.}\ \bibnamefont
  {O'Brien}},\ }\href {\doibase 10.1126/science.114289} {\bibfield  {journal}
  {\bibinfo  {journal} {Science}\ }\textbf {\bibinfo {volume} {318}},\ \bibinfo
  {pages} {1567} (\bibinfo {year} {2007})}\BibitemShut {NoStop}%
\bibitem [{\citenamefont {Loss}\ and\ \citenamefont
  {DiVincenzo}(1998)}]{PhysRevA.57.120}%
  \BibitemOpen
  \bibfield  {author} {\bibinfo {author} {\bibfnamefont {D.}~\bibnamefont
  {Loss}}\ and\ \bibinfo {author} {\bibfnamefont {D.~P.}\ \bibnamefont
  {DiVincenzo}},\ }\href {\doibase 10.1103/PhysRevA.57.120} {\bibfield
  {journal} {\bibinfo  {journal} {Phys. Rev. A}\ }\textbf {\bibinfo {volume}
  {57}},\ \bibinfo {pages} {120} (\bibinfo {year} {1998})}\BibitemShut
  {NoStop}%
\bibitem [{\citenamefont {Wang}\ \emph
  {et~al.}(2019{\natexlab{a}})\citenamefont {Wang}, \citenamefont {He},
  \citenamefont {Chung}, \citenamefont {Hu}, \citenamefont {Yu}, \citenamefont
  {Chen}, \citenamefont {Ding}, \citenamefont {Chen}, \citenamefont {Qin},
  \citenamefont {Yang}, \citenamefont {Liu}, \citenamefont {Duan},
  \citenamefont {Li}, \citenamefont {Gerhardt}, \citenamefont {Winkler},
  \citenamefont {Jurkat}, \citenamefont {Wang}, \citenamefont {Gregersen},
  \citenamefont {Huo}, \citenamefont {Dai}, \citenamefont {Yu}, \citenamefont
  {H\"{o}fling}, \citenamefont {Lu},\ and\ \citenamefont
  {Pan}}]{wang2019towards}%
  \BibitemOpen
  \bibfield  {author} {\bibinfo {author} {\bibfnamefont {H.}~\bibnamefont
  {Wang}}, \bibinfo {author} {\bibfnamefont {Y.-M.}\ \bibnamefont {He}},
  \bibinfo {author} {\bibfnamefont {T.-H.}\ \bibnamefont {Chung}}, \bibinfo
  {author} {\bibfnamefont {H.}~\bibnamefont {Hu}}, \bibinfo {author}
  {\bibfnamefont {Y.}~\bibnamefont {Yu}}, \bibinfo {author} {\bibfnamefont
  {S.}~\bibnamefont {Chen}}, \bibinfo {author} {\bibfnamefont {X.}~\bibnamefont
  {Ding}}, \bibinfo {author} {\bibfnamefont {M.-C.}\ \bibnamefont {Chen}},
  \bibinfo {author} {\bibfnamefont {J.}~\bibnamefont {Qin}}, \bibinfo {author}
  {\bibfnamefont {X.}~\bibnamefont {Yang}}, \bibinfo {author} {\bibfnamefont
  {R.-Z.}\ \bibnamefont {Liu}}, \bibinfo {author} {\bibfnamefont {Z.-C.}\
  \bibnamefont {Duan}}, \bibinfo {author} {\bibfnamefont {J.-P.}\ \bibnamefont
  {Li}}, \bibinfo {author} {\bibfnamefont {S.}~\bibnamefont {Gerhardt}},
  \bibinfo {author} {\bibfnamefont {K.}~\bibnamefont {Winkler}}, \bibinfo
  {author} {\bibfnamefont {J.}~\bibnamefont {Jurkat}}, \bibinfo {author}
  {\bibfnamefont {L.-J.}\ \bibnamefont {Wang}}, \bibinfo {author}
  {\bibfnamefont {N.}~\bibnamefont {Gregersen}}, \bibinfo {author}
  {\bibfnamefont {Y.-H.}\ \bibnamefont {Huo}}, \bibinfo {author} {\bibfnamefont
  {Q.}~\bibnamefont {Dai}}, \bibinfo {author} {\bibfnamefont {S.}~\bibnamefont
  {Yu}}, \bibinfo {author} {\bibfnamefont {S.}~\bibnamefont {H\"{o}fling}},
  \bibinfo {author} {\bibfnamefont {C.-Y.}\ \bibnamefont {Lu}}, \ and\ \bibinfo
  {author} {\bibfnamefont {J.-W.}\ \bibnamefont {Pan}},\ }\href
  {https://www.nature.com/articles/s41566-019-0494-3} {\bibfield  {journal}
  {\bibinfo  {journal} {Nature Photonics}\ }\textbf {\bibinfo {volume} {13}},\
  \bibinfo {pages} {770} (\bibinfo {year} {2019}{\natexlab{a}})}\BibitemShut
  {NoStop}%
\bibitem [{\citenamefont {Senellart~P.}(2017)}]{senellart}%
  \BibitemOpen
  \bibfield  {author} {\bibinfo {author} {\bibfnamefont {S.~G. . W.~A.}\
  \bibnamefont {Senellart~P.}},\ }\href {\doibase
  https://doi.org/10.1038/nnano.2017.218} {\bibfield  {journal} {\bibinfo
  {journal} {Nature Nanotech}\ }\textbf {\bibinfo {volume} {12}},\ \bibinfo
  {pages} {1026–} (\bibinfo {year} {2017})}\BibitemShut {NoStop}%
\bibitem [{\citenamefont {Waks}\ \emph {et~al.}(2002)\citenamefont {Waks},
  \citenamefont {Inoue}, \citenamefont {Santori}, \citenamefont {Fattal},
  \citenamefont {Vuckovic}, \citenamefont {Solomon},\ and\ \citenamefont
  {Yamamoto}}]{waks2002quantum}%
  \BibitemOpen
  \bibfield  {author} {\bibinfo {author} {\bibfnamefont {E.}~\bibnamefont
  {Waks}}, \bibinfo {author} {\bibfnamefont {K.}~\bibnamefont {Inoue}},
  \bibinfo {author} {\bibfnamefont {C.}~\bibnamefont {Santori}}, \bibinfo
  {author} {\bibfnamefont {D.}~\bibnamefont {Fattal}}, \bibinfo {author}
  {\bibfnamefont {J.}~\bibnamefont {Vuckovic}}, \bibinfo {author}
  {\bibfnamefont {G.~S.}\ \bibnamefont {Solomon}}, \ and\ \bibinfo {author}
  {\bibfnamefont {Y.}~\bibnamefont {Yamamoto}},\ }\href {\doibase
  10.1038/420762a} {\bibfield  {journal} {\bibinfo  {journal} {Nature}\
  }\textbf {\bibinfo {volume} {420}},\ \bibinfo {pages} {762} (\bibinfo {year}
  {2002})}\BibitemShut {NoStop}%
\bibitem [{\citenamefont {Bozzio}\ \emph {et~al.}(2022)\citenamefont {Bozzio},
  \citenamefont {Vyvlecka}, \citenamefont {Cosacchi}, \citenamefont {Nawrath},
  \citenamefont {Seidelmann}, \citenamefont {Loredo}, \citenamefont
  {Portalupi}, \citenamefont {Axt}, \citenamefont {Michler},\ and\
  \citenamefont {Walther}}]{Bozzio2022}%
  \BibitemOpen
  \bibfield  {author} {\bibinfo {author} {\bibfnamefont {M.}~\bibnamefont
  {Bozzio}}, \bibinfo {author} {\bibfnamefont {M.}~\bibnamefont {Vyvlecka}},
  \bibinfo {author} {\bibfnamefont {M.}~\bibnamefont {Cosacchi}}, \bibinfo
  {author} {\bibfnamefont {C.}~\bibnamefont {Nawrath}}, \bibinfo {author}
  {\bibfnamefont {T.}~\bibnamefont {Seidelmann}}, \bibinfo {author}
  {\bibfnamefont {J.~C.}\ \bibnamefont {Loredo}}, \bibinfo {author}
  {\bibfnamefont {S.~L.}\ \bibnamefont {Portalupi}}, \bibinfo {author}
  {\bibfnamefont {V.~M.}\ \bibnamefont {Axt}}, \bibinfo {author} {\bibfnamefont
  {P.}~\bibnamefont {Michler}}, \ and\ \bibinfo {author} {\bibfnamefont
  {P.}~\bibnamefont {Walther}},\ }\href {\doibase 10.1038/s41534-022-00626-z}
  {\bibfield  {journal} {\bibinfo  {journal} {npj Quantum Information}\
  }\textbf {\bibinfo {volume} {8}},\ \bibinfo {pages} {104} (\bibinfo {year}
  {2022})}\BibitemShut {NoStop}%
\bibitem [{\citenamefont {Zwerger}\ \emph {et~al.}(2016)\citenamefont
  {Zwerger}, \citenamefont {Briegel},\ and\ \citenamefont
  {D{\"u}r}}]{zwerger2016measurement}%
  \BibitemOpen
  \bibfield  {author} {\bibinfo {author} {\bibfnamefont {M.}~\bibnamefont
  {Zwerger}}, \bibinfo {author} {\bibfnamefont {H.}~\bibnamefont {Briegel}}, \
  and\ \bibinfo {author} {\bibfnamefont {W.}~\bibnamefont {D{\"u}r}},\ }\href
  {\doibase 10.1007/s00340-015-6285-8} {\bibfield  {journal} {\bibinfo
  {journal} {Applied Physics B}\ }\textbf {\bibinfo {volume} {122}},\ \bibinfo
  {pages} {50} (\bibinfo {year} {2016})}\BibitemShut {NoStop}%
\bibitem [{\citenamefont {Lindner}\ and\ \citenamefont
  {Rudolph}(2009)}]{lindner2009proposal}%
  \BibitemOpen
  \bibfield  {author} {\bibinfo {author} {\bibfnamefont {N.~H.}\ \bibnamefont
  {Lindner}}\ and\ \bibinfo {author} {\bibfnamefont {T.}~\bibnamefont
  {Rudolph}},\ }\href {\doibase 10.1103/PhysRevLett.103.113602} {\bibfield
  {journal} {\bibinfo  {journal} {Physical Review Letters}\ }\textbf {\bibinfo
  {volume} {103}},\ \bibinfo {pages} {113602} (\bibinfo {year}
  {2009})}\BibitemShut {NoStop}%
\bibitem [{\citenamefont {Wang}\ \emph {et~al.}(2017)\citenamefont {Wang},
  \citenamefont {He}, \citenamefont {Li}, \citenamefont {Su}, \citenamefont
  {Li}, \citenamefont {Huang}, \citenamefont {Ding}, \citenamefont {Chen},
  \citenamefont {Liu}, \citenamefont {Qin}, \citenamefont {Li}, \citenamefont
  {He}, \citenamefont {Schneider}, \citenamefont {Kamp}, \citenamefont {Peng},
  \citenamefont {H\"{o}fling}, \citenamefont {Lu},\ and\ \citenamefont
  {Pan}}]{wang2017high}%
  \BibitemOpen
  \bibfield  {author} {\bibinfo {author} {\bibfnamefont {H.}~\bibnamefont
  {Wang}}, \bibinfo {author} {\bibfnamefont {Y.}~\bibnamefont {He}}, \bibinfo
  {author} {\bibfnamefont {Y.-H.}\ \bibnamefont {Li}}, \bibinfo {author}
  {\bibfnamefont {Z.-E.}\ \bibnamefont {Su}}, \bibinfo {author} {\bibfnamefont
  {B.}~\bibnamefont {Li}}, \bibinfo {author} {\bibfnamefont {H.-L.}\
  \bibnamefont {Huang}}, \bibinfo {author} {\bibfnamefont {X.}~\bibnamefont
  {Ding}}, \bibinfo {author} {\bibfnamefont {M.-C.}\ \bibnamefont {Chen}},
  \bibinfo {author} {\bibfnamefont {C.}~\bibnamefont {Liu}}, \bibinfo {author}
  {\bibfnamefont {J.}~\bibnamefont {Qin}}, \bibinfo {author} {\bibfnamefont
  {J.-P.}\ \bibnamefont {Li}}, \bibinfo {author} {\bibfnamefont {Y.-M.}\
  \bibnamefont {He}}, \bibinfo {author} {\bibfnamefont {C.}~\bibnamefont
  {Schneider}}, \bibinfo {author} {\bibfnamefont {M.}~\bibnamefont {Kamp}},
  \bibinfo {author} {\bibfnamefont {C.-Z.}\ \bibnamefont {Peng}}, \bibinfo
  {author} {\bibfnamefont {S.}~\bibnamefont {H\"{o}fling}}, \bibinfo {author}
  {\bibfnamefont {C.-Y.}\ \bibnamefont {Lu}}, \ and\ \bibinfo {author}
  {\bibfnamefont {J.-W.}\ \bibnamefont {Pan}},\ }\href {\doibase
  10.1038/nphoton.2017.63} {\bibfield  {journal} {\bibinfo  {journal} {Nature
  Photonics}\ }\textbf {\bibinfo {volume} {11}},\ \bibinfo {pages} {361}
  (\bibinfo {year} {2017})}\BibitemShut {NoStop}%
\bibitem [{\citenamefont {Brassard}\ \emph {et~al.}(2000)\citenamefont
  {Brassard}, \citenamefont {L{\"u}tkenhaus}, \citenamefont {Mor},\ and\
  \citenamefont {Sanders}}]{brassard2000limitations}%
  \BibitemOpen
  \bibfield  {author} {\bibinfo {author} {\bibfnamefont {G.}~\bibnamefont
  {Brassard}}, \bibinfo {author} {\bibfnamefont {N.}~\bibnamefont
  {L{\"u}tkenhaus}}, \bibinfo {author} {\bibfnamefont {T.}~\bibnamefont {Mor}},
  \ and\ \bibinfo {author} {\bibfnamefont {B.~C.}\ \bibnamefont {Sanders}},\
  }\href {\doibase 10.1103/PhysRevLett.85.1330} {\bibfield  {journal} {\bibinfo
   {journal} {Physical Review Letters}\ }\textbf {\bibinfo {volume} {85}},\
  \bibinfo {pages} {1330} (\bibinfo {year} {2000})}\BibitemShut {NoStop}%
\bibitem [{\citenamefont {Rau}\ \emph {et~al.}(2014)\citenamefont {Rau},
  \citenamefont {Heindel}, \citenamefont {Unsleber}, \citenamefont {Braun},
  \citenamefont {Fischer}, \citenamefont {Frick}, \citenamefont {Nauerth},
  \citenamefont {Schneider}, \citenamefont {Vest}, \citenamefont
  {Reitzenstein}, \citenamefont {Kamp}, \citenamefont {Forchel}, \citenamefont
  {H\"{o}fling},\ and\ \citenamefont {Weinfurter}}]{rau2014free}%
  \BibitemOpen
  \bibfield  {author} {\bibinfo {author} {\bibfnamefont {M.}~\bibnamefont
  {Rau}}, \bibinfo {author} {\bibfnamefont {T.}~\bibnamefont {Heindel}},
  \bibinfo {author} {\bibfnamefont {S.}~\bibnamefont {Unsleber}}, \bibinfo
  {author} {\bibfnamefont {T.}~\bibnamefont {Braun}}, \bibinfo {author}
  {\bibfnamefont {J.}~\bibnamefont {Fischer}}, \bibinfo {author} {\bibfnamefont
  {S.}~\bibnamefont {Frick}}, \bibinfo {author} {\bibfnamefont
  {S.}~\bibnamefont {Nauerth}}, \bibinfo {author} {\bibfnamefont
  {C.}~\bibnamefont {Schneider}}, \bibinfo {author} {\bibfnamefont
  {G.}~\bibnamefont {Vest}}, \bibinfo {author} {\bibfnamefont {S.}~\bibnamefont
  {Reitzenstein}}, \bibinfo {author} {\bibfnamefont {M.}~\bibnamefont {Kamp}},
  \bibinfo {author} {\bibfnamefont {A.}~\bibnamefont {Forchel}}, \bibinfo
  {author} {\bibfnamefont {S.}~\bibnamefont {H\"{o}fling}}, \ and\ \bibinfo
  {author} {\bibfnamefont {H.}~\bibnamefont {Weinfurter}},\ }\href {\doibase
  10.1088/1367-2630/16/4/043003} {\bibfield  {journal} {\bibinfo  {journal}
  {New Journal of Physics}\ }\textbf {\bibinfo {volume} {16}},\ \bibinfo
  {pages} {043003} (\bibinfo {year} {2014})}\BibitemShut {NoStop}%
\bibitem [{\citenamefont {Takemoto}\ \emph {et~al.}(2015)\citenamefont
  {Takemoto}, \citenamefont {Nambu}, \citenamefont {Miyazawa}, \citenamefont
  {Sakuma}, \citenamefont {Yamamoto}, \citenamefont {Yorozu},\ and\
  \citenamefont {Arakawa}}]{takemoto2015quantum}%
  \BibitemOpen
  \bibfield  {author} {\bibinfo {author} {\bibfnamefont {K.}~\bibnamefont
  {Takemoto}}, \bibinfo {author} {\bibfnamefont {Y.}~\bibnamefont {Nambu}},
  \bibinfo {author} {\bibfnamefont {T.}~\bibnamefont {Miyazawa}}, \bibinfo
  {author} {\bibfnamefont {Y.}~\bibnamefont {Sakuma}}, \bibinfo {author}
  {\bibfnamefont {T.}~\bibnamefont {Yamamoto}}, \bibinfo {author}
  {\bibfnamefont {S.}~\bibnamefont {Yorozu}}, \ and\ \bibinfo {author}
  {\bibfnamefont {Y.}~\bibnamefont {Arakawa}},\ }\href {\doibase
  10.1038/srep14383} {\bibfield  {journal} {\bibinfo  {journal} {Scientific
  Reports}\ }\textbf {\bibinfo {volume} {5}},\ \bibinfo {pages} {1} (\bibinfo
  {year} {2015})}\BibitemShut {NoStop}%
\bibitem [{\citenamefont {Basso~Basset}\ \emph {et~al.}(2021)\citenamefont
  {Basso~Basset}, \citenamefont {Valeri}, \citenamefont {Roccia}, \citenamefont
  {Muredda}, \citenamefont {Poderini}, \citenamefont {Neuwirth}, \citenamefont
  {Spagnolo}, \citenamefont {Rota}, \citenamefont {Carvacho}, \citenamefont
  {Sciarrino},\ and\ \citenamefont {Trotta}}]{basso2021quantum}%
  \BibitemOpen
  \bibfield  {author} {\bibinfo {author} {\bibfnamefont {F.}~\bibnamefont
  {Basso~Basset}}, \bibinfo {author} {\bibfnamefont {M.}~\bibnamefont
  {Valeri}}, \bibinfo {author} {\bibfnamefont {E.}~\bibnamefont {Roccia}},
  \bibinfo {author} {\bibfnamefont {V.}~\bibnamefont {Muredda}}, \bibinfo
  {author} {\bibfnamefont {D.}~\bibnamefont {Poderini}}, \bibinfo {author}
  {\bibfnamefont {J.}~\bibnamefont {Neuwirth}}, \bibinfo {author}
  {\bibfnamefont {N.}~\bibnamefont {Spagnolo}}, \bibinfo {author}
  {\bibfnamefont {M.~B.}\ \bibnamefont {Rota}}, \bibinfo {author}
  {\bibfnamefont {G.}~\bibnamefont {Carvacho}}, \bibinfo {author}
  {\bibfnamefont {F.}~\bibnamefont {Sciarrino}}, \ and\ \bibinfo {author}
  {\bibfnamefont {R.}~\bibnamefont {Trotta}},\ }\href {\doibase
  10.1126/sciadv.abe6379} {\bibfield  {journal} {\bibinfo  {journal} {Science
  Advances}\ }\textbf {\bibinfo {volume} {7}},\ \bibinfo {pages} {eabe6379}
  (\bibinfo {year} {2021})}\BibitemShut {NoStop}%
\bibitem [{\citenamefont {Schimpf}\ \emph
  {et~al.}(2021{\natexlab{b}})\citenamefont {Schimpf}, \citenamefont {Reindl},
  \citenamefont {Huber}, \citenamefont {Lehner}, \citenamefont {Covre
  Da~Silva}, \citenamefont {Manna}, \citenamefont {Vyvlecka}, \citenamefont
  {Walther},\ and\ \citenamefont {Rastelli}}]{schimpf2021quantum}%
  \BibitemOpen
  \bibfield  {author} {\bibinfo {author} {\bibfnamefont {C.}~\bibnamefont
  {Schimpf}}, \bibinfo {author} {\bibfnamefont {M.}~\bibnamefont {Reindl}},
  \bibinfo {author} {\bibfnamefont {D.}~\bibnamefont {Huber}}, \bibinfo
  {author} {\bibfnamefont {B.}~\bibnamefont {Lehner}}, \bibinfo {author}
  {\bibfnamefont {S.~F.}\ \bibnamefont {Covre Da~Silva}}, \bibinfo {author}
  {\bibfnamefont {S.}~\bibnamefont {Manna}}, \bibinfo {author} {\bibfnamefont
  {M.}~\bibnamefont {Vyvlecka}}, \bibinfo {author} {\bibfnamefont
  {P.}~\bibnamefont {Walther}}, \ and\ \bibinfo {author} {\bibfnamefont
  {A.}~\bibnamefont {Rastelli}},\ }\href {\doibase 10.1126/sciadv.abe890}
  {\bibfield  {journal} {\bibinfo  {journal} {Science Advances}\ }\textbf
  {\bibinfo {volume} {7}},\ \bibinfo {pages} {eabe8905} (\bibinfo {year}
  {2021}{\natexlab{b}})}\BibitemShut {NoStop}%
\bibitem [{\citenamefont {Basso~Basset}\ \emph {et~al.}(2022)\citenamefont
  {Basso~Basset}, \citenamefont {Valeri}, \citenamefont {Neuwirth},
  \citenamefont {Polino}, \citenamefont {Rota}, \citenamefont {Poderini},
  \citenamefont {Pardo}, \citenamefont {Rodari}, \citenamefont {Roccia},
  \citenamefont {Covre~da Silva}, \citenamefont {Ronco}, \citenamefont
  {Spagnolo}, \citenamefont {Rastelli}, \citenamefont {Carvacho}, \citenamefont
  {Sciarrino},\ and\ \citenamefont {Trotta}}]{basso2022daylight}%
  \BibitemOpen
  \bibfield  {author} {\bibinfo {author} {\bibfnamefont {F.}~\bibnamefont
  {Basso~Basset}}, \bibinfo {author} {\bibfnamefont {M.}~\bibnamefont
  {Valeri}}, \bibinfo {author} {\bibfnamefont {J.}~\bibnamefont {Neuwirth}},
  \bibinfo {author} {\bibfnamefont {E.}~\bibnamefont {Polino}}, \bibinfo
  {author} {\bibfnamefont {M.~B.}\ \bibnamefont {Rota}}, \bibinfo {author}
  {\bibfnamefont {D.}~\bibnamefont {Poderini}}, \bibinfo {author}
  {\bibfnamefont {C.}~\bibnamefont {Pardo}}, \bibinfo {author} {\bibfnamefont
  {G.}~\bibnamefont {Rodari}}, \bibinfo {author} {\bibfnamefont
  {E.}~\bibnamefont {Roccia}}, \bibinfo {author} {\bibfnamefont
  {S.}~\bibnamefont {Covre~da Silva}}, \bibinfo {author} {\bibfnamefont
  {G.}~\bibnamefont {Ronco}}, \bibinfo {author} {\bibfnamefont
  {N.}~\bibnamefont {Spagnolo}}, \bibinfo {author} {\bibfnamefont
  {A.}~\bibnamefont {Rastelli}}, \bibinfo {author} {\bibfnamefont
  {G.}~\bibnamefont {Carvacho}}, \bibinfo {author} {\bibfnamefont
  {F.}~\bibnamefont {Sciarrino}}, \ and\ \bibinfo {author} {\bibfnamefont
  {R.}~\bibnamefont {Trotta}},\ }\href {\doibase 10.1088/2058-9565/acae3d}
  {\bibfield  {journal} {\bibinfo  {journal} {Quantum Science and Technology}\
  }\textbf {\bibinfo {volume} {8}},\ \bibinfo {pages} {025002} (\bibinfo {year}
  {2022})}\BibitemShut {NoStop}%
\bibitem [{\citenamefont {Hillery}\ \emph {et~al.}(1999)\citenamefont
  {Hillery}, \citenamefont {Bu\ifmmode~\check{z}\else \v{z}\fi{}ek},\ and\
  \citenamefont {Berthiaume}}]{PhysRevA.59.1829}%
  \BibitemOpen
  \bibfield  {author} {\bibinfo {author} {\bibfnamefont {M.}~\bibnamefont
  {Hillery}}, \bibinfo {author} {\bibfnamefont {V.}~\bibnamefont
  {Bu\ifmmode~\check{z}\else \v{z}\fi{}ek}}, \ and\ \bibinfo {author}
  {\bibfnamefont {A.}~\bibnamefont {Berthiaume}},\ }\href {\doibase
  10.1103/PhysRevA.59.1829} {\bibfield  {journal} {\bibinfo  {journal} {Phys.
  Rev. A}\ }\textbf {\bibinfo {volume} {59}},\ \bibinfo {pages} {1829}
  (\bibinfo {year} {1999})}\BibitemShut {NoStop}%
\bibitem [{\citenamefont {Proietti}\ \emph {et~al.}(2021)\citenamefont
  {Proietti}, \citenamefont {Ho}, \citenamefont {Grasselli}, \citenamefont
  {Barrow}, \citenamefont {Malik},\ and\ \citenamefont
  {Fedrizzi}}]{doi:10.1126/sciadv.abe0395}%
  \BibitemOpen
  \bibfield  {author} {\bibinfo {author} {\bibfnamefont {M.}~\bibnamefont
  {Proietti}}, \bibinfo {author} {\bibfnamefont {J.}~\bibnamefont {Ho}},
  \bibinfo {author} {\bibfnamefont {F.}~\bibnamefont {Grasselli}}, \bibinfo
  {author} {\bibfnamefont {P.}~\bibnamefont {Barrow}}, \bibinfo {author}
  {\bibfnamefont {M.}~\bibnamefont {Malik}}, \ and\ \bibinfo {author}
  {\bibfnamefont {A.}~\bibnamefont {Fedrizzi}},\ }\href {\doibase
  10.1126/sciadv.abe0395} {\bibfield  {journal} {\bibinfo  {journal} {Science
  Advances}\ }\textbf {\bibinfo {volume} {7}},\ \bibinfo {pages} {eabe0395}
  (\bibinfo {year} {2021})}\BibitemShut {NoStop}%
\bibitem [{\citenamefont {Greenberger}\ \emph {et~al.}(1989)\citenamefont
  {Greenberger}, \citenamefont {Horne},\ and\ \citenamefont
  {Zeilinger}}]{Greenberger1989}%
  \BibitemOpen
  \bibfield  {author} {\bibinfo {author} {\bibfnamefont {D.~M.}\ \bibnamefont
  {Greenberger}}, \bibinfo {author} {\bibfnamefont {M.~A.}\ \bibnamefont
  {Horne}}, \ and\ \bibinfo {author} {\bibfnamefont {A.}~\bibnamefont
  {Zeilinger}},\ }\enquote {\bibinfo {title} {Going beyond bell's theorem},}\
  in\ \href {\doibase 10.1007/978-94-017-0849-4_10} {\emph {\bibinfo
  {booktitle} {Bell's Theorem, Quantum Theory and Conceptions of the
  Universe}}},\ \bibinfo {editor} {edited by\ \bibinfo {editor} {\bibfnamefont
  {M.}~\bibnamefont {Kafatos}}}\ (\bibinfo  {publisher} {Springer
  Netherlands},\ \bibinfo {address} {Dordrecht},\ \bibinfo {year} {1989})\ pp.\
  \bibinfo {pages} {69--72}\BibitemShut {NoStop}%
\bibitem [{\citenamefont {Zhong}\ \emph {et~al.}(2018)\citenamefont {Zhong},
  \citenamefont {Li}, \citenamefont {Li}, \citenamefont {Peng}, \citenamefont
  {Su}, \citenamefont {Hu}, \citenamefont {He}, \citenamefont {Ding},
  \citenamefont {Zhang}, \citenamefont {Li}, \citenamefont {Zhang},
  \citenamefont {Wang}, \citenamefont {You}, \citenamefont {Wang},
  \citenamefont {Jiang}, \citenamefont {Li}, \citenamefont {Chen},
  \citenamefont {Liu}, \citenamefont {Lu},\ and\ \citenamefont
  {Pan}}]{zhong201812}%
  \BibitemOpen
  \bibfield  {author} {\bibinfo {author} {\bibfnamefont {H.-S.}\ \bibnamefont
  {Zhong}}, \bibinfo {author} {\bibfnamefont {Y.}~\bibnamefont {Li}}, \bibinfo
  {author} {\bibfnamefont {W.}~\bibnamefont {Li}}, \bibinfo {author}
  {\bibfnamefont {L.-C.}\ \bibnamefont {Peng}}, \bibinfo {author}
  {\bibfnamefont {Z.-E.}\ \bibnamefont {Su}}, \bibinfo {author} {\bibfnamefont
  {Y.}~\bibnamefont {Hu}}, \bibinfo {author} {\bibfnamefont {Y.-M.}\
  \bibnamefont {He}}, \bibinfo {author} {\bibfnamefont {X.}~\bibnamefont
  {Ding}}, \bibinfo {author} {\bibfnamefont {W.}~\bibnamefont {Zhang}},
  \bibinfo {author} {\bibfnamefont {H.}~\bibnamefont {Li}}, \bibinfo {author}
  {\bibfnamefont {L.}~\bibnamefont {Zhang}}, \bibinfo {author} {\bibfnamefont
  {Z.}~\bibnamefont {Wang}}, \bibinfo {author} {\bibfnamefont {L.}~\bibnamefont
  {You}}, \bibinfo {author} {\bibfnamefont {X.-L.}\ \bibnamefont {Wang}},
  \bibinfo {author} {\bibfnamefont {X.}~\bibnamefont {Jiang}}, \bibinfo
  {author} {\bibfnamefont {L.}~\bibnamefont {Li}}, \bibinfo {author}
  {\bibfnamefont {Y.-A.}\ \bibnamefont {Chen}}, \bibinfo {author}
  {\bibfnamefont {N.-L.}\ \bibnamefont {Liu}}, \bibinfo {author} {\bibfnamefont
  {C.-Y.}\ \bibnamefont {Lu}}, \ and\ \bibinfo {author} {\bibfnamefont {J.-W.}\
  \bibnamefont {Pan}},\ }\href {\doibase 10.1103/PhysRevLett.121.250505}
  {\bibfield  {journal} {\bibinfo  {journal} {Physical Review Letters}\
  }\textbf {\bibinfo {volume} {121}},\ \bibinfo {pages} {250505} (\bibinfo
  {year} {2018})}\BibitemShut {NoStop}%
\bibitem [{\citenamefont {Li}\ \emph {et~al.}(2021)\citenamefont {Li},
  \citenamefont {Gu}, \citenamefont {Qin}, \citenamefont {Wu}, \citenamefont
  {You}, \citenamefont {Wang}, \citenamefont {Schneider}, \citenamefont
  {H\"ofling}, \citenamefont {Huo}, \citenamefont {Lu}, \citenamefont {Liu},
  \citenamefont {Li},\ and\ \citenamefont {Pan}}]{PhysRevLett.126.140501}%
  \BibitemOpen
  \bibfield  {author} {\bibinfo {author} {\bibfnamefont {J.-P.}\ \bibnamefont
  {Li}}, \bibinfo {author} {\bibfnamefont {X.}~\bibnamefont {Gu}}, \bibinfo
  {author} {\bibfnamefont {J.}~\bibnamefont {Qin}}, \bibinfo {author}
  {\bibfnamefont {D.}~\bibnamefont {Wu}}, \bibinfo {author} {\bibfnamefont
  {X.}~\bibnamefont {You}}, \bibinfo {author} {\bibfnamefont {H.}~\bibnamefont
  {Wang}}, \bibinfo {author} {\bibfnamefont {C.}~\bibnamefont {Schneider}},
  \bibinfo {author} {\bibfnamefont {S.}~\bibnamefont {H\"ofling}}, \bibinfo
  {author} {\bibfnamefont {Y.-H.}\ \bibnamefont {Huo}}, \bibinfo {author}
  {\bibfnamefont {C.-Y.}\ \bibnamefont {Lu}}, \bibinfo {author} {\bibfnamefont
  {N.-L.}\ \bibnamefont {Liu}}, \bibinfo {author} {\bibfnamefont
  {L.}~\bibnamefont {Li}}, \ and\ \bibinfo {author} {\bibfnamefont {J.-W.}\
  \bibnamefont {Pan}},\ }\href {\doibase 10.1103/PhysRevLett.126.140501}
  {\bibfield  {journal} {\bibinfo  {journal} {Phys. Rev. Lett.}\ }\textbf
  {\bibinfo {volume} {126}},\ \bibinfo {pages} {140501} (\bibinfo {year}
  {2021})}\BibitemShut {NoStop}%
\bibitem [{\citenamefont {Pont}\ \emph {et~al.}(2022)\citenamefont {Pont},
  \citenamefont {Corrielli}, \citenamefont {Fyrillas}, \citenamefont {Agresti},
  \citenamefont {Carvacho}, \citenamefont {Maring}, \citenamefont {Emeriau},
  \citenamefont {Ceccarelli}, \citenamefont {Albiero}, \citenamefont
  {Ferreira}, \citenamefont {Somaschi}, \citenamefont {Senellart},
  \citenamefont {Sagnes}, \citenamefont {Morassi}, \citenamefont {Lemaitre},
  \citenamefont {Senellart}, \citenamefont {Sciarrino}, \citenamefont
  {Liscidini}, \citenamefont {Belabas},\ and\ \citenamefont
  {Osellame}}]{pont2022highfidelity}%
  \BibitemOpen
  \bibfield  {author} {\bibinfo {author} {\bibfnamefont {M.}~\bibnamefont
  {Pont}}, \bibinfo {author} {\bibfnamefont {G.}~\bibnamefont {Corrielli}},
  \bibinfo {author} {\bibfnamefont {A.}~\bibnamefont {Fyrillas}}, \bibinfo
  {author} {\bibfnamefont {I.}~\bibnamefont {Agresti}}, \bibinfo {author}
  {\bibfnamefont {G.}~\bibnamefont {Carvacho}}, \bibinfo {author}
  {\bibfnamefont {N.}~\bibnamefont {Maring}}, \bibinfo {author} {\bibfnamefont
  {P.-E.}\ \bibnamefont {Emeriau}}, \bibinfo {author} {\bibfnamefont
  {F.}~\bibnamefont {Ceccarelli}}, \bibinfo {author} {\bibfnamefont
  {R.}~\bibnamefont {Albiero}}, \bibinfo {author} {\bibfnamefont {P.~H.~D.}\
  \bibnamefont {Ferreira}}, \bibinfo {author} {\bibfnamefont {N.}~\bibnamefont
  {Somaschi}}, \bibinfo {author} {\bibfnamefont {J.}~\bibnamefont {Senellart}},
  \bibinfo {author} {\bibfnamefont {I.}~\bibnamefont {Sagnes}}, \bibinfo
  {author} {\bibfnamefont {M.}~\bibnamefont {Morassi}}, \bibinfo {author}
  {\bibfnamefont {A.}~\bibnamefont {Lemaitre}}, \bibinfo {author}
  {\bibfnamefont {P.}~\bibnamefont {Senellart}}, \bibinfo {author}
  {\bibfnamefont {F.}~\bibnamefont {Sciarrino}}, \bibinfo {author}
  {\bibfnamefont {M.}~\bibnamefont {Liscidini}}, \bibinfo {author}
  {\bibfnamefont {N.}~\bibnamefont {Belabas}}, \ and\ \bibinfo {author}
  {\bibfnamefont {R.}~\bibnamefont {Osellame}},\ }\href@noop {} {\enquote
  {\bibinfo {title} {High-fidelity generation of four-photon ghz states
  on-chip},}\ } (\bibinfo {year} {2022}),\ \Eprint
  {http://arxiv.org/abs/2211.15626} {arXiv:2211.15626 [quant-ph]} \BibitemShut
  {NoStop}%
\bibitem [{\citenamefont {Suprano}\ \emph {et~al.}(2022)\citenamefont
  {Suprano}, \citenamefont {Zia}, \citenamefont {Pont}, \citenamefont
  {Giordani}, \citenamefont {Rodari}, \citenamefont {Valeri}, \citenamefont
  {Piccirillo}, \citenamefont {Carvacho}, \citenamefont {Spagnolo},
  \citenamefont {Senellart}, \citenamefont {Marrucci},\ and\ \citenamefont
  {Sciarrino}}]{suprano2022orbital}%
  \BibitemOpen
  \bibfield  {author} {\bibinfo {author} {\bibfnamefont {A.}~\bibnamefont
  {Suprano}}, \bibinfo {author} {\bibfnamefont {D.}~\bibnamefont {Zia}},
  \bibinfo {author} {\bibfnamefont {M.}~\bibnamefont {Pont}}, \bibinfo {author}
  {\bibfnamefont {T.}~\bibnamefont {Giordani}}, \bibinfo {author}
  {\bibfnamefont {G.}~\bibnamefont {Rodari}}, \bibinfo {author} {\bibfnamefont
  {M.}~\bibnamefont {Valeri}}, \bibinfo {author} {\bibfnamefont
  {B.}~\bibnamefont {Piccirillo}}, \bibinfo {author} {\bibfnamefont
  {G.}~\bibnamefont {Carvacho}}, \bibinfo {author} {\bibfnamefont
  {N.}~\bibnamefont {Spagnolo}}, \bibinfo {author} {\bibfnamefont
  {P.}~\bibnamefont {Senellart}}, \bibinfo {author} {\bibfnamefont
  {L.}~\bibnamefont {Marrucci}}, \ and\ \bibinfo {author} {\bibfnamefont
  {F.}~\bibnamefont {Sciarrino}},\ }\href@noop {} {\enquote {\bibinfo {title}
  {Orbital angular momentum based intra-and inter-particle entangled states
  generated via a quantum dot source},}\ } (\bibinfo {year} {2022}),\ \Eprint
  {http://arxiv.org/abs/2211.05160} {arXiv:2211.05160 [quant-ph]} \BibitemShut
  {NoStop}%
\bibitem [{\citenamefont {Istrati}\ \emph {et~al.}(2020)\citenamefont
  {Istrati}, \citenamefont {Pilnyak}, \citenamefont {Loredo}, \citenamefont
  {Ant{\'o}n}, \citenamefont {Somaschi}, \citenamefont {Hilaire}, \citenamefont
  {Ollivier}, \citenamefont {Esmann}, \citenamefont {Cohen}, \citenamefont
  {Vidro}, \citenamefont {Millet}, \citenamefont {Lemaitre}, \citenamefont
  {Sagnes}, \citenamefont {Harouri}, \citenamefont {Lanco}, \citenamefont
  {Senellart},\ and\ \citenamefont {Eisenberg}}]{istrati2020sequential}%
  \BibitemOpen
  \bibfield  {author} {\bibinfo {author} {\bibfnamefont {D.}~\bibnamefont
  {Istrati}}, \bibinfo {author} {\bibfnamefont {Y.}~\bibnamefont {Pilnyak}},
  \bibinfo {author} {\bibfnamefont {J.}~\bibnamefont {Loredo}}, \bibinfo
  {author} {\bibfnamefont {C.}~\bibnamefont {Ant{\'o}n}}, \bibinfo {author}
  {\bibfnamefont {N.}~\bibnamefont {Somaschi}}, \bibinfo {author}
  {\bibfnamefont {P.}~\bibnamefont {Hilaire}}, \bibinfo {author} {\bibfnamefont
  {H.}~\bibnamefont {Ollivier}}, \bibinfo {author} {\bibfnamefont
  {M.}~\bibnamefont {Esmann}}, \bibinfo {author} {\bibfnamefont
  {L.}~\bibnamefont {Cohen}}, \bibinfo {author} {\bibfnamefont
  {L.}~\bibnamefont {Vidro}}, \bibinfo {author} {\bibfnamefont
  {C.}~\bibnamefont {Millet}}, \bibinfo {author} {\bibfnamefont
  {A.}~\bibnamefont {Lemaitre}}, \bibinfo {author} {\bibfnamefont
  {I.}~\bibnamefont {Sagnes}}, \bibinfo {author} {\bibfnamefont
  {A.}~\bibnamefont {Harouri}}, \bibinfo {author} {\bibfnamefont
  {L.}~\bibnamefont {Lanco}}, \bibinfo {author} {\bibfnamefont
  {P.}~\bibnamefont {Senellart}}, \ and\ \bibinfo {author} {\bibfnamefont
  {H.~S.}\ \bibnamefont {Eisenberg}},\ }\href {\doibase
  10.1038/s41467-020-19341-4} {\bibfield  {journal} {\bibinfo  {journal}
  {Nature Communications}\ }\textbf {\bibinfo {volume} {11}},\ \bibinfo {pages}
  {1} (\bibinfo {year} {2020})}\BibitemShut {NoStop}%
\bibitem [{\citenamefont {Li}\ \emph {et~al.}(2020)\citenamefont {Li},
  \citenamefont {Qin}, \citenamefont {Chen}, \citenamefont {Duan},
  \citenamefont {Yu}, \citenamefont {Huo}, \citenamefont {H\"{o}fling},
  \citenamefont {Lu}, \citenamefont {Chen},\ and\ \citenamefont
  {Pan}}]{li2020multiphoton}%
  \BibitemOpen
  \bibfield  {author} {\bibinfo {author} {\bibfnamefont {J.-P.}\ \bibnamefont
  {Li}}, \bibinfo {author} {\bibfnamefont {J.}~\bibnamefont {Qin}}, \bibinfo
  {author} {\bibfnamefont {A.}~\bibnamefont {Chen}}, \bibinfo {author}
  {\bibfnamefont {Z.-C.}\ \bibnamefont {Duan}}, \bibinfo {author}
  {\bibfnamefont {Y.}~\bibnamefont {Yu}}, \bibinfo {author} {\bibfnamefont
  {Y.}~\bibnamefont {Huo}}, \bibinfo {author} {\bibfnamefont {S.}~\bibnamefont
  {H\"{o}fling}}, \bibinfo {author} {\bibfnamefont {C.-Y.}\ \bibnamefont {Lu}},
  \bibinfo {author} {\bibfnamefont {K.}~\bibnamefont {Chen}}, \ and\ \bibinfo
  {author} {\bibfnamefont {J.-W.}\ \bibnamefont {Pan}},\ }\href
  {https://pubs.acs.org/doi/pdf/10.1021/acsphotonics.0c00192?casa_token=JJPhKAEUCJoAAAAA:StJ9ex7IYQS9t9aUg3JgOEZI8aCFI5j72rFSdom6oLWuRV8qaUGwN6ANDyYlqQ2lnYCOtCFl3iF5RQ}
  {\bibfield  {journal} {\bibinfo  {journal} {ACS Photonics}\ }\textbf
  {\bibinfo {volume} {7}},\ \bibinfo {pages} {1603} (\bibinfo {year}
  {2020})}\BibitemShut {NoStop}%
\bibitem [{\citenamefont {Dan~Cogan}(2023)}]{cogan}%
  \BibitemOpen
  \bibfield  {author} {\bibinfo {author} {\bibfnamefont {O.~K. . D.~G.}\
  \bibnamefont {Dan~Cogan}, \bibfnamefont {Zu-En~Su}},\ }\href {\doibase
  https://doi.org/10.1038/s41566-022-01152-2} {\bibfield  {journal} {\bibinfo
  {journal} {Nature Nanotech}\ }\textbf {\bibinfo {volume} {17}},\ \bibinfo
  {pages} {324} (\bibinfo {year} {2023})}\BibitemShut {NoStop}%
\bibitem [{\citenamefont {De~Greve}\ \emph {et~al.}(2012)\citenamefont
  {De~Greve}, \citenamefont {Yu}, \citenamefont {McMahon}, \citenamefont
  {Pelc}, \citenamefont {Natarajan}, \citenamefont {Kim}, \citenamefont {Abe},
  \citenamefont {Maier}, \citenamefont {Schneider}, \citenamefont {Kamp},
  \citenamefont {H\"{o}fling}, \citenamefont {Hadfield}, \citenamefont
  {Forchel}, \citenamefont {Fejer},\ and\ \citenamefont
  {Yamamoto}}]{de2012quantum}%
  \BibitemOpen
  \bibfield  {author} {\bibinfo {author} {\bibfnamefont {K.}~\bibnamefont
  {De~Greve}}, \bibinfo {author} {\bibfnamefont {L.}~\bibnamefont {Yu}},
  \bibinfo {author} {\bibfnamefont {P.~L.}\ \bibnamefont {McMahon}}, \bibinfo
  {author} {\bibfnamefont {J.~S.}\ \bibnamefont {Pelc}}, \bibinfo {author}
  {\bibfnamefont {C.~M.}\ \bibnamefont {Natarajan}}, \bibinfo {author}
  {\bibfnamefont {N.~Y.}\ \bibnamefont {Kim}}, \bibinfo {author} {\bibfnamefont
  {E.}~\bibnamefont {Abe}}, \bibinfo {author} {\bibfnamefont {S.}~\bibnamefont
  {Maier}}, \bibinfo {author} {\bibfnamefont {C.}~\bibnamefont {Schneider}},
  \bibinfo {author} {\bibfnamefont {M.}~\bibnamefont {Kamp}}, \bibinfo {author}
  {\bibfnamefont {S.}~\bibnamefont {H\"{o}fling}}, \bibinfo {author}
  {\bibfnamefont {R.~H.}\ \bibnamefont {Hadfield}}, \bibinfo {author}
  {\bibfnamefont {A.}~\bibnamefont {Forchel}}, \bibinfo {author} {\bibfnamefont
  {M.~M.}\ \bibnamefont {Fejer}}, \ and\ \bibinfo {author} {\bibfnamefont
  {Y.}~\bibnamefont {Yamamoto}},\ }\href {\doibase 10.1038/nature11577}
  {\bibfield  {journal} {\bibinfo  {journal} {Nature}\ }\textbf {\bibinfo
  {volume} {491}},\ \bibinfo {pages} {421} (\bibinfo {year}
  {2012})}\BibitemShut {NoStop}%
\bibitem [{\citenamefont {Schwartz}\ \emph {et~al.}(2016)\citenamefont
  {Schwartz}, \citenamefont {Cogan}, \citenamefont {Schmidgall}, \citenamefont
  {Don}, \citenamefont {Gantz}, \citenamefont {Kenneth}, \citenamefont
  {Lindner},\ and\ \citenamefont {Gershoni}}]{schwartz2016deterministic}%
  \BibitemOpen
  \bibfield  {author} {\bibinfo {author} {\bibfnamefont {I.}~\bibnamefont
  {Schwartz}}, \bibinfo {author} {\bibfnamefont {D.}~\bibnamefont {Cogan}},
  \bibinfo {author} {\bibfnamefont {E.~R.}\ \bibnamefont {Schmidgall}},
  \bibinfo {author} {\bibfnamefont {Y.}~\bibnamefont {Don}}, \bibinfo {author}
  {\bibfnamefont {L.}~\bibnamefont {Gantz}}, \bibinfo {author} {\bibfnamefont
  {O.}~\bibnamefont {Kenneth}}, \bibinfo {author} {\bibfnamefont {N.~H.}\
  \bibnamefont {Lindner}}, \ and\ \bibinfo {author} {\bibfnamefont
  {D.}~\bibnamefont {Gershoni}},\ }\href
  {https://www.science.org/doi/full/10.1126/science.aah4758} {\bibfield
  {journal} {\bibinfo  {journal} {Science}\ }\textbf {\bibinfo {volume}
  {354}},\ \bibinfo {pages} {434} (\bibinfo {year} {2016})}\BibitemShut
  {NoStop}%
\bibitem [{\citenamefont {Wein}\ \emph {et~al.}(2022)\citenamefont {Wein},
  \citenamefont {Loredo}, \citenamefont {Maffei}, \citenamefont {Hilaire},
  \citenamefont {Harouri}, \citenamefont {Somaschi}, \citenamefont
  {Lema{\^\i}tre}, \citenamefont {Sagnes}, \citenamefont {Lanco}, \citenamefont
  {Krebs}, \citenamefont {Auffeves}, \citenamefont {Simon}, \citenamefont
  {Senellart},\ and\ \citenamefont {Anton-Solanas}}]{wein2022photon}%
  \BibitemOpen
  \bibfield  {author} {\bibinfo {author} {\bibfnamefont {S.~C.}\ \bibnamefont
  {Wein}}, \bibinfo {author} {\bibfnamefont {J.~C.}\ \bibnamefont {Loredo}},
  \bibinfo {author} {\bibfnamefont {M.}~\bibnamefont {Maffei}}, \bibinfo
  {author} {\bibfnamefont {P.}~\bibnamefont {Hilaire}}, \bibinfo {author}
  {\bibfnamefont {A.}~\bibnamefont {Harouri}}, \bibinfo {author} {\bibfnamefont
  {N.}~\bibnamefont {Somaschi}}, \bibinfo {author} {\bibfnamefont
  {A.}~\bibnamefont {Lema{\^\i}tre}}, \bibinfo {author} {\bibfnamefont
  {I.}~\bibnamefont {Sagnes}}, \bibinfo {author} {\bibfnamefont
  {L.}~\bibnamefont {Lanco}}, \bibinfo {author} {\bibfnamefont
  {O.}~\bibnamefont {Krebs}}, \bibinfo {author} {\bibfnamefont
  {A.}~\bibnamefont {Auffeves}}, \bibinfo {author} {\bibfnamefont
  {C.}~\bibnamefont {Simon}}, \bibinfo {author} {\bibfnamefont
  {P.}~\bibnamefont {Senellart}}, \ and\ \bibinfo {author} {\bibfnamefont
  {C.}~\bibnamefont {Anton-Solanas}},\ }\href {\doibase
  10.1038/s41566-022-00979-z} {\bibfield  {journal} {\bibinfo  {journal}
  {Nature Photonics}\ }\textbf {\bibinfo {volume} {16}},\ \bibinfo {pages}
  {374} (\bibinfo {year} {2022})}\BibitemShut {NoStop}%
\bibitem [{\citenamefont {Wang}\ \emph
  {et~al.}(2019{\natexlab{b}})\citenamefont {Wang}, \citenamefont {Hu},
  \citenamefont {Chung}, \citenamefont {Qin}, \citenamefont {Yang},
  \citenamefont {Li}, \citenamefont {Liu}, \citenamefont {Zhong}, \citenamefont
  {He}, \citenamefont {Ding}, \citenamefont {Deng}, \citenamefont {Dai},
  \citenamefont {Huo}, \citenamefont {H\"{o}fling}, \citenamefont {Lu},\ and\
  \citenamefont {Pan}}]{wang2019demand}%
  \BibitemOpen
  \bibfield  {author} {\bibinfo {author} {\bibfnamefont {H.}~\bibnamefont
  {Wang}}, \bibinfo {author} {\bibfnamefont {H.}~\bibnamefont {Hu}}, \bibinfo
  {author} {\bibfnamefont {T.-H.}\ \bibnamefont {Chung}}, \bibinfo {author}
  {\bibfnamefont {J.}~\bibnamefont {Qin}}, \bibinfo {author} {\bibfnamefont
  {X.}~\bibnamefont {Yang}}, \bibinfo {author} {\bibfnamefont {J.-P.}\
  \bibnamefont {Li}}, \bibinfo {author} {\bibfnamefont {R.-Z.}\ \bibnamefont
  {Liu}}, \bibinfo {author} {\bibfnamefont {H.-S.}\ \bibnamefont {Zhong}},
  \bibinfo {author} {\bibfnamefont {Y.-M.}\ \bibnamefont {He}}, \bibinfo
  {author} {\bibfnamefont {X.}~\bibnamefont {Ding}}, \bibinfo {author}
  {\bibfnamefont {Y.-H.}\ \bibnamefont {Deng}}, \bibinfo {author}
  {\bibfnamefont {Q.}~\bibnamefont {Dai}}, \bibinfo {author} {\bibfnamefont
  {Y.-H.}\ \bibnamefont {Huo}}, \bibinfo {author} {\bibfnamefont
  {S.}~\bibnamefont {H\"{o}fling}}, \bibinfo {author} {\bibfnamefont {C.-Y.}\
  \bibnamefont {Lu}}, \ and\ \bibinfo {author} {\bibfnamefont {J.-W.}\
  \bibnamefont {Pan}},\ }\href {\doibase 10.1103/PhysRevLett.122.113602}
  {\bibfield  {journal} {\bibinfo  {journal} {Physical Review Letters}\
  }\textbf {\bibinfo {volume} {122}},\ \bibinfo {pages} {113602} (\bibinfo
  {year} {2019}{\natexlab{b}})}\BibitemShut {NoStop}%
\bibitem [{\citenamefont {Liu}\ \emph {et~al.}(2019)\citenamefont {Liu},
  \citenamefont {Su}, \citenamefont {Wei}, \citenamefont {Yao}, \citenamefont
  {Silva}, \citenamefont {Yu}, \citenamefont {Iles-Smith}, \citenamefont
  {Srinivasan}, \citenamefont {Rastelli}, \citenamefont {Li},\ and\
  \citenamefont {Wang}}]{liu2019solid}%
  \BibitemOpen
  \bibfield  {author} {\bibinfo {author} {\bibfnamefont {J.}~\bibnamefont
  {Liu}}, \bibinfo {author} {\bibfnamefont {R.}~\bibnamefont {Su}}, \bibinfo
  {author} {\bibfnamefont {Y.}~\bibnamefont {Wei}}, \bibinfo {author}
  {\bibfnamefont {B.}~\bibnamefont {Yao}}, \bibinfo {author} {\bibfnamefont
  {S.~F. C.~d.}\ \bibnamefont {Silva}}, \bibinfo {author} {\bibfnamefont
  {Y.}~\bibnamefont {Yu}}, \bibinfo {author} {\bibfnamefont {J.}~\bibnamefont
  {Iles-Smith}}, \bibinfo {author} {\bibfnamefont {K.}~\bibnamefont
  {Srinivasan}}, \bibinfo {author} {\bibfnamefont {A.}~\bibnamefont
  {Rastelli}}, \bibinfo {author} {\bibfnamefont {J.}~\bibnamefont {Li}}, \ and\
  \bibinfo {author} {\bibfnamefont {X.}~\bibnamefont {Wang}},\ }\href {\doibase
  10.1038/s41565-019-0435-9} {\bibfield  {journal} {\bibinfo  {journal} {Nature
  Nanotechnology}\ }\textbf {\bibinfo {volume} {14}},\ \bibinfo {pages} {586}
  (\bibinfo {year} {2019})}\BibitemShut {NoStop}%
\bibitem [{\citenamefont {Fattal}\ \emph {et~al.}(2004)\citenamefont {Fattal},
  \citenamefont {Inoue}, \citenamefont {Vu\ifmmode \check{c}\else
  \v{c}\fi{}kovi\ifmmode~\acute{c}\else \'{c}\fi{}}, \citenamefont {Santori},
  \citenamefont {Solomon},\ and\ \citenamefont
  {Yamamoto}}]{Fattal_bellpol_dot}%
  \BibitemOpen
  \bibfield  {author} {\bibinfo {author} {\bibfnamefont {D.}~\bibnamefont
  {Fattal}}, \bibinfo {author} {\bibfnamefont {K.}~\bibnamefont {Inoue}},
  \bibinfo {author} {\bibfnamefont {J.}~\bibnamefont {Vu\ifmmode \check{c}\else
  \v{c}\fi{}kovi\ifmmode~\acute{c}\else \'{c}\fi{}}}, \bibinfo {author}
  {\bibfnamefont {C.}~\bibnamefont {Santori}}, \bibinfo {author} {\bibfnamefont
  {G.~S.}\ \bibnamefont {Solomon}}, \ and\ \bibinfo {author} {\bibfnamefont
  {Y.}~\bibnamefont {Yamamoto}},\ }\href {\doibase
  10.1103/PhysRevLett.92.037903} {\bibfield  {journal} {\bibinfo  {journal}
  {Phys. Rev. Lett.}\ }\textbf {\bibinfo {volume} {92}},\ \bibinfo {pages}
  {037903} (\bibinfo {year} {2004})}\BibitemShut {NoStop}%
\bibitem [{\citenamefont {Senellart}\ \emph {et~al.}(2017)\citenamefont
  {Senellart}, \citenamefont {Solomon},\ and\ \citenamefont
  {White}}]{senellart2017high}%
  \BibitemOpen
  \bibfield  {author} {\bibinfo {author} {\bibfnamefont {P.}~\bibnamefont
  {Senellart}}, \bibinfo {author} {\bibfnamefont {G.}~\bibnamefont {Solomon}},
  \ and\ \bibinfo {author} {\bibfnamefont {A.}~\bibnamefont {White}},\ }\href
  {\doibase 10.1038/nnano.2017.218} {\bibfield  {journal} {\bibinfo  {journal}
  {Nature Nanotechnology}\ }\textbf {\bibinfo {volume} {12}},\ \bibinfo {pages}
  {1026} (\bibinfo {year} {2017})}\BibitemShut {NoStop}%
\bibitem [{\citenamefont {He}\ \emph {et~al.}(2013)\citenamefont {He},
  \citenamefont {He}, \citenamefont {Wei}, \citenamefont {Wu}, \citenamefont
  {Atat{\"u}re}, \citenamefont {Schneider}, \citenamefont {H{\"o}fling},
  \citenamefont {Kamp}, \citenamefont {Lu},\ and\ \citenamefont
  {Pan}}]{he2013demand}%
  \BibitemOpen
  \bibfield  {author} {\bibinfo {author} {\bibfnamefont {Y.-M.}\ \bibnamefont
  {He}}, \bibinfo {author} {\bibfnamefont {Y.}~\bibnamefont {He}}, \bibinfo
  {author} {\bibfnamefont {Y.-J.}\ \bibnamefont {Wei}}, \bibinfo {author}
  {\bibfnamefont {D.}~\bibnamefont {Wu}}, \bibinfo {author} {\bibfnamefont
  {M.}~\bibnamefont {Atat{\"u}re}}, \bibinfo {author} {\bibfnamefont
  {C.}~\bibnamefont {Schneider}}, \bibinfo {author} {\bibfnamefont
  {S.}~\bibnamefont {H{\"o}fling}}, \bibinfo {author} {\bibfnamefont
  {M.}~\bibnamefont {Kamp}}, \bibinfo {author} {\bibfnamefont {C.-Y.}\
  \bibnamefont {Lu}}, \ and\ \bibinfo {author} {\bibfnamefont {J.-W.}\
  \bibnamefont {Pan}},\ }\href {\doibase 10.1038/nnano.2012.262} {\bibfield
  {journal} {\bibinfo  {journal} {Nature Nanotechnology}\ }\textbf {\bibinfo
  {volume} {8}},\ \bibinfo {pages} {213} (\bibinfo {year} {2013})}\BibitemShut
  {NoStop}%
\bibitem [{\citenamefont {Ollivier}\ \emph {et~al.}(2021)\citenamefont
  {Ollivier}, \citenamefont {Thomas}, \citenamefont {Wein}, \citenamefont
  {de~Buy~Wenniger}, \citenamefont {Coste}, \citenamefont {Loredo},
  \citenamefont {Somaschi}, \citenamefont {Harouri}, \citenamefont {Lemaitre},
  \citenamefont {Sagnes}, \citenamefont {Lanco}, \citenamefont {Simon},
  \citenamefont {Anton}, \citenamefont {Krebs},\ and\ \citenamefont
  {Senellart}}]{ollivier2021hong}%
  \BibitemOpen
  \bibfield  {author} {\bibinfo {author} {\bibfnamefont {H.}~\bibnamefont
  {Ollivier}}, \bibinfo {author} {\bibfnamefont {S.}~\bibnamefont {Thomas}},
  \bibinfo {author} {\bibfnamefont {S.}~\bibnamefont {Wein}}, \bibinfo {author}
  {\bibfnamefont {I.~M.}\ \bibnamefont {de~Buy~Wenniger}}, \bibinfo {author}
  {\bibfnamefont {N.}~\bibnamefont {Coste}}, \bibinfo {author} {\bibfnamefont
  {J.}~\bibnamefont {Loredo}}, \bibinfo {author} {\bibfnamefont
  {N.}~\bibnamefont {Somaschi}}, \bibinfo {author} {\bibfnamefont
  {A.}~\bibnamefont {Harouri}}, \bibinfo {author} {\bibfnamefont
  {A.}~\bibnamefont {Lemaitre}}, \bibinfo {author} {\bibfnamefont
  {I.}~\bibnamefont {Sagnes}}, \bibinfo {author} {\bibfnamefont
  {L.}~\bibnamefont {Lanco}}, \bibinfo {author} {\bibfnamefont
  {C.}~\bibnamefont {Simon}}, \bibinfo {author} {\bibfnamefont
  {C.}~\bibnamefont {Anton}}, \bibinfo {author} {\bibfnamefont
  {O.}~\bibnamefont {Krebs}}, \ and\ \bibinfo {author} {\bibfnamefont
  {P.}~\bibnamefont {Senellart}},\ }\href {\doibase
  10.1103/PhysRevLett.126.063602} {\bibfield  {journal} {\bibinfo  {journal}
  {Physical Review Letters}\ }\textbf {\bibinfo {volume} {126}},\ \bibinfo
  {pages} {063602} (\bibinfo {year} {2021})}\BibitemShut {NoStop}%
\bibitem [{\citenamefont {Clauser}\ \emph {et~al.}(1969)\citenamefont
  {Clauser}, \citenamefont {Horne}, \citenamefont {Shimony},\ and\
  \citenamefont {Holt}}]{clauser1969proposed}%
  \BibitemOpen
  \bibfield  {author} {\bibinfo {author} {\bibfnamefont {J.~F.}\ \bibnamefont
  {Clauser}}, \bibinfo {author} {\bibfnamefont {M.~A.}\ \bibnamefont {Horne}},
  \bibinfo {author} {\bibfnamefont {A.}~\bibnamefont {Shimony}}, \ and\
  \bibinfo {author} {\bibfnamefont {R.~A.}\ \bibnamefont {Holt}},\ }\href
  {\doibase 10.1103/PhysRevLett.23.880} {\bibfield  {journal} {\bibinfo
  {journal} {Physical Review Letters}\ }\textbf {\bibinfo {volume} {23}},\
  \bibinfo {pages} {880} (\bibinfo {year} {1969})}\BibitemShut {NoStop}%
\bibitem [{\citenamefont {Somaschi}\ \emph {et~al.}(2016)\citenamefont
  {Somaschi}, \citenamefont {Giesz}, \citenamefont {De~Santis}, \citenamefont
  {Loredo}, \citenamefont {Almeida}, \citenamefont {Hornecker}, \citenamefont
  {Portalupi}, \citenamefont {Grange}, \citenamefont {Anton}, \citenamefont
  {Demory}, \citenamefont {Gomez}, \citenamefont {Sagnes}, \citenamefont
  {Lanzillotti-Kimura}, \citenamefont {Lemaitre}, \citenamefont {Auffeves},
  \citenamefont {White}, \citenamefont {Lanco},\ and\ \citenamefont
  {Senellart}}]{somaschi2016near}%
  \BibitemOpen
  \bibfield  {author} {\bibinfo {author} {\bibfnamefont {N.}~\bibnamefont
  {Somaschi}}, \bibinfo {author} {\bibfnamefont {V.}~\bibnamefont {Giesz}},
  \bibinfo {author} {\bibfnamefont {L.}~\bibnamefont {De~Santis}}, \bibinfo
  {author} {\bibfnamefont {J.}~\bibnamefont {Loredo}}, \bibinfo {author}
  {\bibfnamefont {M.~P.}\ \bibnamefont {Almeida}}, \bibinfo {author}
  {\bibfnamefont {G.}~\bibnamefont {Hornecker}}, \bibinfo {author}
  {\bibfnamefont {S.~L.}\ \bibnamefont {Portalupi}}, \bibinfo {author}
  {\bibfnamefont {T.}~\bibnamefont {Grange}}, \bibinfo {author} {\bibfnamefont
  {C.}~\bibnamefont {Anton}}, \bibinfo {author} {\bibfnamefont
  {J.}~\bibnamefont {Demory}}, \bibinfo {author} {\bibfnamefont
  {C.}~\bibnamefont {Gomez}}, \bibinfo {author} {\bibfnamefont
  {I.}~\bibnamefont {Sagnes}}, \bibinfo {author} {\bibfnamefont {N.~D.}\
  \bibnamefont {Lanzillotti-Kimura}}, \bibinfo {author} {\bibfnamefont
  {A.}~\bibnamefont {Lemaitre}}, \bibinfo {author} {\bibfnamefont
  {A.}~\bibnamefont {Auffeves}}, \bibinfo {author} {\bibfnamefont {A.~G.}\
  \bibnamefont {White}}, \bibinfo {author} {\bibfnamefont {L.}~\bibnamefont
  {Lanco}}, \ and\ \bibinfo {author} {\bibfnamefont {P.}~\bibnamefont
  {Senellart}},\ }\href {\doibase 10.1038/nphoton.2016.23} {\bibfield
  {journal} {\bibinfo  {journal} {Nature Photonics}\ }\textbf {\bibinfo
  {volume} {10}},\ \bibinfo {pages} {340} (\bibinfo {year} {2016})}\BibitemShut
  {NoStop}%
\bibitem [{\citenamefont {Thomas}\ \emph {et~al.}(2021)\citenamefont {Thomas},
  \citenamefont {Billard}, \citenamefont {Coste}, \citenamefont {Wein},
  \citenamefont {Ollivier}, \citenamefont {Krebs}, \citenamefont
  {Taza{\"\i}rt}, \citenamefont {Harouri}, \citenamefont {Lemaitre},
  \citenamefont {Sagnes}, \citenamefont {Anton}, \citenamefont {Lanco},
  \citenamefont {Somaschi}, \citenamefont {Loredo},\ and\ \citenamefont
  {Senellart}}]{thomas2021bright}%
  \BibitemOpen
  \bibfield  {author} {\bibinfo {author} {\bibfnamefont {S.}~\bibnamefont
  {Thomas}}, \bibinfo {author} {\bibfnamefont {M.}~\bibnamefont {Billard}},
  \bibinfo {author} {\bibfnamefont {N.}~\bibnamefont {Coste}}, \bibinfo
  {author} {\bibfnamefont {S.}~\bibnamefont {Wein}}, \bibinfo {author}
  {\bibfnamefont {H.}~\bibnamefont {Ollivier}}, \bibinfo {author}
  {\bibfnamefont {O.}~\bibnamefont {Krebs}}, \bibinfo {author} {\bibfnamefont
  {L.}~\bibnamefont {Taza{\"\i}rt}}, \bibinfo {author} {\bibfnamefont
  {A.}~\bibnamefont {Harouri}}, \bibinfo {author} {\bibfnamefont
  {A.}~\bibnamefont {Lemaitre}}, \bibinfo {author} {\bibfnamefont
  {I.}~\bibnamefont {Sagnes}}, \bibinfo {author} {\bibfnamefont
  {C.}~\bibnamefont {Anton}}, \bibinfo {author} {\bibfnamefont
  {L.}~\bibnamefont {Lanco}}, \bibinfo {author} {\bibfnamefont
  {N.}~\bibnamefont {Somaschi}}, \bibinfo {author} {\bibfnamefont {J.~C.}\
  \bibnamefont {Loredo}}, \ and\ \bibinfo {author} {\bibfnamefont
  {P.}~\bibnamefont {Senellart}},\ }\href {\doibase
  10.1103/PhysRevLett.126.233601} {\bibfield  {journal} {\bibinfo  {journal}
  {Physical Review Letters}\ }\textbf {\bibinfo {volume} {126}},\ \bibinfo
  {pages} {233601} (\bibinfo {year} {2021})}\BibitemShut {NoStop}%
\bibitem [{\citenamefont {Loredo}\ \emph {et~al.}(2019)\citenamefont {Loredo},
  \citenamefont {Hilaire}, \citenamefont {Harouri}, \citenamefont {Millet},
  \citenamefont {Ollivier}, \citenamefont {Somaschi}, \citenamefont
  {De~Santis}, \citenamefont {Lema{\^\i}tre}, \citenamefont {Sagnes},
  \citenamefont {Lanco}, \citenamefont {Auffeves}, \citenamefont {Krebs},\ and\
  \citenamefont {Senellart}}]{loredo2019generation}%
  \BibitemOpen
  \bibfield  {author} {\bibinfo {author} {\bibfnamefont {J.}~\bibnamefont
  {Loredo}}, \bibinfo {author} {\bibfnamefont {P.}~\bibnamefont {Hilaire}},
  \bibinfo {author} {\bibfnamefont {A.}~\bibnamefont {Harouri}}, \bibinfo
  {author} {\bibfnamefont {C.}~\bibnamefont {Millet}}, \bibinfo {author}
  {\bibfnamefont {H.}~\bibnamefont {Ollivier}}, \bibinfo {author}
  {\bibfnamefont {N.}~\bibnamefont {Somaschi}}, \bibinfo {author}
  {\bibfnamefont {L.}~\bibnamefont {De~Santis}}, \bibinfo {author}
  {\bibfnamefont {A.}~\bibnamefont {Lema{\^\i}tre}}, \bibinfo {author}
  {\bibfnamefont {I.}~\bibnamefont {Sagnes}}, \bibinfo {author} {\bibfnamefont
  {L.}~\bibnamefont {Lanco}}, \bibinfo {author} {\bibfnamefont
  {A.}~\bibnamefont {Auffeves}}, \bibinfo {author} {\bibfnamefont
  {O.}~\bibnamefont {Krebs}}, \ and\ \bibinfo {author} {\bibfnamefont
  {P.}~\bibnamefont {Senellart}},\ }\href {\doibase 10.1038/s41566-019-0506-3}
  {\bibfield  {journal} {\bibinfo  {journal} {Nature Photonics}\ }\textbf
  {\bibinfo {volume} {13}},\ \bibinfo {pages} {803} (\bibinfo {year}
  {2019})}\BibitemShut {NoStop}%
\bibitem [{\citenamefont {Ollivier}\ \emph {et~al.}(2020)\citenamefont
  {Ollivier}, \citenamefont {Maillette~de Buy~Wenniger}, \citenamefont
  {Thomas}, \citenamefont {Wein}, \citenamefont {Harouri}, \citenamefont
  {Coppola}, \citenamefont {Hilaire}, \citenamefont {Millet}, \citenamefont
  {Lemaitre}, \citenamefont {Sagnes}, \citenamefont {Krebs}, \citenamefont
  {Lanco}, \citenamefont {Loredo}, \citenamefont {Anton}, \citenamefont
  {Somaschi},\ and\ \citenamefont {Senellart}}]{ollivier2020reproducibility}%
  \BibitemOpen
  \bibfield  {author} {\bibinfo {author} {\bibfnamefont {H.}~\bibnamefont
  {Ollivier}}, \bibinfo {author} {\bibfnamefont {I.}~\bibnamefont {Maillette~de
  Buy~Wenniger}}, \bibinfo {author} {\bibfnamefont {S.}~\bibnamefont {Thomas}},
  \bibinfo {author} {\bibfnamefont {S.~C.}\ \bibnamefont {Wein}}, \bibinfo
  {author} {\bibfnamefont {A.}~\bibnamefont {Harouri}}, \bibinfo {author}
  {\bibfnamefont {G.}~\bibnamefont {Coppola}}, \bibinfo {author} {\bibfnamefont
  {P.}~\bibnamefont {Hilaire}}, \bibinfo {author} {\bibfnamefont
  {C.}~\bibnamefont {Millet}}, \bibinfo {author} {\bibfnamefont
  {A.}~\bibnamefont {Lemaitre}}, \bibinfo {author} {\bibfnamefont
  {I.}~\bibnamefont {Sagnes}}, \bibinfo {author} {\bibfnamefont
  {O.}~\bibnamefont {Krebs}}, \bibinfo {author} {\bibfnamefont
  {L.}~\bibnamefont {Lanco}}, \bibinfo {author} {\bibfnamefont {J.~C.}\
  \bibnamefont {Loredo}}, \bibinfo {author} {\bibfnamefont {C.}~\bibnamefont
  {Anton}}, \bibinfo {author} {\bibfnamefont {N.}~\bibnamefont {Somaschi}}, \
  and\ \bibinfo {author} {\bibfnamefont {P.}~\bibnamefont {Senellart}},\ }\href
  {\doibase 10.1021/acsphotonics.9b01805} {\bibfield  {journal} {\bibinfo
  {journal} {ACS Photonics}\ }\textbf {\bibinfo {volume} {7}},\ \bibinfo
  {pages} {1050} (\bibinfo {year} {2020})}\BibitemShut {NoStop}%
\bibitem [{\citenamefont {M{\"u}nzberg}\ \emph {et~al.}(2022)\citenamefont
  {M{\"u}nzberg}, \citenamefont {Draxl}, \citenamefont {Covre~da Silva},
  \citenamefont {Karli}, \citenamefont {Manna}, \citenamefont {Rastelli},
  \citenamefont {Weihs},\ and\ \citenamefont {Keil}}]{munzberg2022fast}%
  \BibitemOpen
  \bibfield  {author} {\bibinfo {author} {\bibfnamefont {J.}~\bibnamefont
  {M{\"u}nzberg}}, \bibinfo {author} {\bibfnamefont {F.}~\bibnamefont {Draxl}},
  \bibinfo {author} {\bibfnamefont {S.~F.}\ \bibnamefont {Covre~da Silva}},
  \bibinfo {author} {\bibfnamefont {Y.}~\bibnamefont {Karli}}, \bibinfo
  {author} {\bibfnamefont {S.}~\bibnamefont {Manna}}, \bibinfo {author}
  {\bibfnamefont {A.}~\bibnamefont {Rastelli}}, \bibinfo {author}
  {\bibfnamefont {G.}~\bibnamefont {Weihs}}, \ and\ \bibinfo {author}
  {\bibfnamefont {R.}~\bibnamefont {Keil}},\ }\href {\doibase
  10.1063/5.0091867} {\bibfield  {journal} {\bibinfo  {journal} {APL
  Photonics}\ }\textbf {\bibinfo {volume} {7}},\ \bibinfo {pages} {070802}
  (\bibinfo {year} {2022})}\BibitemShut {NoStop}%
\bibitem [{\citenamefont {Basset}\ \emph {et~al.}(2023)\citenamefont {Basset},
  \citenamefont {Valeri}, \citenamefont {Neuwirth}, \citenamefont {Polino},
  \citenamefont {Rota}, \citenamefont {Poderini}, \citenamefont {Pardo},
  \citenamefont {Rodari}, \citenamefont {Roccia}, \citenamefont {da~Silva},
  \citenamefont {Ronco}, \citenamefont {Spagnolo}, \citenamefont {Rastelli},
  \citenamefont {Carvacho}, \citenamefont {Sciarrino},\ and\ \citenamefont
  {Trotta}}]{Basso_Basset_2023}%
  \BibitemOpen
  \bibfield  {author} {\bibinfo {author} {\bibfnamefont {F.~B.}\ \bibnamefont
  {Basset}}, \bibinfo {author} {\bibfnamefont {M.}~\bibnamefont {Valeri}},
  \bibinfo {author} {\bibfnamefont {J.}~\bibnamefont {Neuwirth}}, \bibinfo
  {author} {\bibfnamefont {E.}~\bibnamefont {Polino}}, \bibinfo {author}
  {\bibfnamefont {M.~B.}\ \bibnamefont {Rota}}, \bibinfo {author}
  {\bibfnamefont {D.}~\bibnamefont {Poderini}}, \bibinfo {author}
  {\bibfnamefont {C.}~\bibnamefont {Pardo}}, \bibinfo {author} {\bibfnamefont
  {G.}~\bibnamefont {Rodari}}, \bibinfo {author} {\bibfnamefont
  {E.}~\bibnamefont {Roccia}}, \bibinfo {author} {\bibfnamefont {S.~F.~C.}\
  \bibnamefont {da~Silva}}, \bibinfo {author} {\bibfnamefont {G.}~\bibnamefont
  {Ronco}}, \bibinfo {author} {\bibfnamefont {N.}~\bibnamefont {Spagnolo}},
  \bibinfo {author} {\bibfnamefont {A.}~\bibnamefont {Rastelli}}, \bibinfo
  {author} {\bibfnamefont {G.}~\bibnamefont {Carvacho}}, \bibinfo {author}
  {\bibfnamefont {F.}~\bibnamefont {Sciarrino}}, \ and\ \bibinfo {author}
  {\bibfnamefont {R.}~\bibnamefont {Trotta}},\ }\href {\doibase
  10.1088/2058-9565/acae3d} {\bibfield  {journal} {\bibinfo  {journal} {Quantum
  Science and Technology}\ }\textbf {\bibinfo {volume} {8}},\ \bibinfo {pages}
  {025002} (\bibinfo {year} {2023})}\BibitemShut {NoStop}%
\end{thebibliography}%


\begin{thebibliography}{9}%
\makeatletter
\providecommand \@ifxundefined [1]{%
 \@ifx{#1\undefined}
}%
\providecommand \@ifnum [1]{%
 \ifnum #1\expandafter \@firstoftwo
 \else \expandafter \@secondoftwo
 \fi
}%
\providecommand \@ifx [1]{%
 \ifx #1\expandafter \@firstoftwo
 \else \expandafter \@secondoftwo
 \fi
}%
\providecommand \natexlab [1]{#1}%
\providecommand \enquote  [1]{``#1''}%
\providecommand \bibnamefont  [1]{#1}%
\providecommand \bibfnamefont [1]{#1}%
\providecommand \citenamefont [1]{#1}%
\providecommand \href@noop [0]{\@secondoftwo}%
\providecommand \href [0]{\begingroup \@sanitize@url \@href}%
\providecommand \@href[1]{\@@startlink{#1}\@@href}%
\providecommand \@@href[1]{\endgroup#1\@@endlink}%
\providecommand \@sanitize@url [0]{\catcode `\\12\catcode `\$12\catcode
  `\&12\catcode `\#12\catcode `\^12\catcode `\_12\catcode `\%12\relax}%
\providecommand \@@startlink[1]{}%
\providecommand \@@endlink[0]{}%
\providecommand \url  [0]{\begingroup\@sanitize@url \@url }%
\providecommand \@url [1]{\endgroup\@href {#1}{\urlprefix }}%
\providecommand \urlprefix  [0]{URL }%
\providecommand \Eprint [0]{\href }%
\providecommand \doibase [0]{https://doi.org/}%
\providecommand \selectlanguage [0]{\@gobble}%
\providecommand \bibinfo  [0]{\@secondoftwo}%
\providecommand \bibfield  [0]{\@secondoftwo}%
\providecommand \translation [1]{[#1]}%
\providecommand \BibitemOpen [0]{}%
\providecommand \bibitemStop [0]{}%
\providecommand \bibitemNoStop [0]{.\EOS\space}%
\providecommand \EOS [0]{\spacefactor3000\relax}%
\providecommand \BibitemShut  [1]{\csname bibitem#1\endcsname}%
\let\auto@bib@innerbib\@empty
\bibitem [{\citenamefont {Li}\ \emph {et~al.}(2021)\citenamefont {Li},
  \citenamefont {Gu}, \citenamefont {Qin}, \citenamefont {Wu}, \citenamefont
  {You}, \citenamefont {Wang}, \citenamefont {Schneider}, \citenamefont
  {H{\"o}fling}, \citenamefont {Huo}, \citenamefont {Lu}, \citenamefont {Liu},
  \citenamefont {Li},\ and\ \citenamefont {Pan}}]{li2021heralded}%
  \BibitemOpen
  \bibfield  {author} {\bibinfo {author} {\bibfnamefont {J.-P.}\ \bibnamefont
  {Li}}, \bibinfo {author} {\bibfnamefont {X.}~\bibnamefont {Gu}}, \bibinfo
  {author} {\bibfnamefont {J.}~\bibnamefont {Qin}}, \bibinfo {author}
  {\bibfnamefont {D.}~\bibnamefont {Wu}}, \bibinfo {author} {\bibfnamefont
  {X.}~\bibnamefont {You}}, \bibinfo {author} {\bibfnamefont {H.}~\bibnamefont
  {Wang}}, \bibinfo {author} {\bibfnamefont {C.}~\bibnamefont {Schneider}},
  \bibinfo {author} {\bibfnamefont {S.}~\bibnamefont {H{\"o}fling}}, \bibinfo
  {author} {\bibfnamefont {Y.-H.}\ \bibnamefont {Huo}}, \bibinfo {author}
  {\bibfnamefont {C.-Y.}\ \bibnamefont {Lu}}, \bibinfo {author} {\bibfnamefont
  {N.-L.}\ \bibnamefont {Liu}}, \bibinfo {author} {\bibfnamefont
  {L.}~\bibnamefont {Li}},\ and\ \bibinfo {author} {\bibfnamefont {J.-W.}\
  \bibnamefont {Pan}},\ }\bibfield  {title} {\bibinfo {title} {Heralded
  nondestructive quantum entangling gate with single-photon sources},\ }\href
  {10.1103/PhysRevLett.126.140501} {\bibfield  {journal} {\bibinfo  {journal}
  {Physical Review Letters}\ }\textbf {\bibinfo {volume} {126}},\ \bibinfo
  {pages} {140501} (\bibinfo {year} {2021})}\BibitemShut {NoStop}%
\bibitem [{\citenamefont {Istrati}\ \emph {et~al.}(2020)\citenamefont
  {Istrati}, \citenamefont {Pilnyak}, \citenamefont {Loredo}, \citenamefont
  {Ant{\'o}n}, \citenamefont {Somaschi}, \citenamefont {Hilaire}, \citenamefont
  {Ollivier}, \citenamefont {Esmann}, \citenamefont {Cohen}, \citenamefont
  {Vidro}, \citenamefont {Millet}, \citenamefont {Lemaitre}, \citenamefont
  {Sagnes}, \citenamefont {Harouri}, \citenamefont {Lanco}, \citenamefont
  {Senellart},\ and\ \citenamefont {Eisenberg}}]{istrati2020sequential}%
  \BibitemOpen
  \bibfield  {author} {\bibinfo {author} {\bibfnamefont {D.}~\bibnamefont
  {Istrati}}, \bibinfo {author} {\bibfnamefont {Y.}~\bibnamefont {Pilnyak}},
  \bibinfo {author} {\bibfnamefont {J.}~\bibnamefont {Loredo}}, \bibinfo
  {author} {\bibfnamefont {C.}~\bibnamefont {Ant{\'o}n}}, \bibinfo {author}
  {\bibfnamefont {N.}~\bibnamefont {Somaschi}}, \bibinfo {author}
  {\bibfnamefont {P.}~\bibnamefont {Hilaire}}, \bibinfo {author} {\bibfnamefont
  {H.}~\bibnamefont {Ollivier}}, \bibinfo {author} {\bibfnamefont
  {M.}~\bibnamefont {Esmann}}, \bibinfo {author} {\bibfnamefont
  {L.}~\bibnamefont {Cohen}}, \bibinfo {author} {\bibfnamefont
  {L.}~\bibnamefont {Vidro}}, \bibinfo {author} {\bibfnamefont
  {C.}~\bibnamefont {Millet}}, \bibinfo {author} {\bibfnamefont
  {A.}~\bibnamefont {Lemaitre}}, \bibinfo {author} {\bibfnamefont
  {I.}~\bibnamefont {Sagnes}}, \bibinfo {author} {\bibfnamefont
  {A.}~\bibnamefont {Harouri}}, \bibinfo {author} {\bibfnamefont
  {L.}~\bibnamefont {Lanco}}, \bibinfo {author} {\bibfnamefont
  {P.}~\bibnamefont {Senellart}},\ and\ \bibinfo {author} {\bibfnamefont
  {H.~S.}\ \bibnamefont {Eisenberg}},\ }\bibfield  {title} {\bibinfo {title}
  {Sequential generation of linear cluster states from a single photon
  emitter},\ }\href {https://doi.org/10.1038/s41467-020-19341-4} {\bibfield
  {journal} {\bibinfo  {journal} {Nature Communications}\ }\textbf {\bibinfo
  {volume} {11}},\ \bibinfo {pages} {1} (\bibinfo {year} {2020})}\BibitemShut
  {NoStop}%
\bibitem [{\citenamefont {Li}\ \emph {et~al.}(2020)\citenamefont {Li},
  \citenamefont {Qin}, \citenamefont {Chen}, \citenamefont {Duan},
  \citenamefont {Yu}, \citenamefont {Huo}, \citenamefont {H\"{o}fling},
  \citenamefont {Lu}, \citenamefont {Chen},\ and\ \citenamefont
  {Pan}}]{li2020multiphoton}%
  \BibitemOpen
  \bibfield  {author} {\bibinfo {author} {\bibfnamefont {J.-P.}\ \bibnamefont
  {Li}}, \bibinfo {author} {\bibfnamefont {J.}~\bibnamefont {Qin}}, \bibinfo
  {author} {\bibfnamefont {A.}~\bibnamefont {Chen}}, \bibinfo {author}
  {\bibfnamefont {Z.-C.}\ \bibnamefont {Duan}}, \bibinfo {author}
  {\bibfnamefont {Y.}~\bibnamefont {Yu}}, \bibinfo {author} {\bibfnamefont
  {Y.}~\bibnamefont {Huo}}, \bibinfo {author} {\bibfnamefont {S.}~\bibnamefont
  {H\"{o}fling}}, \bibinfo {author} {\bibfnamefont {C.-Y.}\ \bibnamefont {Lu}},
  \bibinfo {author} {\bibfnamefont {K.}~\bibnamefont {Chen}},\ and\ \bibinfo
  {author} {\bibfnamefont {J.-W.}\ \bibnamefont {Pan}},\ }\bibfield  {title}
  {\bibinfo {title} {Multiphoton graph states from a solid-state single-photon
  source},\ }\href
  {https://pubs.acs.org/doi/pdf/10.1021/acsphotonics.0c00192?casa_token=JJPhKAEUCJoAAAAA:StJ9ex7IYQS9t9aUg3JgOEZI8aCFI5j72rFSdom6oLWuRV8qaUGwN6ANDyYlqQ2lnYCOtCFl3iF5RQ}
  {\bibfield  {journal} {\bibinfo  {journal} {ACS Photonics}\ }\textbf
  {\bibinfo {volume} {7}},\ \bibinfo {pages} {1603} (\bibinfo {year}
  {2020})}\BibitemShut {NoStop}%
\bibitem [{\citenamefont {Ollivier}\ \emph {et~al.}(2021)\citenamefont
  {Ollivier}, \citenamefont {Thomas}, \citenamefont {Wein}, \citenamefont
  {de~Buy~Wenniger}, \citenamefont {Coste}, \citenamefont {Loredo},
  \citenamefont {Somaschi}, \citenamefont {Harouri}, \citenamefont {Lemaitre},
  \citenamefont {Sagnes}, \citenamefont {Lanco}, \citenamefont {Simon},
  \citenamefont {Anton}, \citenamefont {Krebs},\ and\ \citenamefont
  {Senellart}}]{ollivier2021hong}%
  \BibitemOpen
  \bibfield  {author} {\bibinfo {author} {\bibfnamefont {H.}~\bibnamefont
  {Ollivier}}, \bibinfo {author} {\bibfnamefont {S.}~\bibnamefont {Thomas}},
  \bibinfo {author} {\bibfnamefont {S.}~\bibnamefont {Wein}}, \bibinfo {author}
  {\bibfnamefont {I.~M.}\ \bibnamefont {de~Buy~Wenniger}}, \bibinfo {author}
  {\bibfnamefont {N.}~\bibnamefont {Coste}}, \bibinfo {author} {\bibfnamefont
  {J.}~\bibnamefont {Loredo}}, \bibinfo {author} {\bibfnamefont
  {N.}~\bibnamefont {Somaschi}}, \bibinfo {author} {\bibfnamefont
  {A.}~\bibnamefont {Harouri}}, \bibinfo {author} {\bibfnamefont
  {A.}~\bibnamefont {Lemaitre}}, \bibinfo {author} {\bibfnamefont
  {I.}~\bibnamefont {Sagnes}}, \bibinfo {author} {\bibfnamefont
  {L.}~\bibnamefont {Lanco}}, \bibinfo {author} {\bibfnamefont
  {C.}~\bibnamefont {Simon}}, \bibinfo {author} {\bibfnamefont
  {C.}~\bibnamefont {Anton}}, \bibinfo {author} {\bibfnamefont
  {O.}~\bibnamefont {Krebs}},\ and\ \bibinfo {author} {\bibfnamefont
  {P.}~\bibnamefont {Senellart}},\ }\bibfield  {title} {\bibinfo {title}
  {Hong-ou-mandel interference with imperfect single photon sources},\ }\href
  {https://doi.org/10.1103/PhysRevLett.126.063602} {\bibfield  {journal}
  {\bibinfo  {journal} {Physical Review Letters}\ }\textbf {\bibinfo {volume}
  {126}},\ \bibinfo {pages} {063602} (\bibinfo {year} {2021})}\BibitemShut
  {NoStop}%
\bibitem [{\citenamefont {Kim}\ \emph {et~al.}(2021)\citenamefont {Kim},
  \citenamefont {Kwon},\ and\ \citenamefont {Moon}}]{HOM_laser}%
  \BibitemOpen
  \bibfield  {author} {\bibinfo {author} {\bibfnamefont {H.}~\bibnamefont
  {Kim}}, \bibinfo {author} {\bibfnamefont {O.}~\bibnamefont {Kwon}},\ and\
  \bibinfo {author} {\bibfnamefont {H.}~\bibnamefont {Moon}},\ }\bibfield
  {title} {\bibinfo {title} {Two-photon interferences of weak coherent
  lights},\ }\href {https://doi.org/10.1038/s41598-021-99804-w} {\bibfield
  {journal} {\bibinfo  {journal} {Scientific Reports}\ }\textbf {\bibinfo
  {volume} {48}},\ \bibinfo {pages} {20555} (\bibinfo {year}
  {2021})}\BibitemShut {NoStop}%
\bibitem [{\citenamefont {Altepeter}\ \emph {et~al.}(2005)\citenamefont
  {Altepeter}, \citenamefont {Jeffrey},\ and\ \citenamefont
  {Kwiat}}]{altepeter2005photonic}%
  \BibitemOpen
  \bibfield  {author} {\bibinfo {author} {\bibfnamefont {J.~B.}\ \bibnamefont
  {Altepeter}}, \bibinfo {author} {\bibfnamefont {E.~R.}\ \bibnamefont
  {Jeffrey}},\ and\ \bibinfo {author} {\bibfnamefont {P.~G.}\ \bibnamefont
  {Kwiat}},\ }\bibfield  {title} {\bibinfo {title} {Photonic state
  tomography},\ }\href {https://doi.org/10.1016/S1049-250X(05)52003-2}
  {\bibfield  {journal} {\bibinfo  {journal} {Advances in Atomic, Molecular,
  and Optical physics}\ }\textbf {\bibinfo {volume} {52}},\ \bibinfo {pages}
  {105} (\bibinfo {year} {2005})}\BibitemShut {NoStop}%
\bibitem [{\citenamefont {Hradil}\ \emph {et~al.}(2004)\citenamefont {Hradil},
  \citenamefont {{\v{R}}eh{\'a}{\v{c}}ek}, \citenamefont {Fiur{\'a}{\v{s}}ek},\
  and\ \citenamefont {Je{\v{z}}ek}}]{hradil20043}%
  \BibitemOpen
  \bibfield  {author} {\bibinfo {author} {\bibfnamefont {Z.}~\bibnamefont
  {Hradil}}, \bibinfo {author} {\bibfnamefont {J.}~\bibnamefont
  {{\v{R}}eh{\'a}{\v{c}}ek}}, \bibinfo {author} {\bibfnamefont
  {J.}~\bibnamefont {Fiur{\'a}{\v{s}}ek}},\ and\ \bibinfo {author}
  {\bibfnamefont {M.}~\bibnamefont {Je{\v{z}}ek}},\ }\bibfield  {title}
  {\bibinfo {title} {3 maximum-likelihood methods in quantum mechanics},\
  }\href {https://link.springer.com/chapter/10.1007/978-3-540-44481-7_3}
  {\bibfield  {journal} {\bibinfo  {journal} {Quantum state estimation}\ ,\
  \bibinfo {pages} {59}} (\bibinfo {year} {2004})}\BibitemShut {NoStop}%
\bibitem [{\citenamefont {Horodecki}\ \emph {et~al.}(1995)\citenamefont
  {Horodecki}, \citenamefont {Horodecki},\ and\ \citenamefont
  {Horodecki}}]{horodecki1995violating}%
  \BibitemOpen
  \bibfield  {author} {\bibinfo {author} {\bibfnamefont {R.}~\bibnamefont
  {Horodecki}}, \bibinfo {author} {\bibfnamefont {P.}~\bibnamefont
  {Horodecki}},\ and\ \bibinfo {author} {\bibfnamefont {M.}~\bibnamefont
  {Horodecki}},\ }\bibfield  {title} {\bibinfo {title} {Violating bell
  inequality by mixed spin-12 states: necessary and sufficient condition},\
  }\href {https://doi.org/10.1016/0375-9601(95)00214-N} {\bibfield  {journal}
  {\bibinfo  {journal} {Physics Letters A}\ }\textbf {\bibinfo {volume}
  {200}},\ \bibinfo {pages} {340} (\bibinfo {year} {1995})}\BibitemShut
  {NoStop}%
\bibitem [{\citenamefont {Hildebrand}(2007)}]{hildebrand2007concurrence}%
  \BibitemOpen
  \bibfield  {author} {\bibinfo {author} {\bibfnamefont {R.}~\bibnamefont
  {Hildebrand}},\ }\bibfield  {title} {\bibinfo {title} {Concurrence
  revisited},\ }\href {https://doi.org/doi.org/10.1063/1.2795840} {\bibfield
  {journal} {\bibinfo  {journal} {Journal of Mathematical Physics}\ }\textbf
  {\bibinfo {volume} {48}},\ \bibinfo {pages} {102108} (\bibinfo {year}
  {2007})}\BibitemShut {NoStop}%
\end{thebibliography}%

%

\end{document}


\title{Supplementary Information}

\author{Mauro Valeri}
\address{Dipartimento di Fisica, Sapienza Universit\`{a} di Roma, Piazzale Aldo Moro 5, I-00185 Roma, Italy}

\author{Paolo Barigelli}
\address{Dipartimento di Fisica, Sapienza Universit\`{a} di Roma, Piazzale Aldo Moro 5, I-00185 Roma, Italy}

\author{Beatrice Polacchi}
\address{Dipartimento di Fisica, Sapienza Universit\`{a} di Roma, Piazzale Aldo Moro 5, I-00185 Roma, Italy}

\author{Giovanni Rodari}
\address{Dipartimento di Fisica, Sapienza Universit\`{a} di Roma, Piazzale Aldo Moro 5, I-00185 Roma, Italy}

\author{Gianluca De Santis}
\address{Dipartimento di Fisica, Sapienza Universit\`{a} di Roma, Piazzale Aldo Moro 5, I-00185 Roma, Italy}

\author{Taira Giordani}
\address{Dipartimento di Fisica, Sapienza Universit\`{a} di Roma, Piazzale Aldo Moro 5, I-00185 Roma, Italy}

\author{Gonzalo Carvacho}
\email{gonzalo.carvacho@uniroma1.it}
\address{Dipartimento di Fisica, Sapienza Universit\`{a} di Roma, Piazzale Aldo Moro 5, I-00185 Roma, Italy}

\author{Nicol\`o Spagnolo}
\address{Dipartimento di Fisica, Sapienza Universit\`{a} di Roma, Piazzale Aldo Moro 5, I-00185 Roma, Italy}

\author{Fabio Sciarrino}

\address{Dipartimento di Fisica, Sapienza Universit\`{a} di Roma, Piazzale Aldo Moro 5, I-00185 Roma, Italy}

\maketitle

 As mentioned in the main text, only a few papers reported the generation of an entangled state by the exciton of quantum dots (QD) and thus it is important to fully characterize the generated states. Table~\ref{tab:qdsources} provides a summary of the two-qubit entangling gates based on QDs single-photon sources realized so far in the literature. Since a complete description of the entangled state generated is still lacking, it is important identify a model in order to further establish the expected performances as a function of the experimental parameters. 
\begin{table}[ht!]
\begin{adjustbox}{max width=\textwidth}
    \centering
        
            \begin{tabular}{ |c c c c c c c c| } 
    \hline
        \textbf{Pump Type} & \textbf{Encoding} & \textbf{$\lambda\, [nm]$} & \textbf{Coincidence Rate} &  \textbf{Fidelity} & \textbf{Concurrence} & \textbf{Bell} & \textbf{Ref. / year} \\
    \hline
        \textit{off-resonant at 79 MHz} & \textit{Polarization} & \textit{927.8}  & \textit{7 kHz} & (\textit{92}$\pm$\textit{1})\% & 0.819$\pm 0.008$ & \textit{2.58}$\pm$\textit{0.02} & \textit{this work} \\

        \textit{resonant at 79 MHz} & \textit{Polarization} & \textit{928.05}  & \textit{3.5 kHz} & (\textit{95}$\pm$\textit{1})\% & 0.86$\pm 0.01$ & \textit{2.63}$\pm$\textit{0.02} & \textit{this work} \\
        
        resonant at 76 MHz & Polarization  & 893 & - &  $(83.4 \pm 2.4 )$ \% & - & - & \cite{li2021heralded}  \\ 
        
        resonant at 81 MHz & Polarization  & 925 & - &  $(82 \pm 1 )$\% & - & - & \cite{istrati2020sequential}  \\ 
        
        resonant at 76 MHz & Polarization & 893 & 45 kHz &  92 \% & - & - & \cite{li2020multiphoton}  \\ 
        
    \hline
             \end{tabular} \end{adjustbox}
           
    \caption{Summary of the entangled photon sources realized using the exciton emission by quantum dot devices. The fidelity shown in  \cite{li2020multiphoton} is reported without an error.}
    \label{tab:qdsources}
\end{table}
\section{Model of the prepared state}
In this section, we provide theoretical insights into the different noise and losses in our apparatus. Ideally, the quantum dot generates one single photon for each excitation pulse with a  repetition rate of $R_{\text{QD}} \sim 79$ MHz, corresponding to a distance of $\tau \sim 12$ ns between consecutive emissions. However, a non-unitary probability $p_1$ of one-photon emission is due to some possible events of non-excitation and the presence of residual laser light or emission of more than one photon from the QD, occurring with probability $p_0$ and $p_2$, respectively, so that $p_0+p_1+p_2=1 $. 

The value of $p_2$ can be estimated through the second-order auto-correlation function $g^{(2)}(0)$. Indeed, considering the general quantum state composed of up to two-photon states, that is $\hat{\rho}_{2}=p_{0}\ket{0}\bra{0}+p_{1}\ket{1}\bra{1}+p_{2}\ket{2}\bra{2}+\sum_{j>i=0}^2 p_{ij}\ket{i}\bra{j}$ for some coefficients $p_{ij}$, the second-order auto-correlation function provides:
\begin{equation}
\label{eq:g2p2}
    g^{(2)}(0) = \frac{\braket{\hat{n}(\hat{n}-1)}}{\braket{\hat{n}}^2} = \frac{2p_2}{(p_1+2p_2)^2},
\end{equation}
where the expectation value of an operator $\hat{f}$ is computed as $\braket{\hat{f}}=Tr[\hat{\rho}_{2}\hat{f}]$. 
Thus, the probability of emission of the QD source can be estimated from the inversion of Eq.~\eqref{eq:g2p2}, the normalization condition and their relation to the Brightness of the dot itself: $B=p_{1}+p_{2}$. In particular, when using LA configuration, it shows fixed values of $p_{0} = 1-B $ and $p_{1}$,$p_{2}$ are computed from $g^{(2)}(0)$, while the RF configuration has practically $p_{0}=0$, so only Eq.~\eqref{eq:g2p2} is sufficient to estimate the probabilities (adding the normalization condition).
\newline
Besides normalization factors, the general two-photon state exiting the source on the two outputs with zero-time delay that we expect is given by the following expression:
\begin{equation}
\label{eq:model}
    \hat{\rho} \propto c_{11} \hat{\rho}_{11} + c_{02} \hat{\rho}_{02} + c_{12}\hat{\rho}_{12} + c_{l}\hat{\rho}_{l} + O(\eta^3),
\end{equation}
where $\hat{\rho}_{11}$ is the density matrix of the correct generation and $\hat{\rho}_{12} $,$\hat{\rho}_{02}$ and $\hat{\rho}_{l}$ correspond to the density matrix of noise elements (described in the next sections). The parameter $\eta$ is the overall transmission of the system, including the budget losses (Table I of main text), the transmission efficiency of the polarization measurement platform ($80\%$) and the detector efficiency ($35\%$). The term $ O(\eta^3)$ states that we consider only two photons arriving at the measurement platform; the reasons for this model are described in the following.

\begin{table}[ht!]\begin{adjustbox}{max width=\textwidth}
    \centering        
\begin{tabular}{l c l l}
\hline
 & &  Longitudinal-Acoustic phonon (LA) & Resonant Fluorescence (RF) \\
\hline \hline 
overall losses & $\qquad$ & $\eta = 0.00829$ & $\eta = 0.00504$ \\
HOM visibility & & \begin{tabular}{l} $v_m=0.904 \pm 0.003$ (measured) \\ $v=0.927 \pm 0.003$ (corrected by $g^{(2)}(0)$) \end{tabular} & \begin{tabular}{l} $v_m=0.918 \pm 0.003$ (measured) \\ $v=0.949 \pm 0.003$ (corrected by $g^{(2)}(0)$) \end{tabular} \\
second-order correlation function & & $g^{(2)}(0) = 0.012 \pm 0.001$ & $g^{(2)}(0) = 0.016 \pm 0.002$ \\

detector dark-counts probability & & $p_{dd} \sim 1.5\cdot10^{-6}$ & $p_{dd} \sim 1.5\cdot10^{-6}$ \\
 
\hline \hline
beam splitting ratios & \multicolumn{3}{c}{\begin{tabular}{l} $51.7/48.3 \pm 0.2$ (first BS) \\ $52.5/47.5 \pm 0.2$ (second BS) \end{tabular}} \\
\hline 
\end{tabular}\end{adjustbox}
\label{tab:parameters}
    \caption{Experimental parameters of our setup when considering the two possible excitation schemes i.e. Longitudinal-Acoustic phonon (LA) and Resonant Fluorescence (RF).}
\end{table} 

The interferometer architecture allows the interference of two photons emitted after a delay $\tau$ at most. Events resulting from two consecutive single-photon (Fig.~\ref{fig:combinations}a) have an emission probability $p^{2}_1 $ and are revealed with probability $\eta^2$. The same detection probability $\eta^2$ is associated with the coincidence event deriving from only one single two-photon emission (Fig.~\ref{fig:combinations}c) which happens at the source with probability $p_0p_2$, i.e. when no photon is emitted after or before. Differently, events resulting from the combination of three photons, given by a single emission and single emission plus noise photon (Fig.~\ref{fig:combinations}b), have an emission probability of $p_1 p_2$ being revealed with probabilities $\eta^2(1-\eta)$ and $\eta^3$, respectively, when only two photons or all three photons survive. As reported in the main text, the overall transmission is $\eta < 10^{-2}$, so contributions depending on $\eta^3$ are negligible. This is the case, for example, of three-photon terms associated with the probability $p_1 p_2$. Similarly, the emission of four photons, with probability of generation $p^{2}_{2}$ and revealed with probability $\eta^{2}(1-\eta)^{2}$, generates different noise terms depending on which photons survive up to the measurement process (Fig.~\ref{fig:combinations}d).
\newline
The other possible detectable events are given by coincidences with detector dark counts or environmental light, which occur with probability $p_{dd}$. The dark count probability associated to the detector can be computed as $p_{dd}=R_{dd} \cdot \Delta t_{cw}$, where $R_{dd}$ is the rate of dark count signal (measured by switching off the QD emission) and $\Delta t_{cw}$ the adopted coincidence window. Tab. 2 of  \ref{tab:parameters} provides the measured values of such terms in our experimental apparatus and the order of magnitude of other interesting experimental parameters. 
\begin{figure*}[htp]
    \centering
    \includegraphics[width=0.9\textwidth]{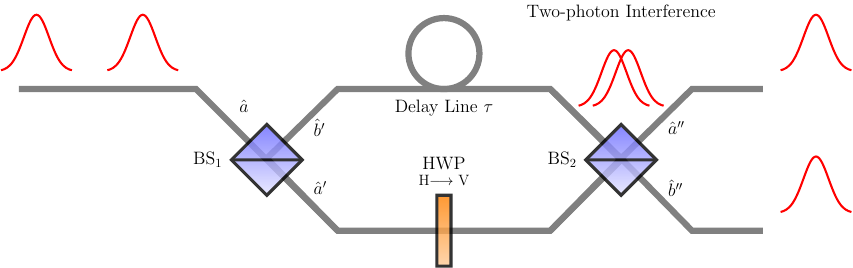}
    \caption{Both the photons at time $t$ and $t+\tau$ arrive at the same input branch of the first BS. The photons in the transmitted path $\hat{a}'$ have their polarization switched, while the ones in the reflected branch $\hat{b}'$ are delayed by the time $\tau$. The photons arriving at the second beam-splitter simultaneously interfere and create coincidences.}
    \label{fig:mzi_scheme}
\end{figure*}

\begin{figure*}[htp]
    \centering
    \includegraphics[width=0.9\textwidth]{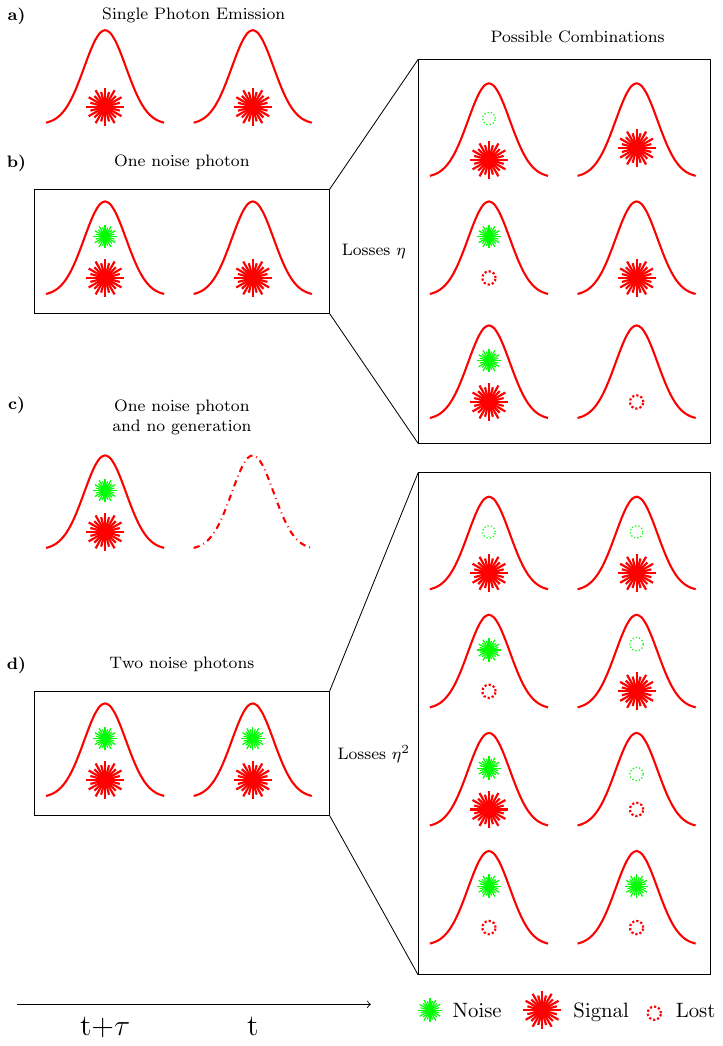}
    \caption{Two-photon interference can be achieved by different kinds of generation: (a) two consecutive emitted single photons, (b) the combination of two photons (signal and noise) and single emission,(c) a double emission followed (or preceded) by a non-excitation and (d) the combination of two pulses having two photons each, The events (b) and (d)  can realize a two-photon interference depending on which of the three or four photons are lost.}
    \label{fig:combinations}
\end{figure*}


In order to characterize properly Eq.~\eqref{eq:model}, we evolve the quantum state of each term according to the second quantization rules applied to our experimental interferometer (Fig.~\ref{fig:mzi_scheme}).  
We consider the following notation: all the creation(annihilation) operators referring to the transmitted arm of the interferometer are indicated as $\hat{a}$($\hat{a}^{\dagger}$), while the reflected arm is indicated as $\hat{b}$($\hat{b}^{\dagger}$); the two possible states of polarization as H (horizontal) and V (vertical), while all the further possible internal degrees of freedom are indicated with a relative greek letter subscripts ($\psi$,$\phi$,$\gamma$,$\ldots$). For example, the initial input state of the interferometer deriving from consecutive emissions is defined by:
\begin{equation}
    \hat{a}^{\dagger}_{\psi,H}(t) \hat{a}^{\dagger}_{\phi,H}(t+\tau)\ket{0}_{a}\ket{0}_{b}.
\end{equation}
where the arrival time on the apparatus is indicated considering the first photon at a time t and the second at $t+\tau$. 
The interferometer is composed of two external beam-splitters (BSs) in sequence $\hat{BS}_1$ and $\hat{BS}_2$. Then, in the internal arms, a delay line $\hat{D}(\tau)$ acts on reflected arm i.e. upper path in Suppl. Fig.1 ($\hat{b}_{\psi,H}^{\dagger}(t) \xrightarrow[]{\hat{D}(\tau)} \hat{b}_{\psi,H}^{\dagger}(t+\tau)$) and a polarization switch $\hat{S}$ on the transmitted one ($\hat{a}^{\dagger}_{\psi,H}(t) \xrightarrow[]{\hat{S}} \hat{a}^{\dagger}_{\psi,V}(t)$) . The action of a $\hat{BS}$ with transmissivity $t$ and reflectivity $r$ on the two inputs ($\hat{a},\hat{b}$), is described by:
\begin{equation}
\begin{pmatrix}
\hat{a}' \\
\hat{b}'
\end{pmatrix}=
\begin{pmatrix}
t & ir\\
ir & t
\end{pmatrix}
\begin{pmatrix}
\hat{a} \\
\hat{b}
\end{pmatrix}
\end{equation}
where $t^{2}+r^{2}=1$. This corresponds to the transformation: 
\begin{equation}
     \hat{a}^{\dagger}_{\psi,H}(t) \xrightarrow[]{\hat{BS}} t \hat{a}^{\dagger}_{\psi,H}(t) +i r \hat{b}^{\dagger}_{\psi,H}(t) \quad \text{and} \quad 
     \hat{b}^{\dagger}_{\psi,H}(t) \xrightarrow[]{\hat{BS}} ir\hat{a}^{\dagger}_{\psi,H}(t) + t\hat{b}^{\dagger}_{\psi,H}(t)
\end{equation}
Therefore, the total operation of the interferometer is described by:
\begin{equation}\label{eq:u}
    \hat{U} = \hat{BS}_2 \hat{D}(\tau) \hat{S} \hat{BS}_1.
\end{equation}
Eq.~\eqref{eq:u} can be applied to different initial states in order to recover the density matrices $\hat{\rho}_{11},\hat{\rho}_{12},\hat{\rho}_{02},\hat{\rho}_{l}$. 

Finally, it is important to note the difference between the LA and RF excitation scheme. Apart from the different probability of being excited per excitation pulse ($p_1$), the RF excitation case shows exciton emission in coherent superposition with the vacuum state: 
\begin{equation}\label{eq:RF}
    \bigl( \sqrt{1-q} +  e^{i \phi_q}\sqrt{q}\hat{a}^{\dagger}(t)\bigr)\ket{0}
\end{equation}
with $q\in [0,1],\phi_q \in \mathbb{R}$. In RF excitation, the superposition in Eq.~\eqref{eq:RF} affects the density matrices of the coincidence events provided by the Eq.~\eqref{eq:model} as well as generating interference effects on the singles photon rate \cite{istrati2020sequential}. However, LA scheme can be retrieved from RF results by simply setting $q=1$.
\subsection{Generation of the correct signal}

We start considering the input state of two consecutive single-photon emissions (Fig.~\ref{fig:combinations}(a)) for LA and RF, that is:
\begin{equation}\label{eq:statein}
\begin{split}
    &\text{LA}: \qquad \hat{a}^{\dagger}_{\phi_2,H}(t+\tau) \hat{a}^{\dagger}_{\phi_1,H}(t) \ket{0}_{a}\ket{0}_{b}, \\ \\
    &\text{RF}: \qquad \bigl[ \sqrt{1-q} + e^{i \phi_q} \sqrt{q}\hat{a}^{\dagger}_{\phi_2,H}(t+\tau)\bigr]  \bigl[ \sqrt{1-q}  + e^{i \phi_q} \sqrt{q}\hat{a}^{\dagger}_{\phi_1,H}(t)\bigr] \ket{0}_{a}\ket{0}_{b}.
\end{split}
\end{equation}
The state after the interferometer can be computed by first applying Eq.~\eqref{eq:u} to Eq.~\eqref{eq:statein}. Then, the result is post-selected to coincidence events with zero-time delay. These are provided by the case in which there is one photon in each path and, at the same time, the two photons arrive at the measurement platform together (corresponding to the operators with the same temporal delay). In this way, the final state $\ket{\psi_{11}}$ reads:
\begin{equation} \label{eq:la_in_state} \begin{split}
    \ket{\psi_{11}}^{\text{LA}}&=ir_{1}t_{1} \bigl[ t^{2}_{2}\hat{b}^{\dagger}_{\phi_1,H}(t+\tau)\hat{a}^{\dagger}_{\phi_2,V}(t+\tau)-r_{2}^{2}\hat{a}^{\dagger}_{\phi_1,H}(t+\tau)\hat{b}^{\dagger}_{\phi_2,V}(t+\tau) \bigr]\ket{0}_{a}\ket{0}_{b} = \\ 
    &= ir_{1}t_{1}(t^{2}_{2}\ket{\phi_2,V}_{a}\ket{\phi_1,H}_{b} - r_{2}^{2}\ket{\phi_1,H}_{a}\ket{\phi_2,V}_{b}), \\ \\ \ket{\psi_{11}}^{\text{RF}}&=q\ket{\psi_{11}}^{\text{LA}}.
\end{split} \end{equation}    
Finally, the density matrix of the final state is given by the definition: $\hat{\rho}'_{11}=\ket{\psi_{11}}\bra{\psi_{11}}$. However, the detectors are not sensitive to the internal degrees of freedom ($\phi_1,\phi_2$), so they must be traced out. Supposing there are discrete sets of orthogonal basis states $\{\ket{i}\}$ and $\{\ket{j}\}$ which completely describe $\ket{\phi_1}$ and $\ket{\phi_2}$, respectively, we have:
\begin{equation}\label{eq:trace}
    \hat{\rho}_{11} = \mathrm{Tr}_{\phi_1,\phi_2}[\hat{\rho}'_{11}] = \sum_{i,j}\bra{i}\bra{j}(\ket{\psi_{11}}\bra{\psi_{11}})\ket{i}\ket{j}
\end{equation}
Considering all the possible terms of $\hat{\rho}'_{11}$, it is possible to distinguish two main contributions:
\begin{equation}
    \begin{split}
       &\sum_{i,j}\braket{i|\phi_1,H}_{a}\braket{j|\phi_2,V}_{b}\braket{\phi_1,H|i}_{a}\braket{\phi_2,V|j}_{b} + \sum_{i,j}\braket{i|\phi_2,V}_{a}\braket{j|\phi_1,H}_{b}\braket{\phi_2,V|i}_{a}\braket{\phi_1,H|j}_{b} =
        \\
        &= \sum_{i,j}\braket{\phi_1|i}_{a}\braket{i|\phi_1}_{a}\braket{\phi_2|j}_{b}\braket{j|\phi_2}_{b}\ket{H}_{a}\ket{V}_{b}\bra{H}_{a}\bra{V}_{b}
        + \sum_{i,j}\braket{\phi_2|i}_{a}\braket{i|\phi_2}_{a}\braket{\phi_1|j}_{b}\braket{j|\phi_1}_{b}\ket{V}_{a}\ket{H}_{b}\bra{V}_{a}\bra{H}_{b} = \\ 
        &= \braket{\phi_1|\phi_1}_{a}\braket{\phi_2|\phi_2}_{b}\ket{H}_{a}\ket{V}_{b}\bra{H}_{a}\bra{V}_{b} + \braket{\phi_2|\phi_2}_{a}\braket{\phi_1|\phi_1}_{b}\ket{V}_{a}\ket{H}_{b}\bra{V}_{a}\bra{H}_{b} = 
        \\ 
        &= \ket{V}_{a}\ket{H}_{b}\bra{V}_{a}\bra{H}_{b} + \ket{H}_{a}\ket{V}_{b}\bra{H}_{a}\bra{V}_{b}
        \\\\
        &\sum_{i,j}\braket{i|\phi_1,H}_{a}\braket{j|\phi_2,V}_{b}\braket{\phi_2,V|i}_{a}\braket{\phi_1,H|j}_{b} + \sum_{i,j}\braket{i|\phi_2,V}_{a}\braket{j|\phi_1,H}_{b}\braket{\phi_1,H|i}_{a}\braket{\phi_2,V|j}_{b} =
        \\
        &= \sum_{i,j}\braket{\phi_2|i}_{a}\braket{i|\phi_1}_{a}\braket{\phi_1|j}_{b}\braket{j|\phi_2}_{b}\ket{H}_{a}\ket{V}_{b}\bra{V}_{a}\bra{H}_{b} + \sum_{i,j}\braket{\phi_1|i}_{a}\braket{i|\phi_2}_{a}\braket{\phi_2|j}_{b}\braket{j|\phi_1}_{b}\ket{V}_{a}\ket{H}_{b}\bra{H}_{a}\bra{V}_{b} =
        \\
        &= \braket{\phi_1|\phi_2}_{a}\braket{\phi_2|\phi_1}_{b}\ket{V}_{a}\ket{H}_{b}\bra{H}_{a}\bra{V}_{b} + \braket{\phi_1|\phi_2}_{a}\braket{\phi_2|\phi_1}_{b}\ket{V}_{a}\ket{H}_{b}\bra{H}_{a}\bra{V}_{b} =
        \\
        &= v (\ket{V}_{a}\ket{H}_{b}\bra{H}_{a}\bra{V}_{b} + \ket{V}_{a}\ket{H}_{b}\bra{H}_{a}\bra{V}_{b}) \\
    \end{split}
\end{equation}
where $|\braket{\phi_i |\phi_i}|=1$ (with $i=1,2$) and we have defined $v=| \braket{\phi_1 |\phi_2} |^2$. As discussed in the main text, two signal photons are characterized by a partial distinguishability, which is expressed in terms of HOM visibility by $v$ (computed considering the $g^{(2)}(0)$ contribution), that is the overlap between the two consecutive photons. Therefore, the final state is:
\begin{equation*}\label{eq:rho} 
\hat{\rho}_{11}^{\text{LA}} = r_{1}^{2}t^{2}_{1}\bigl[t^{4}_{2}\ket{V}_{a}\bra{V}_{a}\otimes\ket{H}_{b}\bra{H}_{b} +r^{4}_{2}\ket{H}_{a}\bra{H}_{a}\otimes\ket{V}_{b}\bra{V}_{b} -vt^{2}_{2}r^{2}_{2}(e^{i\chi}\ket{V}_{a}\bra{H}_{a}\otimes\ket{H}_{b}\bra{V}_{b}+e^{-i\chi}\ket{H}_{a}\bra{V}_{a}\otimes\ket{V}_{b}\bra{H}_{b})     \bigr]=
\end{equation*}
\begin{equation}\label{eq:rho_true}
\begin{split}
&=r^{2}_{1}t^{2}_{1} 
\begin{pmatrix}
0 & 0 & 0 & 0 \\
0 & r^{4}_{2} & -ve^{-i\chi}t^{2}_{2}r^{2}_{2} & 0 \\
0 & -ve^{i\chi}t^{2}_{2}r^{2}_{2} & t^{4}_{2} & 0 \\
0 & 0 & 0 & 0 \\
\end{pmatrix}  \\
\hat{\rho}_{11}^{\text{RF}} &= q^2 \hat{\rho}_{11}^{\text{LA}}    
\end{split}
\end{equation}
where $e^{i\chi}$ is the phase added by the linear crystal to an H-polarized photon going through arm b of the second BS. 
For ideal parameters (i.e. $r_1=t_1, r_2=t_2, v=1,\chi=0$) Eq.\eqref{eq:rho_true} corresponds to the singlet Bell state except from a normalization factor.
\subsection{Interference with Noise Photons}

To determine all the possible noise sources that affect the generation of the state in  Eq.~\eqref{eq:model}, we have to consider the possibility of not emitting a single photon and the presence of residual photons from the laser pump.  
The different combinations of noise photons and signal photons give rise to different terms depending on the particular configuration and their number in each pulse; such cases are shown in Fig.~\ref{fig:combinations}(b,c,d). Dividing the cases in function of the probability $p_{2}$, the first order correction consists of having three photons in total or just two photons in case of no emission. 
\subsubsection*{First order correction }
Here, one pulse has two photons, signal ($\phi_1$) and noise ($\gamma$), while a second one has only the single-photon signal ($\phi_2$). Note that the noise photon derives from laser emission and it is totally distinguishable by the signal photon \cite{ollivier2021hong}, that is $| \braket{\phi_i |\gamma} |=0$ (with $i=1,2$). This corresponds to a three-photon initial state given by:
\begin{equation}
\label{eq:threestate} \begin{split}
    &\text{LA}: \qquad \hat{a}^{\dagger}_{\phi_2,H}(t+\tau) \hat{a}_{\phi_1,H}^{\dagger}(t) \hat{a}_{\gamma,H}^{\dagger}(t) \ket{0}_{a}\ket{0}_{b} 
    \\
    &\text{RF}: \qquad  \bigl[ \sqrt{1-q} + e^{i \theta_{q}}\sqrt{q}\hat{a}^{\dagger}_{\phi_2,H}(t+\tau)\bigr] \bigl[ \sqrt{1-q} + e^{i \theta_{q}}\sqrt{q}\hat{a}^{\dagger}_{\phi_1,H}(t)\bigr] \hat{a}_{\gamma,H}^{\dagger}(t) \ket{0}_{a}\ket{0}_{b}
\end{split}
\end{equation}
In the RF scheme, we assume that the noise photon ($\gamma$) is not emitted in superposition with the vacuum. As aforementioned, we consider the three-photon terms negligible ($O(\eta^3)$). Therefore, two-photon coincidences derive from this case only when two of three photons survive. The contributions to the final density matrix depend on which photon is lost during the transmission, thus identifying three terms corresponding to the combinations reported in Fig.~\ref{fig:combinations} (b). The probability of each event is proportional to $p_{1}^{2}\eta^{2}(1-\eta)$ without considering the normalization factor. Finally, residual laser photons can also give rise to a coincidence when it is emitted alongside a zero-photon emission Fig.~\ref{fig:combinations}(c), that is the term $\hat{\rho}_{02}$ of Eq.~\eqref{eq:model}. In this case, the probability is proportional to $p_{0}p_{2}\eta^{2}$. All these contributions will be analyzed in the following.

\textbf{Two-photon interference from the same pulse}. When the photon in Eq.~\eqref{eq:threestate} corresponding to $\hat{a}^{\dagger}_{\phi_2,H}(t+\tau)$ is lost, the initial state before the interferometer is:
\begin{equation}\begin{split}
    &\text{LA}: \qquad \hat{a}_{\phi_1,H}^{\dagger}(t)\hat{a}_{\gamma,H}^{\dagger}(t)\ket{0}_{a}\ket{0}_{b}
    \\ \\
    &\text{RF}: \qquad \bigl( \sqrt{1-q} + e^{i \theta_{q}} \sqrt{q_{1}}\hat{a}^{\dagger}_{\phi_1,H}\bigr)(t) \hat{a}_{\gamma,H}^{\dagger}(t)\ket{0}_{a}\ket{0}_{b}
\end{split}\end{equation}
Notably, this is also the case of $\hat{\rho}_{02}$. Application of Eq.~\eqref{eq:u} and post-selecting the states in which there are coincidence events with zero-time delay, 
provide the superposition $\ket{\psi_{12}^{(1)}}$:
\begin{equation}\begin{split}
    \ket{\psi_{12}^{(1)}}^{\text{LA}} = ir^{2}_{1}t_{2}r_{2}\bigl(\hat{a}_{\phi_1,H}^{\dagger}(t+\tau)\hat{b}_{\gamma,H}^{\dagger}(t+\tau)+&\hat{b}_{\phi_1,H}^{\dagger}(t+\tau)\hat{a}_{\gamma,H}^{\dagger}(t+\tau) \bigr)\ket{0}_{a}\ket{0}_{b} \\
    +   it^{2}_{1}t_{2}r_{2}\bigl(\hat{a}_{\phi_1,V}^{\dagger}(t)\hat{b}_{\gamma,V}^{\dagger}(t)+&\hat{b}_{\phi_1,V}^{\dagger}(t)\hat{a}_{\gamma,V}^{\dagger}(t) \bigr)\ket{0}_{a}\ket{0}_{b}
    \\ \\
    \ket{\psi_{12}^{(1)}}^{\text{RF}} = \sqrt{q} \ket{\psi_{12}^{(1)}}^{\text{LA}}
\end{split}\end{equation}
In this case, unlike Eq.~\eqref{eq:trace}, for computing the final density matrix the detection not only requires the trace over internal degrees of freedom but also detectors cannot distinguish if the coincidence occurs at time $t$ or $t+\tau$. Therefore, the final state is given by:
\begin{equation}
    \hat{\rho}_{12}^{(1)} = \bra{t}\mathrm{Tr}_{\phi_1,\gamma}[\ket{\psi_{12}^{(1)}}\bra{\psi_{12}^{(1)}}]\ket{t}
    +\bra{t+\tau}\mathrm{Tr}_{\phi_1,\gamma}[\ket{\psi_{12}^{(1)}}\bra{\psi_{12}^{(1)}}]\ket{t+\tau}
\end{equation}
which provides:
    \begin{equation}\label{eq:rho02} 
    \begin{split}
        \hat{\rho}_{12}^{(1)}{}^{\text{LA}} &= 2t^{2}_{2}r^{2}_{2}(r^{4}_{1}\ket{H}_{a}\ket{H}_{b}\bra{H}_{a}\bra{H}_{b}+t^{4}_{1}\ket{V}_{a}\ket{V}_{b}\bra{V}_{a}\bra{V}_{b}) = 2t^{2}_{2}r^{2}_{2} \begin{pmatrix}
    r^{4}_{1} & 0 & 0 & 0 \\
    0 & 0 & 0 & 0 \\
    0 & 0 & 0 & 0 \\
    0 & 0 & 0 & t^{4}_{1} \\
    \end{pmatrix} \\ \\
    \hat{\rho}_{12}^{(1)}{}^{\text{RF}} & = q \hat{\rho}_{12}^{(1)}{}^{\text{LA}}
    \end{split}
    \end{equation}
Notice that this contribution to the final state has the same density matrix of the $\hat{\rho}_{02}$, so we can write $\hat{\rho}_{02}^{\text{LA}}=\hat{\rho}_{12}^{(1)}{}^{\text{LA}}$ and $\hat{\rho}_{02}^{\text{RF}}=\hat{\rho}_{12}^{(1)}{}^{\text{RF}}$.

\textbf{Interference between signal and noise photon}. When one of the two photons inside the multiphoton pulse emission is lost, it gives rise to two possible contributions: 
    The photon $\hat{a}^{\dagger}_{\phi_1,H}(t)$ of Eq.~\eqref{eq:threestate} is lost. The initial state before the interferometer is:
\begin{equation}\begin{split}
    &\text{LA}: \qquad \hat{a}_{\phi_2,H}^{\dagger}(t+\tau)\hat{a}_{\gamma,H}^{\dagger}(t)\ket{0}_{a}\ket{0}_{b} 
    \\ \\
    &\text{RF}: \qquad \bigl[ \sqrt{1-q} + e^{i \theta_{q}}\sqrt{q}\hat{a}^{\dagger}_{\phi_2,H}(t+\tau)\bigr] \hat{a}_{\gamma,H}^{\dagger}(t)\ket{0}_{a}\ket{0}_{b}
\end{split}\end{equation}

This state correspond to Eq.~\eqref{eq:rho} having overlap zero, i.e. $v=|\braket{\phi_2 |\gamma}|=0$ . Therefore, the final density matrix reads:
\begin{equation} \label{ideal_state}
\begin{split}
    \hat{\rho}_{12}^{(2)}{}^{\text{LA}} &= r_{1}^{2}t^{2}_{1}\bigl[t^{4}_{2}\ket{V}_{a}\bra{V}_{a}\otimes\ket{H}_{b}\bra{H}_{b} +r^{4}_{2}\ket{H}_{a}\bra{H}_{a}\otimes\ket{V}_{b}\bra{V}_{b} \bigr] = r^{2}_{1}t^{2}_{1}
\begin{pmatrix}
0 & 0 & 0 & 0 \\
0 & r^{4}_{2} & 0 & 0 \\
0 & 0 & t^{4}_{2} & 0 \\
0 & 0 & 0 & 0 \\
\end{pmatrix}\\   \\
\hat{\rho}_{12}^{(2)}{}^{\text{RF}} &= q\hat{\rho}_{12}^{(2)}{}^{\text{LA}} \end{split}\end{equation}

\textbf{Interference between two signal photons}. The photon $\hat{a}^{\dagger}_{\gamma,H}(t)$ of Eq.~\eqref{eq:threestate} is lost; the initial state is the same of Eq.~\eqref{eq:statein}, thus $\hat{\rho}_{12}^{(3)}{}^{\text{LA}} =  \hat{\rho}_{11}^{\text{LA}}$ and $\hat{\rho}_{12}^{(3)}{}^{\text{RF}} =  \hat{\rho}_{11}^{\text{RF}}$. 

\subsubsection*{Second order correction }
In this case, both pulses have two photons each, one signal photon ($\phi_{1,2}$) and one noise photon ($\gamma_{1,2}$). If the photon derives from laser emission and signals are again totally distinguishable, $| \braket{\phi_i |\gamma_{i}} |=0$ (with $i,j=1,2$). The overlap between laser photon may be different from zero $|\braket{\gamma_{1}|\gamma_{2}}|=v_{l}$; its theoretical value for the same laser is equal to $1/2$ (\cite{HOM_laser}). The initial state of this case corresponds to:
\begin{equation}\label{eq:fourstate} \begin{split}
    &\text{LA}: \qquad \hat{a}^{\dagger}_{\phi_2,H}(t+\tau) 
    \hat{a}^{\dagger}_{\gamma_2,H}(t+\tau)
    \hat{a}_{\phi_1,H}^{\dagger}(t) \hat{a}_{\gamma_1,H}^{\dagger}(t) \ket{0}_{a}\ket{0}_{b} 
    \\
    &\text{RF}: \qquad  \bigl[ \sqrt{1-q} + e^{i \theta_{q}}\sqrt{q}\hat{a}^{\dagger}_{\phi_2,H}(t+\tau)\big] 
    \hat{a}^{\dagger}_{\gamma_2,H}(t+\tau)
    \bigl[ \sqrt{1-q} + e^{i \theta_{q}}\sqrt{q}\hat{a}^{\dagger}_{\phi_1,H}(t)\bigr] \hat{a}_{\gamma_{1},H}^{\dagger}(t) \ket{0}_{a}\ket{0}_{b}.
\end{split}
\end{equation}
Again, we consider the terms containing only two photons thus corresponding to the cases shown in Fig. \ref{fig:combinations}(d); the probability of every event is proportional to $p_{2}^{2}\eta^{2}(1-\eta)^{2}$.
\begin{itemize}
    \item Two photons in the same pulse are lost: this case has the same density matrix as $\hat{\rho}_{20}$, so that $\hat{\rho}_{22}^{(1)\text{LA}} = \hat{\rho}_{20}^{\text{LA}}$ and $\hat{\rho}_{22}^{(1)\text{RF}} = \hat{\rho}_{20}^{\text{RF}}$.
    \item Both the noise photons are lost: the case corresponds to the correct generation term: $\hat{\rho}_{22}^{(2)\text{LA}} = \hat{\rho}_{11}^{\text{LA}}$ and $\hat{\rho}_{22}^{(2)\text{RF}} = \hat{\rho}_{11}^{\text{RF}}$.
    \item one signal photon and one noise photon are lost: again, the case is analogous to $\rho_{12}^{(2)}$: $\hat{\rho}_{22}^{(3)\text{LA}} = \hat{\rho}_{12}^{(2)\text{LA}}$ and $\hat{\rho}_{22}^{(3)\text{RF}} = \hat{\rho}_{12}^{(2)\text{RF}}$.
    \item Two signal photons are lost: in this case two noise photons interact with a degree of indistinguishability given by $v_{l}$; the matrix has the same form of the correct generation, with a different overlap term:
    \begin{equation}\label{eq:rho} 
\begin{split}
    \hat{\rho}_{22}^{(4)\text{LA}} &= r_{1}^{2}t_{1}^{2}
\begin{pmatrix}
0 & 0 & 0 & 0 \\
0 & r^{4}_{2} & -v_{l}e^{-i\chi}t^{2}_{2}r^{2}_{2} & 0 \\
0 & -v_{l}e^{i\chi}t^{2}_{2}r^{2}_{2} & t^{4}_{2} & 0 \\
0 & 0 & 0 & 0 \\
\end{pmatrix} \\ \\
\hat{\rho}_{22}^{(4)\text{RF}} &=\hat{\rho}_{22}^{(4)\text{LA}}
\end{split}
\end{equation}
\end{itemize}
\subsection{Analysis of the final state}
According to the model in Eq.~\eqref{eq:model}, the unnormalized density matrix can be rewritten by grouping together the terms having the same density matrix: 
\begin{equation}
    \hat{\rho}_{\text{exp}} \propto c_{11} \hat{\rho}_{11} + c_{12} \hat{\rho}^{(2)}_{12} + c_{02} \hat{\rho}_{02} + c_{l} \hat{\rho}_{l}
\end{equation}
where the coefficients are given by:
\begin{equation} \label{eq:final_no_norm}
    \begin{split}
    c_{11} &= p_{1}^{2}\eta^{2} + p_{1}p_{2}\eta^{2}(1-\eta) + p_{2}^{2}\eta^{2}(1-\eta)^{2} \\
    c_{12} &= p_{1}p_{2}\eta^{2}(1-\eta) + p_{2}^{2}\eta^{2}(1-\eta)^{2} \\
    c_{02} &= p_{0}p_{2}\eta^{2} + p_{1}p_{2}\eta^{2}(1-\eta) + p_{2}^{2}\eta^{2}(1-\eta)^{2} \\
    c_{l} &= p_{2}^{2}\eta^{2}(1-\eta)^{2}\\
    \end{split}
\end{equation}
 The only difference between the LA and RF case in our model is the presence of $q$. Therefore, considering $q$ in the range $[0,1]$ represents a general model valid for both excitation schemes, where LA is recovered by setting $q=1$. The density matrix considered for the theoretical prediction is the normalized version of \eqref{eq:final_no_norm}, specifically $\hat{\rho}_{exp} \longrightarrow \hat{\rho}_{exp}/Tr[\hat{\rho}_{exp}]$:
\begin{equation}\label{eq:rhofin}\begin{split}
\hat{\rho}_{\text{exp}} &=\frac{1}{\mathcal{N}}(c_{11} \hat{\rho}_{11} + c_{12} \hat{\rho}^{(2)}_{12} + c_{02} \hat{\rho}_{02} + c_{l} \hat{\rho}_{l}) = \frac{p_{1}^{2}\eta^{2} + p_{1}p_{2}\eta^{2}(1-\eta) + p_{2}^{2}\eta^{2}(1-\eta)^{2}}{\mathcal{N}}r_{1}^{2}t_{1}^{2}q^2
\begin{pmatrix}
0 & 0 & 0 & 0 \\
0 & r^{4}_{2} & -vt^{2}_{2}r^{2}_{2}e^{-i\chi} & 0 \\
0 & -vt^{2}_{2}r^{2}_{2}e^{i\chi} & t^{4}_{2} & 0 \\
0 & 0 & 0 & 0 \\
\end{pmatrix} + \\
&+\frac{p_{1}p_{2}\eta^{2}(1-\eta) + p_{2}^{2}\eta^{2}(1-\eta)^{2}}{\mathcal{N}}r^{2}_{1}t^{2}_{1}q
\begin{pmatrix}
0 & 0 & 0 & 0 \\
0 & r^{4}_{2} & 0 & 0 \\
0 & 0 & t^{4}_{2} & 0 \\
0 & 0 & 0 & 0 \\
\end{pmatrix} + \frac{p_{0}p_{2}\eta^{2} + p_{1}p_{2}\eta^{2}(1-\eta) + p_{2}^{2}\eta^{2}(1-\eta)^{2}}{\mathcal{N}}2t^{2}_{2}r^{2}_{2}q
\begin{pmatrix}
    r^{4}_{1} & 0 & 0 & 0 \\
    0 & 0 & 0 & 0 \\
    0 & 0 & 0 & 0 \\
    0 & 0 & 0 & t^{4}_{1} \\
    \end{pmatrix} + \\
    & + \frac{p_{2}^{2}\eta^{2}(1-\eta)^{2}}{\mathcal{N}}r_{1}^{2}t_{1}^{2}
\begin{pmatrix}
0 & 0 & 0 & 0 \\
0 & r^{4}_{2} & -v_{l}t^{2}_{2}r^{2}_{2}e^{-i\chi} & 0 \\
0 & -v_{l}t^{2}_{2}r^{2}_{2}e^{i\chi} & t^{4}_{2} & 0 \\
0 & 0 & 0 & 0 \\
\end{pmatrix}
\end{split}
\end{equation}
where $\mathcal{N}$ is the normalization constant (setting the ideal case $t_{1}=r_{1}=t_{2}=r_{2}=1/\sqrt{2}$):
\begin{equation}
    \mathcal{N} = \frac{\eta^{2}}{8}\bigl[q^{2}p_{1}^{2} + qp_{2}(2p_{0}+p_{1}(1-\eta)(3+q)) + p_{2}^{2}(1-\eta)^{2}(1+q(3+q))
    \bigr]
    \end{equation}
As discussed in the main text, a Bell violation certifies the presence of entanglement anytime the violation is higher than 2 with a maximum value of $S=2\sqrt{2}$ for the ideal singlet Bell state. When our apparatus is set to prepare for such a state, we perform the measures $\hat{A}_0=\hat{\sigma}_z\ $, $\hat{A}_1=\hat{\sigma}_x$ and $\hat{B}_0=(\hat{\sigma}_z+\hat{\sigma}_x)/\sqrt{2}$, $\hat{B}_1=(\hat{\sigma}_z-\hat{\sigma}_x)/\sqrt{2}$, in order to achieve the maximal violation of the ideal state. However, the presence of the experimental noise and losses unavoidably deteriorates and limits the maximum attainable value of $S$. This new limit can be calculated by exploiting the model of equation Eq.~\eqref{eq:rhofin}, that is computing  $S(\hat{\rho})$ with the actual correlators $E_{\rho}(A_i,B_j)=Tr[\hat{\rho} \hat{A}_i \otimes \hat{B}_j]$ (with $i,j=0,1$) on the modelled states. In this way, we obtain:
\begin{equation}\begin{split}\label{eq:models}
S(\hat{\rho}_{\text{exp}})&=\sqrt{2} \left| \frac{(1+v)p_{1}^{2}q^{2} - p_{2}q(2p_{0}+p_{1}(1-\eta)(1-q(1+v)))+p_{2}^{2}(1-\eta)^{2}(1+v_{l}-q(1-q(1+v)))}{p_{1}^{2}q^{2}+p_{2}q(2p_{0}+p_{1}(1-\eta)(3+q))+p_{2}^{2}(1-\eta)^{2}(1+q(3+q))} \right|
\end{split}\end{equation}
The same kind of study can be replicated for quantum state tomography. In particular, the fidelity of the quantum state $\hat{\rho}_1$ with respect to a second state $\hat{\rho}_2$ is computed by $\mathcal{F}(\hat{\rho}_1,\hat{\rho}_2) = Tr[\sqrt{\sqrt{\hat{\rho}_1} \hat{\rho}_2 \sqrt{\hat{\rho}_1}}]$. Thus, the best fidelity achievable between the modelled states with respect to the ideal singlet state $\hat{\rho}_{\psi^-}$ is:
\begin{equation}\label{eq:modelf}
    \mathcal{F}(\hat{\rho}_{\text{exp}},\hat{\rho}_{\psi^-}) = \frac{(1+v)p_{1}^{2}q^{2} +p_{2}q(2p_{0}-p_{1}(1-\eta)(1-q(1+v)))+p_{2}^{2}(1-\eta^{2})(1+v_{l}+q+q^{2}(1+v))}
    {p_{1}^{2}q^{2}+p_{2}q(2p_{0}+p_{1}(1-\eta)(3+q))+p_{2}^{2}(1-\eta)^{2}(1+q(3+q))} . \\
\end{equation}
These relations, i.e. Eqs.~\eqref{eq:models},~\eqref{eq:modelf}, can be used to identify the main experimental parameters that most limit the best performance obtainable from our source. 
The results show as the main variations are due to two parameters: the HOM visibility $v$ and the presence of noise photons quantified by $g^{(2)}(0)$. The other parameters show no significant contribution when varied around their values. In particular, looking at parameter $q$, the model shows no substantial difference between adopting LA or RF excitation schemes; even the balance of the final beam-splitter turns out to be irrelevant.
 To take into account any other source of noise or experimental imperfection, we consider the Werner state:
 \begin{equation}
     \hat{\rho}_{wn} = c_{wn}\hat{\rho}_{exp} + \frac{1-c_{wn}}{4}\mathbf{I}
 \end{equation}
 where the coefficient $c_{wn}$ determines the fraction of white noise in the experimental state. If the Bell's Parameter can be easily computed from \eqref{eq:models} with $S(\hat{\rho}_{wn})=c_{wn}S(\hat{\rho}_{exp})$, the fidelity shows a more complex relation: 
 \begin{equation}\begin{split}
     \mathcal{F}_{wn}(\hat{\rho}_{wn},\hat{\rho}_{\psi^-}) &= \frac{1}
    {4[p_{1}^{2}q^{2}+p_{2}q[2p_{0}+p_{1}(1-\eta)(3+q)]+p_{2}^{2}(1-\eta)^{2}(1+q(3+q))]} \times \\
    & \times \bigl[p_{1}^{2}q^{2}(1+c_{wn}(1+2v))+ p_{2}^{2}(1-\eta)^{2}[1+c_{wn}(1+2v_{l})+q(3+q)-c_{wn}q(1-q(1+2v))] +  \\
    &+ p_{2}q[2p_{0}(1-c_{wn})+p_{1}(1-\eta)(3+q-c_{wn}(1-q(1-2v))]\bigr] .
    \end{split}
 \end{equation}
On the one hand, this study allows reasonable research for parameter values that make the model compatible with experimental data. On the other hand, it allows us to define the best attainable entangled state in our experimental setup. We note that variations of considered parameters can occur in the experimental setup for instability of employed components or any error during their estimation procedure.

\section{Estimation of concurrence and maximal Bell violation}

As pointed out in the main text, throughout the experiment we periodically evaluated the density matrix of the experimentally generated state by performing quantum state tomography (QST) \cite{altepeter2005photonic}. On a two-qubit polarization encoded state QST is performed by evaluating the frequency two-photon coincidence events over the eigenvectors of the set of 9 operators:
\begin{equation}
    \hat{A} \otimes \hat{B} = \sigma_i \otimes \sigma_j; \qquad \forall \sigma_i,\sigma_j \in \{\sigma_x, \sigma_y, \sigma_z\}
\end{equation}
Note that a direct estimation of the density matrix elements from the experimentally determined frequencies could lead to a non-physical result due to the presence of experimental noise. To avoid this behaviour, we employed maximum likelihood techniques as in ref.\cite{hradil20043} to reconstruct physically meaningful density matrices from the observed data.

From the experimentally evaluated density matrices, we can compute directly a meaningful metric which quantifies the entanglement of bipartite density matrices describing the mixed states of a two-qubit system: the \emph{concurrence} \cite{horodecki1995violating,hildebrand2007concurrence}. Such quantity is an \emph{entanglement monotone}, i.e. a non-negative function $\mathcal{C}(\rho) \rightarrow \mathbb{R}^+$ related to the degree of entanglement of the density matrix $\rho$. Formally, the concurrence is defined as:

\begin{equation}
    \mathcal{C}(\rho) = \max \{0, \lambda_1 - \lambda_2 - \lambda_3 - \lambda_4 \} \in [0,1]
\end{equation}
where $\lambda_1 \geq \lambda_2 \geq \lambda_3 \geq  \lambda_4$ are the eigenvalues in decreasing order of the Hermitian matrix obtained as:
\begin{equation}
    R = \sqrt{\sqrt{\rho} \Tilde{\rho} \sqrt{\rho}} \qquad  \Tilde{\rho} = (\sigma_y \otimes \sigma_y) \rho^* (\sigma_y \otimes \sigma_y)
\end{equation}
$\mathcal{C}(\rho)$ as an entanglement monotone is then equal to 1 for maximally entangled states, such as the usual Bell states, while it is equal to 0 for fully separable states. On our generated states, we computed a mean concurrence of $0.819 \pm 0.008$ ($0.85 \pm 0.05$) under LA (RF) excitation schemes, that is in both cases the experimentally generated states exhibit a high degree of entanglement.

\begin{figure}[h]
    \centering
    \includegraphics[width=.99\textwidth]{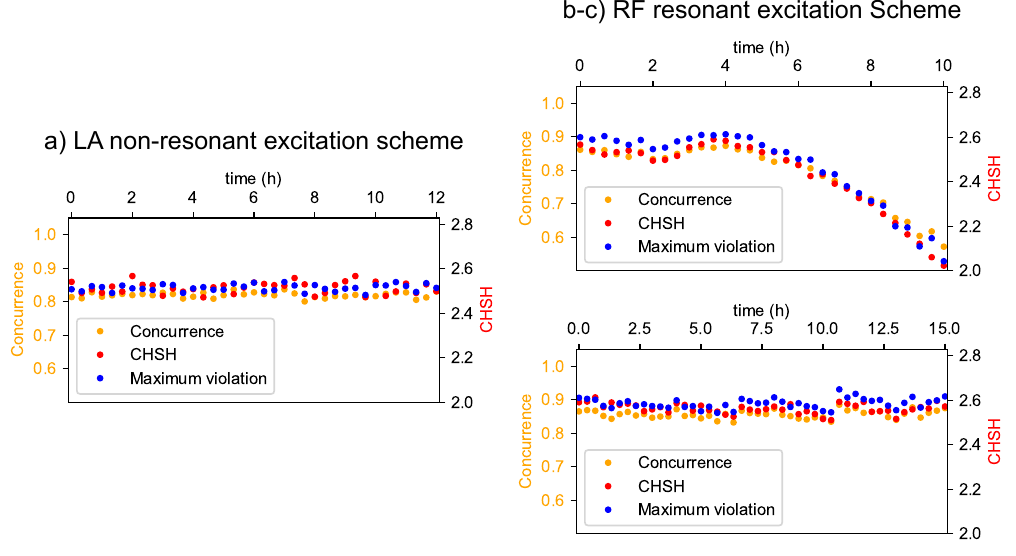}
    \caption{In these plots, we report the concurrence computed from the experimentally evaluated density matrices, through quantum state tomography measurements, together with the maximal Bell-CHSH violation computed through the Horodecki criterion. Panel a) refers to Figure 5 of the main text, while panels b-c) refer to Figure 6 of the main text.}
    \label{fig:concurrence}
\end{figure}

Moreover, from the experimentally evaluated density matrices, we can also evaluate the maximally reachable violation of the Bell-CHSH inequality by means of the so-called Horodecki criterion \cite{horodecki1995violating}. Formally, the Horodecki criterion states that given a two-qubit density matrix $\rho$, one can construct a matrix $T(\rho)$ such that:
\begin{equation}
    T_{ij}(\rho) = \mathrm{Tr}{\rho \sigma_i \sigma_j}; \qquad \sigma_i, \sigma_j \in \{\sigma_x, \sigma_y, \sigma_z\}
\end{equation}
and compute the maxima of the Bell-CHSH quantity as:
\begin{equation}
    B^{\textrm{max}}_\rho = 2\sqrt{\lambda_1^2 + \lambda_2^2}
\end{equation}
where $(\lambda_1, \lambda_2)$ are the highest eigenvalues of the matrix $T_\rho \cdot T^T_\rho$, where $T^T_\rho$ denotes the transpose of $T_\rho$.
The behaviour of the state concurrence and maximally achievable Bell-CHSH violation, spanning the same timeframe reported in the main text and for both the phonon-assisted (LA) and resonant (RF) excitation schemes, are reported in Fig.~\ref{fig:concurrence}.

%